\documentclass{article} 
\usepackage{amsmath,amssymb,amsthm,graphicx}
\usepackage{yhmath}
\usepackage{enumerate}

\newcommand{\E}{\mathcal{E}}

\newcommand{\G}{\mathcal{G}}
\newcommand{\Gs}{\mathcal{G}^\star}
\newcommand{\I}{\mathcal{I}}
\renewcommand{\L}{\mathcal{L}}

\renewcommand{\P}{\mathcal{P}}
\newcommand{\T}{\mathcal{T}}
\newcommand{\V}{\mathcal{V}}

\newcommand{\x}{\>x}

\newcommand{\xx}{\>x}

\newcommand{\h}{{\bf h}}
\newcommand{\J} {{\bf J}}
\newcommand{\K} {{\bf J}^{Bethe}}
\newcommand{\m} {{\bf m}}
\newcommand{\hm}{\hat m}
\newcommand{\hp}{\hat p}
\newcommand{\hC}{{\hat \chi}}
\newcommand{\ExB}{\mathbb{E}_{Bethe}}
\newcommand{\hP}{{\hat P}}

\newcommand{\bh} {{\bf h}}
\newcommand{\bJ} {{\bf J}}
\newcommand{\bm} {{\bf m}}

\makeatletter

\@addtoreset{equation}{section}

\@addtoreset{figure}{section}
\makeatother

\newcommand\egaldef{\stackrel{\mbox{\upshape\tiny def}}{=}}

\newcommand\1{\leavevmode\hbox{\rm \small1\kern-0.35em\normalsize1}}
\newcommand\ind[1]{\1_{\{#1\}}}

\def\DD{\displaystyle}

\DeclareMathOperator{\atanh}{atanh}

\DeclareMathOperator*{\argmin}{argmin}

\DeclareMathOperator{\Tr}{Tr}
\DeclareMathOperator{\var}{Var}
\DeclareMathOperator{\cov}{Cov}
\DeclareMathOperator{\diag}{Diag}
\newcommand\pth[1]{\wideparen{#1}}

\title{Pairwise MRF Calibration by Perturbation of the Bethe Reference Point}

\author{Cyril Furtlehner\and Yufei Han\and Jean-Marc Lasgouttes\and Victorin Martin}

\begin{document}

\maketitle

\begin{abstract}
We investigate different ways of generating approximate solutions to the
pairwise Markov random field (MRF) selection problem. We focus mainly on the inverse Ising problem, but discuss also
the somewhat related inverse Gaussian problem because both types of MRF are suitable 
for inference tasks with the belief propagation algorithm (BP) under certain conditions.
Our approach consists in to take a Bethe mean-field solution obtained with 
a maximum spanning tree (MST) of pairwise mutual information,
referred to as the \emph{Bethe reference point}, for further perturbation procedures. 
We consider three different ways 
following this idea: in the first one, we select and calibrate iteratively the optimal 
links to be added starting from  the Bethe reference point; 
the second one is based on the observation that the natural gradient can be computed analytically
at the Bethe point; in the third one, assuming no local field and using low temperature expansion
we develop a dual loop joint model based on a well chosen fundamental cycle basis. 
We indeed  identify a subclass of planar models, which we refer to as \emph{Bethe-dual graph models},  having possibly many loops, 
but characterized by a singly connected dual factor graph, 
for which the partition function and the linear response  can be computed exactly in respectively O(N)
and $O(N^2)$ operations,
thanks to a dual weight propagation (DWP) message passing procedure that we set up. 
When restricted to this subclass of models,
the inverse Ising problem being convex, becomes tractable at any temperature. 
Experimental tests on various datasets  with refined $L_0$ or $L_1$ regularization procedures 
indicate that these approaches may be  competitive and useful alternatives to existing ones.
\end{abstract}

\tableofcontents

\section{Introduction}
The problem at stake is a model selection problem, in the MRF families, 
where $N$ variables are observed pair by pair. The optimal solution 
is the MRF with maximal entropy obeying moment constraints. For binary variables,
this happens then to be 
the Ising model with highest log-likelihood. It is a difficult problem, where both 
the graph structure and the values of the fields and couplings have to be found. In addition, 
we wish to ensure that the model is compatible with the fast inference 
algorithm ``belief propagation'' (BP) to be useful at large scale for real-time inference tasks.
This leads us to look  at least for a good trade-off between likelihood and sparsity.

Concerning the Inverse Ising Problem (IIP), the existing approaches fall mainly in the following categories:
\begin{itemize}
\item Purely computational efficient approaches rely on various optimization schemes 
of the log likelihood~\cite{LeGaKo} or on pseudo-likelihood~\cite{HoTi} along with 
sparsity constraints to select the only relevant features.     
\item Common analytical approaches are based on the Plefka expansion~\cite{Plefka} of the Gibbs
free energy by making the assumption that the coupling constants $J_{ij}$ are small. The picture is then 
of a weakly correlated uni-modal probability measure. For example, the recent approach proposed in~\cite{CoMo}
is based on this assumption. 
\item Another possibility is 
to assume that relevant coupling $J_{ij}$ have locally a tree-like structure. The
Bethe approximation~\cite{YeFrWe} can then be used with possibly 
loop corrections. Again this corresponds to having a weakly correlated uni-modal probability measure 
and these kind of approaches are referred to as pseudo-moment matching methods in the 
literature for the reason explained in the previous section. For example
the approaches proposed in~\cite{KaRo,WeTe,MoMe,YaTa} are based on this assumption. 
\item In the case where a multi-modal distribution is expected, then a model with many 
attraction basins is to be found and Hopfield-like models~\cite{Hopfield,CoMoSe} are likely to be more relevant.
To be mentioned also is a recent  mean-field methods~\cite{NgBe2} 
which allows one to find in some simple cases the Ising couplings
of a low temperature model, i.e. displaying multiple probabilistic modes.
\end{itemize}
On the side of inverse Gaussian problem, not surprisingly similar methods have been developed
by explicitly performing $L_0$ and $L_1$ matrix norm penalizations on the inverse covariance matrices, so as to determine 
sparse non-zero couplings in estimated inverse covariance matrices for large-scale statistical 
inference applications~\cite{FrHaTi,HSDR} where direct inversion is not amenable. 
In our context the goal is a bit different. In general cases, the underlying inverse covariance 
matrix is not necessarily sparse. What we aim to find is a good sparse approximation to the exact 
inverse covariance matrix. Furthermore, sparsity constraint is not enough for constructing graph 
structure to be used in conjunction with BP. 
Known sufficient conditions celled  walk-summability~\cite{MaJoWi} (WS) 
are likely to be imposed instead of (or in addition to) the sparsity constraint. To
the best of our knowledge not much  work is taking this point into consideration at the
noticeable exception of~\cite{AnTaHuWi} by restricting the class of learned graph
structures. To complete this overview, let us mention also that some authors proposed
information based structure learning methods~\cite{NeBaSaSh} quite in line 
with some approaches to be discussed in the present paper.

In some preceding work dealing with a road traffic inference application, with large scale and 
real time specifications~\cite{FuLaFo,FuLaAu,FuHaLa}, we have noticed that these methods could not be used blindly without 
drastic adjustment, in particular to be compatible with belief propagation. This led 
us to develop some heuristic models related to the Bethe approximation. 
The present work is an attempt to give a theoretical basis and firmer ground to these heuristics and to 
develop new ones. 

The paper is organized as follows: in Section~\ref{sec:prelim} we review some standard 
statistical physics  approaches to the IIP, mainly based on perturbation expansions, and derive 
some new and useful expressions of  susceptibility coefficients in Section~\ref{sec:cumulant}. In Section~\ref{sec:iip}
we explain what we mean by the \emph{Bethe reference point} and develop an iterative proportional scaling (IPS) 
based method to incrementally, link by link, refine this approximate solution, both for the 
inverse Ising and inverse GMRF problems. Section~\ref{sec:perturbation} explores a second way to refine the 
Bethe reference point, based on the Plefka's expansion and on results of Section~\ref{sec:cumulant}.
A third method, based on strong coupling expansion and leading to a dual weight propagation algorithm (DWP)
is presented in Section~\ref{sec:dwp}. Merits of these methods differs, which makes them complementary to each other.
The first one is particularly useful when the underlying graph structure is not given; the second 
one, by giving explicitly the natural gradient direction, 
can be used to reduce the number of parameters to tune; finally the third one can be fast and exact for given very
sparse structure, assuming in addition no local fields. For sake of comparison, we explain 
in Section~\ref{sec:penal} how to adapt standard $L_0$ and $L_1$ norm-penalized sparsity inducing optimization 
framework for finding relevant approximate solutions to the inverse GMRF problem in our context. Comparison is then made in 
Section~\ref{sec:experiments} with our own IPS based approach, in addition to other numerical 
experiments illustrating performances of the other methods.

\section{Preliminaries}\label{sec:prelim}
\subsection{Inverse Ising problem}
In this section we consider  binary variables ($x_i\in\{0,1\}$), which at our convenience may be 
also written as spin variables $s_i = 2x_i-1\in\{-1,1\}$.
We assume that from a set of historical observations, 
the empirical mean $\hat m_i$  (resp.\ covariance $\hat \chi_{ij}$) is
given for each variable $s_i$ (resp.\ each pair of variable $(s_i,s_j)$).
In this case, from Jayne's maximum entropy principle~\cite{Jaynes}, imposing these 
moments to the joint distribution leads to a model pertaining to the exponential 
family, the so-called Ising model:
\begin{equation}\label{def:ising}
\P({\bf s}) = \frac{1}{Z[\bJ,\bh]}\exp\bigl(\sum_ih_is_i+\sum_{i,j}J_{ij}s_is_j\bigr)
\end{equation}
where the local fields $\bh=\{h_i\}$ and the coupling constants $\bJ=\{J_{ij}\}$ are the Lagrange 
multipliers associated respectively  to mean and covariance constraints when maximizing the entropy of 
$\P$.
They are obtained as minimizers of the dual optimization problem, namely 
\begin{equation}\label{eq:iipb}
({\bh}^\star,{\bJ}^\star) = \argmin_{({\bh},{\bJ})} \L[\bh,\bJ],
\end{equation}
with 
\begin{equation}\label{def:LL}
\L[\bh,\bJ] = \log Z[\bh,\bJ]-\sum_ih_i\hm_i-\sum_{ij}J_{ij}\hat m_{ij}
\end{equation}
the log likelihood. 
This leads to  invert the  linear response equations:
\begin{align}
\frac{\partial \log Z}{\partial h_i}[\bh,\bJ] &= \hm_i\label{eq:cei}\\[0.2cm]
\frac{\partial \log Z}{\partial J_{ij}}[\bh,\bJ] &= \hat m_{ij},\label{eq:ccij}
\end{align}
$\hat m_{ij} = \hm_i\hm_j+\hC_{ij}$ being the empirical expectation of $s_is_j$.
As noted e.g.\ in~\cite{CoMo},
the solution is minimizing the cross entropy, a Kullback-Leibler distance between the 
empirical distribution $\hP$ based on historical data and the Ising model:
\begin{equation}\label{eq:dkl}
D_{KL} [\hat{\cal P}\Vert {\cal P}] = \log Z[\bh,\bJ]-\sum_ih_i\hm_i-\sum_{i<j}J_{ij}\hat m_{ij} - S(\hat{\cal P}).
\end{equation}
The set of equations (\ref{eq:cei},\ref{eq:ccij}) cannot be solved exactly in general
because the computational cost of $Z$ is exponential.
Approximations resorting to various mean field methods can be used  to evaluate
$Z[\bh,\bJ]$.

\paragraph{Plefka's expansion}\label{sec:plefka}
To simplify the problem, it is customary to make use of the Gibbs free energy,
i.e.\ the Legendre transform of the free energy, to impose the individual expectations $\bm = \{\hm_i\}$ for 
each variable:
\[
G[\bm,\bJ] = \bh^{T}(\bm)\bm + F[\bh(\bm),\bJ]
\]
(with $F[\bh,\bJ] \egaldef -\log Z[\bh,\bJ]$, $\bh^{T}\bm$ is the ordinary scalar product) 
where $\bh(\bm)$ depends implicitly on $\bm$ 
through the set of  constraints
\begin{equation}
\frac{\partial F}{\partial h_i} = -m_i.\label{eq:constraints}
\end{equation}
Note that by duality we have
\begin{equation}\label{eq:dual1}
\frac{\partial G}{\partial m_i} = h_i(\bm), 
\end{equation}
and 
\begin{equation}\label{eq:dual2}
\Bigl[\frac{\partial^2 G}{\partial m_i\partial m_j}\Bigr] = \left[\frac{d\h}{d{\bm}}\right]_{ij} 
=\left[\frac{d\m}{d{\bh}}\right]^{-1}_{ij}  
= -\Bigl[\frac{\partial^2 F}{\partial h_i\partial h_j}\Bigr]^{-1} = \bigl[\chi^{-1}\bigr]_{ij}.
\end{equation}
i.e.\ the inverse susceptibility matrix.
Finding a set of $J_{ij}$ satisfying this last relation along with (\ref{eq:dual1}) 
yields a solution to the inverse Ising problem 
since the $m$'s and $\chi$'s are given. Still a way to connect the couplings directly with the covariance matrix
is given by the relation
\begin{equation}\label{eq:eq2}
\frac{\partial G}{\partial J_{ij}} = - m_{ij}.
\end{equation}
The Plefka expansion is used to expand the Gibbs free energy in power of the coupling $J_{ij}$ assumed to be small. 
Multiplying all coupling $J_{ij}$ by $\alpha$ yields the following cluster expansion:
\begin{align}
G[\bm,\alpha\bJ] &= \bh^{T}(\bm,\alpha)\bm + F[\bh(\bm,\alpha),\alpha\bJ]\label{eq:plefka1}\\[0.2cm]
&= G_0[\bm] + \sum_{n=0}^\infty\frac{\alpha^n}{n!}G_n[\bm,\bJ]\label{eq:plefka2}
\end{align}
where each term $G_n$ corresponds to cluster  contributions of size $n$ in the number of links $J_{ij}$ involved, 
and $\bh(\bm,\alpha)$ depends implicitly on $\alpha$ in order to always fulfill (\ref{eq:constraints}).   
This precisely is the Plefka expansion, and each term of the expansion (\ref{eq:plefka2}) can be
obtained by successive derivation of (\ref{eq:plefka1}).  
We have 
\[
G_0[\bm] = \sum_i\frac{1+m_i}{2}\log\frac{1+m_i}{2}+\frac{1-m_i}{2}\log\frac{1-m_i}{2}.
\]
Letting 
\[
H_J \egaldef \sum_{i<j} J_{ij}s_is_j,
\]
using (\ref{eq:constraints}), the two first derivatives of (\ref{eq:plefka1}) w.r.t $\alpha$ read
\begin{align}\label{eq:dG1}
\frac{dG[\bm,\alpha\bJ]}{d\alpha} &= -{\mathbb E}_\alpha\bigl(H_J\bigr), \\
\label{eq:dG2}
\frac{d^2G[\bm,\alpha\bJ]}{d\alpha^2} &= -\var_\alpha\bigl(H_J\bigr)
-\sum_i\frac{dh_i(\bm,\alpha)}{d\alpha}\cov_\alpha\bigl(H_J,s_i\bigr),
\end{align}
where subscript $\alpha$ indicates that expectations, variance and covariance are taken at given $\alpha$. 
To get successive derivatives of $\bh(\bm,\alpha)$
one can  use (\ref{eq:dual1}). Another possibility is to express the fact that $\bm$ is fixed, 
\begin{align*}
\frac{dm_i}{d\alpha} = 0 & = -\frac{d}{d\alpha}\frac{\partial F[\bh(\alpha),\alpha\bJ]}{\partial h_i}\nonumber\\[0.2cm]
& = \sum_{j}h'_j(\alpha)\cov_\alpha(s_i,s_j)+\cov_\alpha(H_J,s_i), 
\end{align*}
giving
\begin{equation}\label{eq:dha}
h'_i(\alpha) = - \sum_j[\chi_\alpha^{-1}]_{ij}\cov_\alpha(H_J,s_j),
\end{equation}
where $\chi_\alpha$ is the susceptibility delivered by the model when $\alpha\ne 0$.
To get the first two  terms in the Plefka expansion, we need to compute these 
quantities at $\alpha=0$:  
\begin{align*}
\var\bigl(H_J\bigr) &= \sum_{i<k,j} J_{ij}J_{jk}m_im_k(1-m_j^2)+\sum_{i<j}J_{ij}^2(1-m_i^2m_j^2),\\[0.2cm]
\cov\bigl(H_J,s_i\bigr) &= \sum_{j}J_{ij}m_j(1-m_i^2),\\[0.2cm]
h'_i(0) &= - \sum_{j} J_{ij}m_j,
\end{align*}
(by convention $J_{ii}=0$ in these sums).
The first and second orders then finally read:
\[
G_1[\bm,\bJ] = -\sum_{i<j}J_{ij}m_im_j,\qquad\qquad G_2[\bm,\bJ] = -\sum_{i<j}J_{ij}^2(1-m_i^2)(1-m_j^2),
\]
and correspond respectively to the mean field and to the TAP approximation. Higher order 
terms have been computed in~\cite{GeYe}.

At this point we are in position to find an  approximate solution to the inverse Ising problem, either 
by inverting equation (\ref{eq:dual2}) or  (\ref{eq:eq2}). To get a solution at a given order $n$
in the coupling, solving  (\ref{eq:eq2}) requires $G$ at order $n+1$, while it is needed at order 
$n$ in (\ref{eq:dual2}).

Taking the expression of $G$ up to second order gives 
\[
\frac{\partial G}{\partial J_{ij}} = -m_im_j-J_{ij}(1-m_i^2)(1-m_j^2),
\]
and (\ref{eq:eq2}) leads directly to the basic mean-field solution:
\begin{equation}\label{eq:JMF}
J_{ij}^{MF} = \frac{\hC_{ij}}{(1-\hm_i^2)(1-\hm_j^2)}.
\end{equation}
At this level of approximation for $G$, using (\ref{eq:dual1}) we also have
\[
h_i = \frac{1}{2}\log\frac{1+m_i}{1-m_i}-\sum_jJ_{ij}m_j+\sum_j J_{ij}^2m_i(1-m_j^2)
\]
which corresponds precisely to the TAP equations. Using now (\ref{eq:dual2}) gives
\[
\frac{\partial h_i}{\partial m_j}= [\chi^{-1}]_{ij} = 
\delta_{ij}\bigl(\frac{1}{1-m_i^2}+\sum_kJ_{ik}^2(1-m_k^2)\bigr) - J_{ij}-2J_{ij}^2m_im_j.  
\]
Ignoring the diagonal terms, the TAP solution is conveniently expressed in terms of the 
inverse empirical susceptibility, 
\begin{equation}\label{eq:JTAP}
J_{ij}^{TAP} = -\frac{2[\hat \chi^{-1}]_{ij}}{1+\sqrt{1-8\hm_i\hm_j[\hat \chi^{-1}]_{ij}}},
\end{equation}
where the branch corresponding to a vanishing coupling in the limit of small correlation 
i.e.\ small $\hat \chi_{ij}$ and $[\hat \chi^{-1}]_{ij}$ for $i\ne j$, has been chosen.

\paragraph{Bethe approximate solution}
When the graph formed by the pairs $(i,j)$ for which the 
correlations $\hat \chi_{ij}$ are given by some observations is a tree, 
the following form of the joint probability 
corresponding to the Bethe approximation:
\begin{equation}\label{eq:ansatz}
{\cal P}(\xx) = \prod_{i<j}\frac{\hp_{ij}(x_i,x_j)}{\hp(x_i)\hp(x_j)}\prod_i\hp_i(x_i), 
\end{equation}
yields actually an exact solution to the inverse problem (\ref{eq:iipb}),
where the $\hp$ are the single and pair variables empirical marginal given by the observations. 
Using the following identity 
\begin{align*}
\log\frac{\hat p_{ij}(s_i,s_j)}{\hat p_i(s_i)\hat p_j(s_j)} &= 
\frac{(1+s_i)(1+s_j)}{2}\log\frac{\hat p_{ij}^{11}}{\hat p_i^1\hat p_j^1}
+\frac{(1+s_i)(1-s_j)}{2}\log\frac{\hat p_{ij}^{10}}{\hat p_i^1\hat p_j^0}\\[0.2cm]
&+\frac{(1-s_i)(1+s_j)}{2}\log\frac{\hat p_{ij}^{01}}{\hat p_i^0\hat p_j^1}
+\frac{(1-s_i)(1-s_j)}{2}\log\frac{\hat p_{ij}^{00}}{\hat p_i^0\hat p_j^0}
\end{align*}
where the following parametrization of the $\hat p$'s
\begin{align}
\hp_i^{x} &\egaldef \hp\bigl(\frac{1+s_i}{2} = x\bigr) = \frac{1}{2}(1+\hat m_i(2x-1)),\label{eq:pi}\\[0.2cm]
\hp_{ij}^{xy} &\egaldef \hp\bigl(\frac{1+s_i}{2}=x,\frac{1+s_j}{2}=y\bigr)\nonumber\\[0.2cm] 
&= \frac{1}{4}(1+\hat m_i(2x-1)+\hat m_j(2y-1)+\hat m_{ij}(2x-1)(2y-1)\label{eq:pij}
\end{align}
relating the empirical frequency statistics to the empirical  ``magnetizations'' $m\equiv \hat m$,
can be used.
Summing up the different terms gives us the mapping onto an Ising model
(\ref{def:ising}) with
\begin{align}
h_i &= \frac{1-d_i}{2}\log\frac{\hp_i^1}{\hp_i^0}
+\frac{1}{4}\sum_{j\in\partial i}\log\Bigl(\frac{\hp_{ij}^{11}\ 
\hp_{ij}^{10}}{\hp_{ij}^{01}\ \hp_{ij}^{00}}\Bigr),\label{eq:hi}\\[0.2cm]
J_{ij} &= \frac{1}{4}\log\Bigl(\frac{\hp_{ij}^{11}\ \hp_{ij}^{00}}{\hp_{ij}^{01}\ \hp_{ij}^{10}}\Bigr),
\qquad\forall\ (i,j)\in\E,\label{eq:Jij}
\end{align}
where $d_i$ is the number of neighbors of $i$, using the notation $j\in\partial i$
for ``$j$ neighbor of $i$''. The partition function is then explicitly given by
\begin{equation}\label{eq:zbethe}
Z_{Bethe}[\hp] =  \exp\biggl[-\frac{1}{4}\sum_{(i,j)\in\E}\log\bigl(\hp_{ij}^{00}\ \hp_{ij}^{01}\ \hp_{ij}^{10}\ \hp_{ij}^{11}\bigr)
-\sum_i\frac{1-d_i}{2}\log(\hp_i^0\ \hp_i^1)\biggr]
\end{equation}
The corresponding Gibbs free energy can then be written explicitly using (\ref{eq:hi}--\ref{eq:zbethe}).
With fixed magnetizations $m_i$'s, and given a set of couplings $\{J_{ij}\}$, the parameters $m_{ij}$
are implicit function
\[
m_{ij} = m_{ij}(m_i,m_j,J_{ij}),
\]
obtained by inverting the relations (\ref{eq:Jij}).
For the linear response, we get from (\ref{eq:hi}) a result derived first in~\cite{WeTe}:
\begin{align*}
\frac{\partial h_i}{\partial m_j} &= \Bigl[\frac{1-d_i}{1-m_i^2}\\[0.2cm] 
&+ \frac{1}{16}\sum_{k\in \partial i}\Bigl(
\bigl(\frac{1}{\hp_{ik}^{11}}+\frac{1}{\hp_{ik}^{01}}\bigr)\bigl(1+\frac{\partial m_{ik}}{\partial m_i}\bigr)+
\bigl(\frac{1}{\hp_{ik}^{00}}+
\frac{1}{\hp_{ik}^{10}}\bigr)\bigl(1-\frac{\partial m_{ik}}{\partial m_i}\bigr)\Bigr)\Bigr]\delta_{ij}\\[0.2cm]
&+\frac{1}{16}\Bigl(
\bigl(\frac{1}{\hp_{ij}^{11}}+\frac{1}{\hp_{ij}^{10}}\bigr)\bigl(1+\frac{\partial m_{ij}}{\partial m_i}\bigr)+
\bigl(\frac{1}{\hp_{ij}^{00}}+
\frac{1}{\hp_{ij}^{01}}\bigr)\bigl(1-\frac{\partial m_{ij}}{\partial m_i}\bigr)\Bigr)\Bigr]\delta_{j\in\partial i}.
\end{align*}
Using (\ref{eq:Jij}), we can also express
\[
\frac{\partial m_{ij}}{\partial m_i} = 
-\frac{\frac{1}{\hp_{ij}^{11}}+\frac{1}{\hp_{ij}^{01}}-\frac{1}{\hp_{ij}^{10}}-\frac{1}{\hp_{ij}^{00}}}
{\frac{1}{\hp_{ij}^{11}}+\frac{1}{\hp_{ij}^{01}}+\frac{1}{\hp_{ij}^{10}}+\frac{1}{\hp_{ij}^{00}}}, 
\]
so that with little assistance of Maple, we may finally reach the expression given in~\cite{NgBe}   
\begin{align}
[\hat \chi^{-1}]_{ij} = \Bigl[\frac{1-d_i}{1-m_i^2} &+\sum_{k\in \partial i}\frac{1-m_k^2}{(1-m_i^2)(1-m_k^2)
-\hat \chi_{ik}^2}\Bigr]\delta_{ij}\nonumber\\[0.2cm]
&-\frac{\hat \chi_{ij}}{(1-m_i^2)(1-m_j^2)-\hat \chi_{ij}^2}\ \delta_{j\in \partial i},\label{eq:invchis}
\end{align}
equivalent to the original one derived in~\cite{WeTe}, albeit written in a different form, 
more suitable to discuss the inverse Ising problem. This expression is  quite paradoxical since
the inverse of the $[\chi]_{ij}$ matrix, which coefficients appear on the right hand side of 
this equation, should coincide  with the left hand side, given as input of the inverse Ising problem.
The existence of an exact solution can therefore be checked directly as a self-consistency 
property of the input data $\hat\chi_{ij}$: for a given pair $(i,j)$
either:
\begin{itemize}
\item  $[\hat\chi^{-1}]_{ij}\ne 0$, then 
this self-consistency relation (\ref{eq:invchis}) has to hold and $J_{ij}$ is given by (\ref{eq:Jij})
using  $\hat m_{ij}= \hat m_i\hat m_j+\hat \chi_{ij}$.
\item $[\hat\chi^{-1}]_{ij} =  0$ then $J_{ij}=0$ but  $\hat\chi_{ij}$, which can be non-vanishing, is obtained by
inverting $[\hat\chi^{-1}]$ defined by (\ref{eq:invchis}).
\end{itemize}
Finally, complete consistency of the solution is checked on the diagonal elements in (\ref{eq:invchis}).
If full consistency is not verified, this equation can nevertheless be used to find approximate solutions.
Remark that, if we restrict the set of equations (\ref{eq:invchis}), e.g.\ by some thresholding procedure,
in such a way that the corresponding 
graph is a spanning tree, then, by construction, $\chi_{ij} \equiv \hat\chi_{ij}$ will be solution 
on this restricted set of edges, simply because the BP equations are exact on a tree.
The various methods proposed for example in~\cite{MoMe,YaTa} actually correspond to different 
heuristics for finding approximate solutions to this set of constraints.
As noted in~\cite{NgBe}, a direct way to proceed is to eliminate $\chi_{ij}$ in the equations obtained  
from (\ref{eq:Jij}) and (\ref{eq:invchis}):
\begin{align*}
\chi_{ij}^2+2\chi_{ij}(m_im_j-\coth(2J_{ij}))+(1-m_i^2)(1-m_j^2) = 0\\[0.2cm]
\chi_{ij}^2-\frac{\chi_{ij}}{[\chi^{-1}]_{ij}}-(1-m_i^2)(1-m_j^2) = 0.
\end{align*}
This leads directly to 
\begin{equation}\label{eq:JBETHE}
J_{ij}^{Bethe} = -\frac{1}{2}\atanh\Bigl(\frac{2[\hat\chi^{-1}]_{ij}}{\sqrt{1+4(1-\hm_i^2)
(1-\hm_j^2)[\hat\chi^{-1}]_{ij}^2}-2\hm_i\hm_j[\hat\chi^{-1}]_{ij}}\Bigr),
\end{equation}
while the corresponding computed of $\chi_{ij}$, instead of the observed one $\hat \chi_{ij}$,  
has to be inserted in (\ref{eq:hi}) to be fully consistent.
Note that $J_{ij}^{Bethe}$ and $J_{ij}^{TAP}$ coincide at second order in $[\hat\chi^{-1}]_{ij}$.

\paragraph{Hopfield model}
As mentioned in the introduction when the distribution to be modeled is multi-modal, 
the situation corresponds to finding an Ising model in the low temperature  
phase with many modes, referred to as Mattis states in the physics literature. 
Previous methods assume implicitly a high temperature where only one single mode, 
``the paramagnetic state'' is selected. The Hopfield model, introduced originally to 
model auto-associative memories, is a special case of an Ising model, where the coupling 
matrix is of low rank $p\le N$ and corresponds to the sum of outer products of $p$ given spin vectors $\{\xi^k,k=1\ldots p\}$, 
each one representing a specific attractive pattern:
\[
J_{ij} = \frac{1}{p}\sum_{k=1}^N\xi_i^k\xi_j^k
\] 
In our inference context, these patterns are not given directly, the input of the model being 
the covariance matrix.
In~\cite{CoMoSe} these couplings are interpreted  as the contribution stemming from the $p$ 
largest principle axes of the correlation matrix. This lead in particular the authors
to propose an extension of the Hopfield model by introducing repulsive patterns to take
as well into account the smallest principal axes. Assuming small patterns
coefficients $|\xi^k| <1/\sqrt{N}$, 
they come up with the following couplings with highest likelihood:
\[
J_{ij}^{Hopfield} \equiv \frac{1}{\sqrt{(1-m_i^2)(1-m_j^2)}}
\sum_{k=1}^p\Bigl(\bigl(1-\frac{1}{\lambda_k}\bigr)v_i^kv_j^k - \bigl(\frac{1}{\lambda_{N-k}}-1\bigr)v_i^{N-k}v_j^{N-k}\Bigr)
\]
at first order of the perturbation.
At this order of approximation the local fields are given by 
\[
h_i = \tanh(m_i)-\sum_j J_{ij}^{Hopfield}m_j.
\]
In a previous study~\cite{FuLaAu} we found a connection between the plain direct BP method with the Hopfield model, 
by considering a $1$-parameter deformation of the Bethe inference model~(\ref{eq:ansatz})
\begin{equation}\label{eq:bpalpha}
{\cal P}(\xx) = \prod_{i<j}\Bigl(\frac{\hp_{ij}(x_i,x_j)}{\hp(x_i)\hp(x_j)}\Bigr)^\alpha\prod_i\hp_i(x_i),
\end{equation}
with $\alpha\in[0,1]$. We observed indeed that when the data corresponds to some multi-modal
measure with well separated components, this measure coincides asymptotically 
with an Hopfield model made only of attractive pattern, representative of each component of the underlying measure.
$\alpha$ represents  basically the inverse temperature of the model 
and is easy to calibrate in practice.

\subsection{More on the Bethe susceptibility}\label{sec:cumulant}
\begin{figure}[ht]
\centerline{\resizebox*{\textwidth}{!}{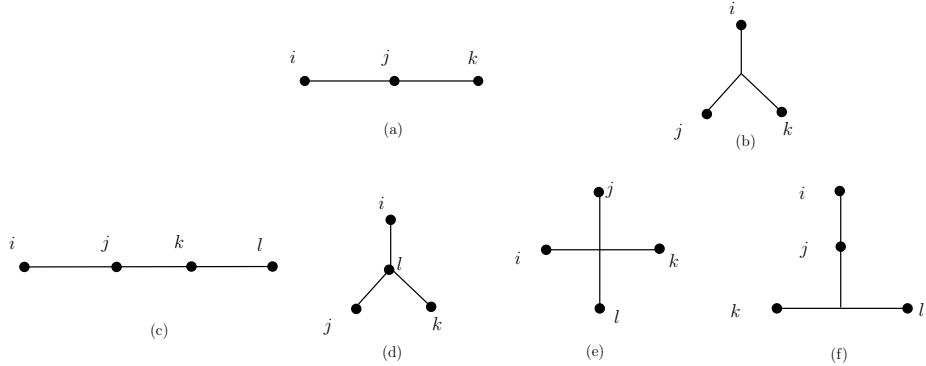}}
\caption{\label{fig:vertex}Various cumulant topologies of order three (a,b) and four (c-f).}
\end{figure}
The explicit relation (\ref{eq:invchis}) between susceptibility and inverse susceptibility  
coefficients is not the only one that can be obtained. In fact, it is the specific property of
a singly connected factor graph that two variables $x_i$ and $x_j$, conditionally to 
a variable $x_k$ are independent if $k$ is on the path between $i$ and
$j$ along the tree:
\[
p(x_i,x_j,x_k) = p(x_i\vert x_k)p(x_j\vert x_k)p(x_k) = \frac{p(x_i,x_k)p(x_j,x_k)}{p(x_k)}
\]
Using the parametrization (\ref{eq:pi},\ref{eq:pij}) with $m_{ij}=m_im_j+\chi_{ij}$
yields immediately 
\begin{equation}\label{eq:chirec}
\chi_{ij} = \frac{\chi_{ik}\chi_{jk}}{1-m_k^2},\qquad\forall\ k\in(i,j)\ \text{along}\ \T.
\end{equation}
By recurrence we get, as noticed in e.g.\ \cite{Mora}, 
given the path $i_0=i,i_1,\ldots,i_{n+1}=j$ between $i$ and $j$ along the 
tree $\T$
\[
\chi_{ij} = \frac{\prod_{a=0}^n\chi_{i_ai_{a+1}}}{\prod_{a=1}^n(1-m_{i_a}^2)},
\]
reflecting the factorization of the joint measure. This expression actually coincides 
with (\ref{eq:invchis}) only on a tree. On a loopy graph, this last expression 
should be  possibly replaced by  a sum over paths.

Higher order susceptibility coefficients are built as well in terms of elementary 
building blocks given by the pairwise susceptibility coefficients $\chi_{ij}$.
The notations generalize into the following straightforward manner:
\begin{align*}
m_{ijk} &\egaldef {\mathbb E}(s_is_js_k) \egaldef m_im_jm_k + m_i\chi_{jk}+m_j\chi_{ik}+m_k\chi_{ij}+\chi_{ijk} \\[0.2cm]
m_{ijkl} &\egaldef {\mathbb E}(s_is_js_ks_l) \egaldef m_im_jm_km_l \\[0.2cm] 
&+m_im_j\chi_{kl}+m_im_k\chi_{jl}+m_im_l\chi_{jk}+m_jm_k\chi_{il}+m_jm_l\chi_{ik}+m_km_l\chi_{ij}\\[0.2cm]
&+m_i\chi_{jkl}+m_j\chi_{ikl}+m_k\chi_{ijl}+m_l\chi_{ijk}+\chi_{ijkl}, 
\end{align*}
where $\chi_{ijk}$ and $\chi_{ijkl}$ are respectively three and four point susceptibilities.
The quantities being related to the corresponding marginals similarly to (\ref{eq:pi},\ref{eq:pij}):
\begin{align*}
p(s_i,s_j,s_k) &= \frac{1}{8}\bigl(1+m_is_i+m_js_j+m_ks_k\\[0.2cm]
&+m_{ij}s_is_j+m_{ik}s_is_k+m_{jk}s_js_k+m_{ijk}s_is_js_k\bigr)\\[0.2cm]
p(s_i,s_j,s_k,s_l) &= \frac{1}{16}\bigl(1+m_is_i+m_js_j+m_ks_k+m_ls_l\\[0.2cm]
&+m_{ij}s_is_j+m_{ik}s_is_k+m_{il}s_is_l+m_{jk}s_js_k+m_{jl}s_js_l+m_{kl}s_ks_l\\[0.2cm]
&+m_{ijk}s_is_js_k+m_{ijl}s_is_js_l+m_{ikl}s_is_ks_l+m_{jkl}s_js_ks_l+m_{ijkl}s_is_js_ks_l\bigr)
\end{align*}
Using the basic fact that, on the tree 
\[
p(s_i,s_j,s_k) = \frac{p(s_i,s_j)p(s_j,s_k)}{p(s_i)}
\]
when $j$ is on the path $\pth{ik}$ given by $\T$,
and  
\[
p(s_i,s_j,s_k) = \sum_{s_l}\frac{p(s_i,s_l)p(s_j,s_l)p(s_k,s_l)}{p(s_l)^2}
\]
when path $\wideparen{ij}$, $\wideparen{ik}$ and $\pth{jk}$ along $\T$ intersect on vertex $l$,
we obtain
\begin{align*}
\chi_{ijk} = \begin{cases}
\DD -2\frac{m_l}{(1-m_l^2)^2}\chi_{il}\chi_{jl}\chi_{kl} & \text{with}\ \{l\} =
(i,j)\cap (i,k)\cap(j,k)
\ \text{along}\ \T,\\[0.2cm]
\DD -2m_j\chi_{ik}& \text{if}\ j\in(i,k)\ \text{along}\ \T.
\end{cases}
\end{align*}
For the fourth order, more cases have to be distinguished. 
When $i$, $j$, $k$ and $l$ are aligned as on Figure~\ref{fig:vertex}.c, 
in this order on the path $\pth{il}$ along $\T$
we have 
\[
p(s_i,s_j,s_k,s_l) = \frac{p(s_i,s_j)p(s_j,s_k,s_l)}{p(s_j)^2}
\]
which leads to 
\[
\chi_{ijkl} = 4m_km_j\chi_{il} - \chi_{ik}\chi_{jl}-\chi_{il}\chi_{jk}.
\]
When path $\pth{ij}$, $\pth{ik}$ and $\pth{jk}$ along $\T$ intersect on vertex $l$ as on Figure~\ref{fig:vertex}.d,
we obtain instead~\footnote{This apparently non-symmetric expression can be symmetrized with help of (\ref{eq:chirec}).}
\[
\chi_{ijkl} = 2\frac{3m_l^2-1}{1-m_l^2}\chi_{ij}\chi_{kl}.
\]
For the situation corresponding to Figure~\ref{fig:vertex}.e, we have 
\[
p_(s_i,s_j,s_k,s_l) = \sum_{s_q}\frac{p(s_i,s_j,s_q)p(s_q,s_k,s_l)}{p(s_q)^2}
\]
which leads to 
\[
\chi_{ijkl} = 2\frac{3m_q^2-1}{1-m_q^2}\chi_{ij}\chi_{kl}.
\]
Finally, for the situation corresponding to Figure~\ref{fig:vertex}.f, we have
\[
p_(s_i,s_j,s_k,s_l) = \sum_{s_q}\frac{p(s_i,s_j)p(s_j,s_k,s_l)}{p(s_j)^2}
\]
leading to 
\[
\chi_{ijkl} = -\chi_{ik}\chi_{jl}-\chi_{jk}\chi_{il}+4\frac{m_km_q}{1-m_q^2}\chi_{ij}\chi_{lq}.
\]

\subsection{Sparse inverse estimation of covariance matrix}
Let us leave the Ising modeling issue aside for a while and introduce another related graph selection problem, 
named sparse inverse covariance estimation, which is defined on real continuous random variables.
This method aims at constructing a sparse factor graph structure by 
identifying conditionally independent pairs of nodes in the graph, given 
empirical covariances of random variables. Assuming that all nodes in the graph follow a joint
multi-variate Gaussian distribution with mean $\mu$ and covariance
matrix $C$, the existing correlation between the nodes $i$ and $j$
given all the other nodes in the graph are indicated by the non-zero $ij$th entry of the precision matrix
$C^{-1}$, while zero entries correspond to independent pairs of variables. Therefore, under the joint normal distribution
assumption, selection of factor graph structures amounts to finding the
sparse precision matrix that best describes the underlying data distribution,
given the fixed empirical covariance matrix. When the derived inverse estimation is sparse, 
it becomes easier to compute marginal distribution of each random 
variable and conduct statistical inference.
To achieve that goal, optimizations methods have been developed based on 
$L_0$ or $L_1$ norm penalty for the estimation of $C^{-1}$, to
enhance its sparse structure constraint on the estimated inverse of covariance matrix and 
discover underlying conditionally independent parts.

Let $\hat C \in \mathbb{R}^{n \times n}$ be the empirical covariance matrix
of $n$ random variables (represented as the nodes in the graph model).
The sparsity penalized maximum likelihood estimation $A$ of the
precision matrix $C^{-1}$ can be derived by solving the following
positive definite cone program:
\begin{equation}\label{eq:logdetopt}
 A = \argmin_{ X \succ 0}   -{\cal L}(X) + \lambda P(X)
\end{equation}
where
\begin{equation}\label{def:LLcov}
{\cal L}(A) \egaldef \log\det(A)-\Tr(A\hat C),
\end{equation}
is the log likelihood of the distribution defined by $A$,
$\log\det$ being the logarithm of the determinant, and $P(A)$ is
a sparsity inducing regularization term~\cite{FrHaTi}.
$\lambda$ is the regularization coefficient balancing the data-fitting
oriented likelihood and sparsity penalty. Since the precision matrix
of joint normal distribution should be positive definite, any feasible
solution to this optimization problem is thus required to locate
within a positive definite cone. The penalty term $P(A)$ is typically
constructed using sparsity inducing matrix norm, also known as
sparse learning in the domain of statistical learning. There are two
typical configurations of $P(A)$:
\begin{itemize}
\item The $L_0$ norm $\|A\|_{L_0}$ of the matrix $X$,
  which counts the number of non-zero elements in the matrix. It is
  also known as the cardinality or the non-zero support of the matrix.
  Given its definition, it is easy to find that $L_0$ norm based
  penalty is a strong and intuitive appeal for sparsity
  structure of the estimated precision matrix. However, it is
  computationally infeasible to solve exact $L_0$-norm minimization directly,
  due to the fact that exact $L_0$ norm penalty is discontinuous and
  non-differentiable. In practice, one either uses a continuous
  approximation to the form of the $L_0$-penalty, or  solve it
  using a greedy method. Due to the non-convexity of the exact $L_0$ norm
  penalty, only a local optimum of the feasible
  solution can be guaranteed. Nevertheless, $L_0$ norm penalty usually leads to much
  sparser structure in the estimation, while local optimum is good
  enough for most practical cases.
  
 \item The $L_1$ matrix norm $\|A\|_{L_1} = \sum_{i ,j}^n |A_{ij}|$.
  $L_1$ norm penalty was firstly introduced into the standard least
  square estimation problem by Tibshirani~\cite{TibLASSO}, under the
  name "Lasso regression''. Minimizing the $L_1$ norm based penalty
  encourages sparse non-zero entries in the estimated precision matrix $A$,
  which achieves a selection of informative variables for regression
  and reduces complexity of regression model efficiently. Further
  extension of the basic $L_1$-norm penalty function allows one assigning
  different non-negative weight values $\lambda_{ij}$ to different
  entries $A_{ij}$, as $\sum_{i ,j}^n\lambda_{ij}|A_{ij}|$. This
  weighted combination can constrain the sparse penalties only on the
  off-diagonal entries, so as to avoid unnecessary sparsity on the
  diagonal elements. Furthermore, this extension allows us to
  introduce prior knowledge about the conditional independence
  structure of the graph into the joint combination problem.
\end{itemize}

\begin{figure}
	\centering
	\includegraphics[width=0.8\textwidth]{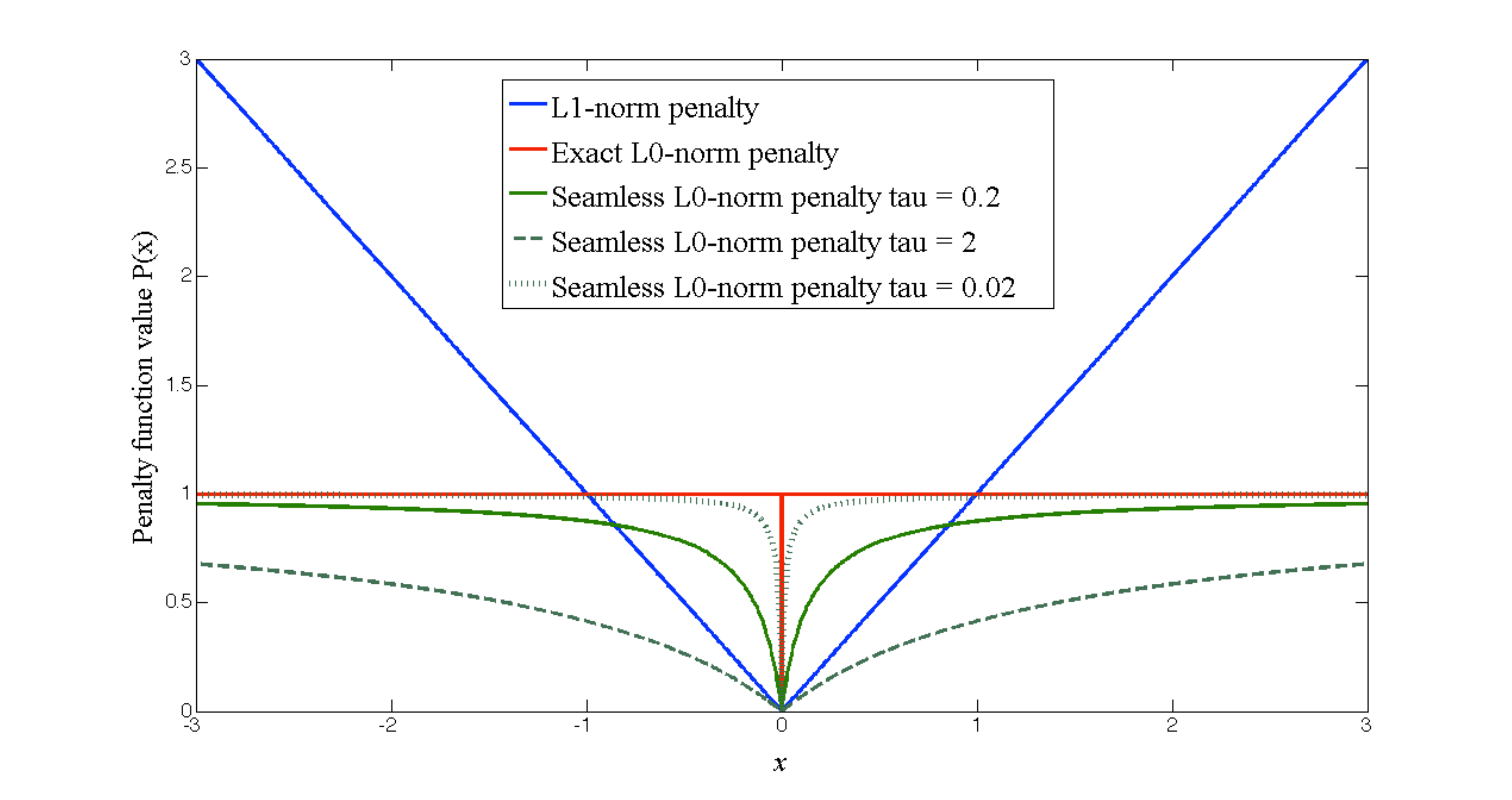}
	\caption{Demonstration of $L_1$ norm, exact $L_0$ norm and Seamless $L_0$ norm based penalty function.}
 	\label{fig:L0L1penfuncf}
\end{figure}

For further understanding of the relation between the exact $L_0$
and $L_1$ norm penalty, we illustrate them with respect to one scalar
variable in Figure~\ref{fig:L0L1penfuncf}. As we can see, within $[-1,1]$, $L_1$-norm
penalty plays as a convex envelope of the exact $L_0$-norm penalty. Due to the convexity
property of $L_1$ norm penalty, the global optimum of the convex programming problem can be achieved with even linear
computational complexity~\cite{TibLASSO,FrHaTi}. However,
although $L_1$ norm based penalty leads to computationally sound
solutions to the original issue, it also introduces modeling bias into
the penalized maximum likelihood estimation. As illustrated in the
figure, when the underlying true values of the matrix entries are
sufficiently large, the corresponding $L_1$ norm based regularization
performs linearly increased penalty to those entries, which thus
results in a severe bias w.r.t.\ the maximum likelihood
estimation~\cite{FanASA}. In contrast, $L_0$ norm based penalty avoids
such issue by constraining the penalties of non-zero elements to be
constant. It has been proved in~\cite{TibLASSO} that the $L_1$ norm
penalty discovers the underlined sparse structure when some suitable
assumptions are satisfied. However, in general cases, the quality
of the solutions is not clear.

\section{Iterative proportional scaling from the Bethe reference point}\label{sec:iip}
\subsection{Bethe reference point and optimal $1$-link correction}\label{sec:onelink}
As observed in the previous section, 
when using the Bethe approximation to find an approximate solution to the IIP, 
the consistency check should then be that either the factor graph be sparse,
nearly a tree, either the coupling are small. There are then two distinct ways of 
using the Bethe approximation:
\begin{itemize}
\item the direct way, where the form of the joint distribution (\ref{eq:ansatz}) is assumed with a complete graph.
There is then by construction a belief propagation fixed point for which the beliefs satisfy all the constraints.   
This solution to be meaningful requires small correlations, so that the belief propagation fixed point 
be stable and unique, allowing the corresponding log likelihood to be well approximated. Otherwise, this 
solution is not satisfactory, but a pruning procedure, which amounts to select a sub-graph based on mutual 
information, can be used. The first step is to find the maximum spanning tree (MST) with mutual information 
taken as edges weights. 
How to add new links to this baseline solution in a consistent way will be the subject of the next section.     
\item the indirect way consists in first inverting the potentially non-sparse correlation matrix. 
If the underlying interaction matrix is actually a tree, this will be visible in the inverse correlation 
matrix, indicated directly by the non-zero entries. Corresponding couplings are then determined
through (\ref{eq:invchis}). This procedure seems to work better than the previous 
one also when no sparsity but weak coupling is assumed. It corresponds in fact to the equations solved iteratively 
by the susceptibility propagation algorithm~\cite{MoMe}.
\end{itemize} 

Let us first justify the intuitive assertion concerning the optimal model with tree like factor graphs, 
valid for any type of MRF.
Suppose that we are given a set of single and pairwise empirical marginals $\hat p_i$ and $\hat p_{ij}$
for a set of $N$ real variables $\{x_i,i=1\ldots N\}$. 
If we start from an empty graph with no link, the joint probability distribution 
is simply the product form
\[
\P^{(0)}(x_i) = \prod_{i=1}^N \hat p_i(x_i).
\]
Adding a link $(ij)$ to the empty graph is optimally done by multiplying $\P^{(0)}$ by $\hat p_{ij}/\hat p_i\hat p_j$.
The gain in log likelihood is then simply the mutual information between $x_i$ and $x_j$.
Thus, as long as no loop get closed by the procedure, the best candidate link corresponds to  the pair
of variables with maximum mutual information and the measure reads
after $n$ steps
\[
 \P^{(n)}(\x) = \prod_{(ij)\in\G^{(n)}} \frac{p_{ij}(x_i,x_j)}{p_i(x_i)p_j(x_j)} \prod_{i=1}^n
 p_i(x_i).
\]
This suggests that a good initialization point 
for the algorithm is the maximum spanning tree with edges weights
given by the relevant mutual information. This corresponds to the classical results of~\cite{ChoLee} concerning inference
using dependence trees.
It is optimal in the class of singly connected graphical models.
In the following, we will refer in the text to this specific approximate
solution as the \emph{Bethe reference point}. 

Starting from this point, we want now to add
one factor $\psi_{ij}$ to produce the distribution 
\begin{equation}\label{eq:1link}
\P^{(n+1)}(\x) = \P^{(n)}(\x)\times \frac{\psi_{ij}(x_i,x_j)}{Z_\psi}
\end{equation}
with 
\[
Z_\psi = \int dx_idx_j p_{ij}^{(n)}(x_i,x_j)\psi_{ij}(x_i,x_j).
\]
The log likelihood corresponding to this new distribution reads
\[
\L' = \L + \int d\x \hat\P(\x)\log\psi_{ij}(x_i,x_j)-\log Z_\psi.
\]
Since the the functional derivative w.r.t.\ $\psi$ is
\[
\frac{\partial\L'}{\partial \psi_{ij}(x_i,x_j)} = \frac{\hat p(x_i,x_j)}{\psi_{
ij}(x_i,x_j)} - 
\frac{p_{ij}^{(n)}(x_i,x_j)}{Z_\psi},
\] 
$\forall (x_i,x_j)\in \Omega^2$, the maximum is attained for
\begin{equation}\label{eq:1linkfactor}
\psi_{ij}(x_i,x_j) = \frac{\hat p_{ij}(x_i,x_j)}{p_{ij}^{(n)}(x_i,x_j)}\text{ 
with }  Z_\psi=1,
\end{equation}
where $p^{(n)}(x_i,x_j)$ is the reference marginal distribution obtained from
$\P^{(n)}$. The correction to the log-likelihood can then be rewritten as
\begin{equation}\label{eq:onelink}
\Delta \L = D_{KL}(\hat p_{ij}\Vert p_{ij}^{(n)}).
\end{equation}
Sorting all the links w.r.t.\ this quantity yields the (exact) optimal $1$-link 
correction to be made. 
The interpretation is therefore immediate: the best candidate is the one for 
which the current model yields the joint marginal $p_{ij}^{(n)}$ that
is most distant target $\hat p_{ij}$. Note that the update mechanism
is indifferent to whether the link 
has to be added or simply modified.

\subsection{Iterative proportional scaling for GMRF model selection} 
In the statistics literature this procedure is referred to as the \emph{iterative proportional scaling} (IPS)
procedure, originally proposed for contingency table estimations 
and extended further to MRF maximum likelihood estimation~\cite{DaRa,DelPieLa}.
Assuming the structure of the graph is known, it appears not to be very efficient~\cite{Malouf} 
when compared to other gradient based methods. 
The problem of the method is that it requires the knowledge of all pairwise marginals
$\{p_{ij},(ij)\notin \G^{(n)}\}$
at each iteration step $n$. In the Ising case this is done by a sparse matrix inversion through equation (\ref{eq:invchis}), 
which potentially renders the method a bit expensive and rapidly inaccurate after many links have been added.
For Gaussian MRF, the situation is different, because in that case 
the correction to the log likelihood can be evaluated directly by another means.
Making full use of the incremental characteristics of IPS we 
propose to construct the sparse GMRF graph structure link by link based on IPS in an efficient way. 

Indeed, the correction factor
(\ref{eq:1linkfactor}) reads in that case
\[
\psi_{ij}(x_i,x_j) = \exp\Bigl(-\frac{1}{2}(x_i,x_j)\bigl(\hat
C_{\{ij\}}^{-1}-C_{\{ij\}}^{-1}\bigr)(x_i,x_j)^T\Bigr),
\]
where $C_{\{ij\}}$ (resp.\ $\hat C_{\{ij\}}$) represents the restricted $2\times 2$ 
covariance matrix corresponding to the pair $(x_i,x_j)$ of 
the current model specified by precision 
matrix $A = C^{-1}$ (resp.\ the reference model). Let  $[C_{\{ij\}}]$
denote the
$N\times N$ matrix formed by completing $C_{\{ij\}}$ with zeros. The new model
obtained after adding or changing 
link $(ij)$ reads 
\begin{equation}\label{eq:perturbation}
A' = A + [\hat C_{\{ij\}}^{-1}]-[C_{\{ij\}}^{-1}] \egaldef A+[V].  
\end{equation}
with a log likelihood variation given by: 
\[
\Delta \L = \frac{C_{ii}\hat C_{jj}+C_{jj}\hat C_{ii}-2C_{ij}\hat
C_{ij}}{\det(C_{\{ij\}})}- 2 -\log\frac{\det(\hat
C_{\{ij\}})}{\det(C_{\{ij\}})}.
\]
Let us notice the following useful formula:
\begin{align}\label{eq:inverse}
(A+[V])^{-1} &=
A^{-1}-A^{-1}[V](1+A^{-1}[V])^{-1}A^{-1}
\nonumber\\[0.2cm]
&= A^{-1}-A^{-1}[V](1+[C_{\{ij\}}][V])^{-1}A^{-1},
\end{align}
valid for a $2\times 2$ perturbation matrix $V = V_{\{ij\}}$.
Using this formula, the new covariance matrix reads
\begin{equation}\label{eq:covupdate}
C' = A'^{-1} = A^{-1}- A^{-1}[C_{\{ij\}}^{-1}]\bigl(1-[\hat
C_{\{ij\}}][C_{\{ij\}}^{-1}]\bigr)A^{-1}.
\end{equation}
The number of operations needed to maintain the covariance matrix  
after each addition is therefore ${\cal O}(N^2)$. This technical point is determinant 
to render our approach useful in practice.

Let us now examine under which condition adding or modifying links in this way lets
the covariance matrix remain positive semi-definite. By adding a $2\times 2$
matrix, we expect a quadratic correction to the determinant:
\begin{align*}
\det(A') &= \det(A)\det(1+A^{-1}V)\\[0.2cm]
&= \det A \times \frac{\det(C_{\{ij\}})}{\det(\hat C_{\{ij\}})},
\end{align*}
which is obtained directly because $A^{-1}V$ has non zero entries only on 
column $i$ and $j$. Multiplying $V$ by some parameter $\alpha\ge 0$, define
\[
P(\alpha) \egaldef \det(1+\alpha A^{-1}V) 
= \alpha^2\det\bigl(C_{\{ij\}}\hat
C_{\{ij\}}^{-1}-\frac{\alpha-1}{\alpha}\bigr).
\]
$P(\alpha)$ is proportional to  the characteristic polynomial of the matrix
$C_{\{ij\}}\hat C_{\{ij\}}^{-1}$ of argument $(\alpha-1)/\alpha$. So $P$ will have
no root in $[0,1]$, and $A'$ be definite positive, iff
$C_{\{ij\}}\hat C_{\{ij\}}^{-1}$ is definite positive. Since the product of
eigenvalues, $\det(C_{\{ij\}}\hat C^{-1}_{\{ij\}})$, is positive, one has to
check their sum, given by $\Tr(C_{\{ij\}}\hat C_{\{ij\}}^{-1})$:
\begin{equation}\label{eq:tracepos}
C_{ii}\hat C_{jj}+C_{jj}\hat C_{ii}-2C_{ij}\hat C_{ij} > 0.
\end{equation}
$C_{\{ij\}}$ and $\hat C_{\{ij\}}$ are individually positive definite:
\[
C_{ii}C_{jj}-{C_{ij}}^2 > 0 \qquad\text{and}\qquad \hat C_{ii}\hat C_{jj}-\hat 
C_{ij}^2 > 0,
\] 
from which we deduce that
\[
\Bigl(\frac{C_{ii}\hat C_{jj}+C_{jj}\hat C_{ii}}{2}\Bigr)^2 > C_{ii}\hat C_{jj}
C_{jj}\hat C_{ii} > {C_{ij}}^2 {\hat C_{ij}}^2,
\]
giving finally that (\ref{eq:tracepos}) is always fulfilled  
when both $C_{\{ij\}}$ and $\hat C_{\{ij\}}$ are non-degenerate.

\paragraph{Removing one link}

To use this in an algorithm, it is also  desirable to be able to remove 
links, so that, with help 
of a penalty coefficient per link, the  model can be optimized with a desired 
connectivity level.

For the Gaussian model, if $A$ is the coupling matrix, removing the link $(i,j)$
amounts to chose a factor  $\psi_{ij}$ in (\ref{eq:1link}) of the form:
\[
\psi_{ij}(x_i,x_j) = \exp\bigl(A_{ij} x_i x_j\bigr)
\]
($x_i$ and $x_j$ are assumed centered as in the preceding section).
Again, let $V$ denote the perturbation in the precision matrix such that $A'
=A+V$
is the new one.
The corresponding change in the log likelihood reads
\[
\Delta \L = \log\det(1+A^{-1}V)-\Tr(V\hat C).
\]
Rearranging this expression leads to
\[
\Delta \L = \log\bigl(1-2A_{ij}C_{ij}-A_{ij}^2\det(C_{\{ij\}})\bigr) +
2A_{ij}\hat
C_{ij},
\]
and using again (\ref{eq:inverse}), we get for the new covariance matrix
\begin{equation}\label{eq:covupdate2}
C' = C - \frac{A_{ij}}{1-2A_{ij}C_{ij}-A_{ij}^2\det(C_{\{ij\}})}\ C [B_{\{ij\}}]
C, 
\end{equation}
with 
\[
 B_{\{ij\}} \egaldef \left[
\begin{matrix} 
A_{ij}C_{jj} & 1-A_{ij}C_{ij}\\[0.2cm]
1-A_{ij}C_{ij} & A_{ij}C_{ii}
\end{matrix}\right].
\]
To check for the positive-definiteness property of $A'$, let us observe first 
that 
\[
\det(A') = \det(A)\times P(A_{ij}), 
\]
with 
\[
P(x) = \bigl(1-x(C_{ij}-\sqrt{C_{ii}C_{jj}})\bigr)\bigl(1-x(C_{ij}+\sqrt{C_{ii}C_{jj}})\bigr).
\]
When $x$ varies from $0$ to $A_{ij}$,
$P(x)$ should remains strictly positive to insure that $A'$ is definite positive. 
This results in the following condition: 
\[
\frac{1}{ C_{ij} - \sqrt{C_{ii}C_{jj}}} < A_{ij} < \frac{1}{\sqrt{C_{ii}C_{jj}}
+ C_{ij}}.
\]

\subsection{Imposing walk-summability}

Having a sparse GMRF gives no guarantee about its compatibility with GaBP. 
In order to be able to use the GaBP algorithm for inference, more strict constraints on graph connectivity need to be imposed. The most
precise condition known for convergence and validity of GaBP is 
called walk-summability (WS) and is extensively described by~\cite{MaJoWi}.
The two necessary and sufficient conditions for WS that we consider here are:
\begin{enumerate}[(i)]
\item The matrix $\diag(A) - |R(A)|$ is definite positive, where $R(A) \egaldef A
- \diag( A)$ contains the off-diagonal terms of $A$.
 \item $\rho(|R'(A)|) <1$, with $R'(A)_{ij}\egaldef
\frac{R(A)_{ij}}{\sqrt{A_{ii}A_{jj}}}$,
\end{enumerate}
with $\rho(A)$, called spectral radius of $A$, denotes the maximal modulus of
the eigenvalues of the matrix $A$.

\paragraph{Adding or Modifying one link}
If we  wish to increase the likelihood by adding or modifying a link
$(i,j)$ to the graph, the model is perturbed according to
\eqref{eq:perturbation}. Assuming that $A$ is WS, we want to express conditions
under which $A'$ is also WS. 

Using the definition (i) of walk summability we can derive a sufficient condition
for WS by imposing that $W(A') \egaldef \diag(A') - |R(A')|$ remains definite
positive. For $\alpha \in [0,1]$ we have
\[
 W(A+\alpha V) = W(A) + [\phi(\alpha V,A)],
\]
with 
\[
\phi(V,A) \egaldef \left[\begin{array}{cc}
                 V_{ii} & |A_{ij}|-|V_{ij}+A_{ij}|\\
                 |A_{ij}|-|V_{ji}+A_{ji}| & V_{jj}\\ 
                 \end{array}\right]
\]
We express the determinant of $W(A+\alpha V)$ as
\begin{align*}
 \det (W(A+\alpha V)) &=  \det \left( W\right)\det \left(1+W^{-1}[\phi(\alpha V,
A)]
\right)\\
  &= \det \left( W\right) \Theta(\alpha).
\end{align*}
shortening $W(A)$ in $W$. $\Theta(\alpha)$ is a degree $2$ polynomial by parts:
\begin{align*}
 \Theta(\alpha) \egaldef &\det (W^{-1}_{\{ij\}})\left(
\alpha^2V_{ii}V_{jj}-B(\alpha)^2\right)\\
&+ \alpha\left(W^{-1}_{ii}V_{ii}+W^{-1}_{jj}V_{jj}\right) + 2B(\alpha)W_{ij},
\end{align*}
with $B(\alpha) \egaldef |A_{ij}|-|\alpha V_{ji}+A_{ji}|$. Our sufficient condition
for WS of $A'$ is simply
\begin{equation}\label{eq:WSSC}
 \Theta(\alpha)>0,\forall \alpha \in [0,1],
\end{equation}
which boils down in the worst case to compute the roots of $2$
polynomials of degree $2$. 
Note that checking this sufficient condition imposes to keep track of the
matrix $W^{-1}$ which requires $\mathcal{O}(N^2)$ operations at each
step, using again~\eqref{eq:inverse}. 

\paragraph{Removing one link}
Removing one link of the graph will change the matrix $A$ in $A'$ such as $|R'(A')|
\leq |R'(A)|$ where $\leq$ denotes element-wise comparison. Then,
using elementary results on positive matrices,
$\rho(|R'(A')|) \leq \rho(|R'(A)|)$ and thus removing one
link of a WS model yields a new WS model.

\smallskip

As we shall see, WS might be a too strong constraint to impose since
it is easy to find non WS models that are compatible with GaBP\@. The framework
we present here allows us to impose weaker spectral constraint, e.g.\ we could
only ensure that $\diag(A) - R(A)$ remains definite positive. This is
equivalent to replace (i) by $\rho(R'(A))<1$.
\subsection{IPS based GMRF model selection algorithms}\label{sec:greedyalgo}
We are now equipped to define algorithms based on additions/modifications/deletions
of links.
\paragraph{Algorithm 1: Incremental graph construction by link addition}
\begin{itemize}
\item[S1] INPUT: the MST graph, and corresponding covariance matrix $C$.
\item[S2]: select the link with highest $\Delta\L$, compatible with
\eqref{eq:WSSC} if we wish to impose WS. Update $C$ according
to~(\ref{eq:covupdate}).
\item[S3]: repeat S2 until convergence~(i) or until a target connectivity is 
reached~(ii).
\item[S4]: if~(ii) repeat S2 until convergence by restricting the link selection
 in the set of existing ones. 
\end{itemize}
Each single addition/modification involves $N^2$ operations due to the covariance matrix 
update~(\ref{eq:covupdate}). Without S4, the complexity is therefore
${\cal O}(KN^3)$ if $K$ is the final mean connectivity, i.e.\ ${\cal O}(N^3)$ 
in the sparse domain and ${\cal O}(N^4)$ in the dense one. In practice, if many links
are added, S4 has to be run for several intermediate values of $K$ with a few updates for 
each existing link, so its contribution to the complexity is also 
of the same order ${\cal O}(KN^3)$.

\paragraph{Algorithm 2: IPS with backtracking  by link addition/deletion}

\begin{itemize}
\item[S1] INPUT: the MST graph, and corresponding covariance matrix $C$, a link 
penalty coefficient $\nu$.
\item[S2]: select the modification with highest $\Delta\L-s\nu$, 
with $s=+1$ for an addition, $s=-1$ for a deletion and $s=0$ for a modification, compatible with the WS
preserving condition of $A'$.
Update $C$ according to (\ref{eq:covupdate}) and (\ref{eq:covupdate2}) respectively
for an addition/modification  and a deletion.
\item[S3]: repeat S2 until convergence.
\item[S4]: repeat S2 until convergence by restricting the link selection
 in the set of existing ones. 
\end{itemize}

In absence of penalty ($\nu=0$) the algorithm will simply generate a model 
for any value of the mean connectivity, hence delivering an almost continuous 
Pareto set of solutions, with all possible trade-offs between sparsity and 
likelihood, as long as walk summability is satisfied.
With a fixed penalty, the algorithm convergences instead 
towards a solution with the connectivity depending implicitly on $\nu$; 
it corresponds roughly to the point $K^\star$ where the slope is  
\[
\frac{\Delta\L}{N\Delta K}(K^\star) = \nu.
\]

If we want to use the backtracking mechanism allowed by the penalty
term without converging to a specific connectivity, we may also let 
$\nu$ be adapted dynamically. A simple way is to adapt $\nu$ with the 
rate of information gain by letting
\[
\nu = \eta\Delta\L_{\text{add}},\quad\text{with}\quad\eta\in[0,1[,
\]
where $\Delta\L_{\text{add}}$ corresponds to the gain of the last link addition. 
With such a setting, $\nu$ is always kept just below the information gain 
per link,
allowing thus the algorithm to carry on toward higher connectivity. 
This heuristic of course 
assumes a concave Pareto front.

\section{Perturbation theory near the Bethe point}\label{sec:perturbation}
\subsection{Linear response of the Bethe reference point}
The approximate Boltzmann machines described in the introduction are obtained either by perturbation 
around the trivial point corresponding to a model of independent variables, the first order yielding the 
Mean-field solution and the  second order the TAP one, either by using the 
linear response delivered in the Bethe approximation. We propose to combine in a way the two procedures,
by computing the perturbation around the Bethe model associated to the MST
with weights given by mutual information. 
We denote by ${\cal T}\subset{\cal E}$, the subset of links 
corresponding to the MST, considered as given as well as the susceptibility 
matrix $[\chi_{Bethe}]$ given explicitly by its inverse through (\ref{eq:invchis}),
in term of the empirically observed ones $\hat\chi$.
Following the same lines as the one given in Section~\ref{sec:prelim}, 
we consider again the Gibbs free energy to impose the individual expectations $\m = \{\hm_i\}$ 
given for each variable. 
Let $\K = \{K_{ij},(i,j)\in{\cal T}\}$ the set of Bethe-Ising couplings, i.e.\ the set of coupling 
attached to the MST s.t. corresponding susceptibilities are fulfilled and $\J = \{J_{ij},(i,j)\in\E\}$
a set of Ising coupling corrections.     
The Gibbs free energy reads now
\[
G[\m,\J] = \h^{T}(\m)\m + F\bigl[\h(\m),\K+\J\bigr]
\]
where $\h(\m)$  depends implicitly on $\m$
through the same set of  constraints (\ref{eq:constraints}) as before.
The only difference resides in the choice of the reference point. We 
start from the Bethe solution given by the set of coupling $\K$ 
instead of starting with an independent model. 

The Plefka expansion is used again to expand the Gibbs free energy in power of the coupling $J_{ij}$ assumed to be small. 
Following the same lines as in Section~\ref{sec:plefka}, but with $G_0$ now replaced by 
\[
G_{Bethe}[\m] = \h^{T}(\m)\m  - \log Z_{Bethe}\bigl[\h(\m),\J^{Bethe}\bigr],
\]
and $h_i$, $J^{Bethe}$ and $Z_{Bethe}$ given respectively by
(\ref{eq:hi},\ref{eq:Jij},\ref{eq:zbethe})  
where $\E$ is now replaced by $\T$,
letting again 
\[
H^1 \egaldef \sum_{i<j} J_{ij}s_is_j,
\]
and following the same steps (\ref{eq:dG1},\ref{eq:dG2},\ref{eq:dha}) leads 
to the following modification of the local fields
\[
h_i = h_i^{Bethe} - \sum_j[\chi_{Bethe}^{-1}]_{ij}\cov_{Bethe}(H^1,s_i)\qquad\forall i\in\V  
\]
to get the following Gibbs free energy at second order in $\alpha$ (after replacing $H^1$ by $\alpha H^1$):
\begin{align*}
G[\m,\alpha J] &= G_{Bethe}(\m)-\alpha{\mathbb E}_{Bethe}(H^1)\\[0.2cm]
-\frac{\alpha^2}{2}\Bigl(&\var_{Bethe}(H^1)-
\sum_{ij}[\chi_{Bethe}^{-1}]_{ij}\cov_{Bethe}(H^1,s_i)\cov_{Bethe}(H^1,s_j)\Bigr)+o(\alpha^2).
\end{align*}
This is the general expression for the linear response near the Bethe reference point that we now use.
\begin{align}\label{eq:gblr}
G_{BLR}[\bJ] &\egaldef -{\mathbb E}_{Bethe}(H^1)\\[0.2cm]
&-\frac{1}{2}\Bigl(\var_{Bethe}(H^1)-
\sum_{i,j}[\chi_{Bethe}^{-1}]_{ij}\cov_{Bethe}(H^1,s_i)\cov_{Bethe}(H^1,s_j)\Bigr).
\end{align}
represents the Gibbs free energy at this order of approximation.
It it is given explicitly through 
\begin{align*}
\ExB(H^1) & = \sum_{i<j} J_{ij}m_{ij}\\[0.2cm]
\var_{Bethe}(H^1) & = \sum_{i<j,k<l} J_{ij}J_{kl}\bigl(m_{ijkl}- m_{ij}m_{kl}\bigl) \\[0.2cm]
\cov_{Bethe}(H^1,s_k) & = \sum_{i<j} J_{ij}(m_{ijk} - m_{ij}m_k)
\end{align*}
where 
\begin{align*}
m_i   &\egaldef \ExB(s_i),\qquad
m_{ij}  \egaldef \ExB(s_is_j)\\[0.2cm]
m_{ijk} &\egaldef \ExB(s_is_js_k),\qquad
m_{ijkl}  \egaldef \ExB(s_is_js_ks_l)
\end{align*}
are the moments delivered by the Bethe approximation. With the material given in Section~\ref{sec:cumulant}
these are given in closed form in terms of the Bethe susceptibility  coefficients $\chi_{Bethe}$. 
Concerning the log-likelihood, it is given now by:
\begin{equation}\label{eq:llo2}
\L[\bJ] = -G_{Bethe}(\m)-G_{BLR}[\bJ]-\sum_{ij}(J_{ij}^{Bethe}+J_{ij})\hat m_{ij}+o(J^2).
\end{equation}
$G_{BLR}$ is at most quadratic in the $J$'s and contains the local projected Hessian
of the log likelihood onto the magnetization constraints (\ref{eq:constraints}) with
respect to this set of parameters. 
This is nothing else than the Fisher information matrix associated to these parameter
$J$ which is known 
to be positive-semidefinite, which means that the log-likelihood associated to this parameter space 
is convex. Therefore it makes sense to use the quadratic approximation (\ref{eq:llo2}) to 
find the optimal point.

\subsection{Line search along the natural gradient in a reduced space}
Finding the corresponding couplings still amounts 
to solve a linear problem of size $N^2$ in the number of variables which will hardly scale up for 
large system sizes. We have to resort to some simplifications which 
amounts to reduce the size of the problem, i.e. the number of independent couplings. 
To reduce the problem size we can take a reduced number of link into consideration, i.e.
the one associated with a  large mutual information or to partition them in a way which remains 
to decide, into a small number $q$ of group 
$\G_\nu, \nu=1,\ldots q$. Then, to each group $\nu$ is associated a parameter $\alpha_\nu$ 
with a global perturbation of the form
\[
H^1 = \sum_{\nu=1}^q \alpha_\nu H_\nu
\]
where each $H_\nu$ involves the link only present in $\G_\nu$. 
\[
H_\nu \egaldef \sum_{(i,j)\in\G_\nu} J_{ij} s_is_j,
\]
and the $J_{ij}$ are fixed in some way to be discussed soon.
The corresponding constraints, which ultimately insures a max log-likelihood in this reduced parameter space are then 
\[
\frac{\partial G_{BLR}}{\partial \alpha_\nu} = -\hat{\mathbb E}(H_\nu).
\]
This leads to the solution:
\[
\alpha_\mu = \sum_{\nu=1}^q\ \I^{-1}_{\mu\nu}\bigl(\hat{\mathbb E}(H_\nu) - \ExB(H_\nu)\bigr)
\]
where the Fisher information matrix $\I$ has been introduced and which reads in the present case
\begin{equation}\label{eq:Fisher}
\I_{\mu\nu} = \bigl[\cov_{Bethe}(H_\mu,H_\nu)-\sum_{i\ne j\atop(i,j)\in\T} 
[\chi_{Bethe}^{-1}]_{ij}\cov_{Bethe}(H_\mu,s_i)\cov_{Bethe}(H_\nu,s_j)\bigl]
\end{equation}
The interpretation of this solution is to look in the direction of the natural gradient~\cite{Amari,ArAuHaOl} 
of the log likelihood.
The exact computation of  the entries of the Fisher matrix involves up to 4th order moments 
and can be computed using results of Section~\ref{sec:cumulant}.
At this point, the way of choosing the groups of edges and the perturbation couplings
$J_{ij}$ of the corresponding links, leads to various possible algorithms.  
For example, to connect this approach to the one proposed in Section~\ref{sec:onelink}, 
the first group of links can be given by the MST, with parameter $\alpha_0$
and their actual couplings $J_{ij}=J_{ij}^{Bethe}$ at the Bethe approximation; 
making a short list of the $q-1$ best links candidates to be added to the graph, 
according to the information criteria~\ref{eq:onelink}, defines the other groups as singletons. 
It is then reasonable to attach them    
the value 
\[
J_{ij} = \frac{1}{4}\log\frac{\hat p_{ij}^{11}}{p_{ij}^{11}}\frac{\hat p_{ij}^{00}}{p_{ij}^{00}}\frac{p_{ij}^{01}}{\hat p_{ij}^{01}}
\frac{p_{ij}^{10}}{\hat p_{ij}^{10}},
\]
of the coupling according to (\ref{eq:1linkfactor}), while the modification of the 
local fields as a consequence of (\ref{eq:1linkfactor})  can be dropped since 
the Gibbs free energy take  it already into account implicitly,  
in order to maintain single variable magnetization $m_i=\hat m_i$ correctly imposed.

\subsection{Reference point at low temperature}
Up to now we have considered the case where the reference model is supposed to be a tree
and is represented by a single BP fixed point. From the point of view of the Ising model 
this corresponds to perturb a high temperature model in the paramagnetic phase.
In practice the data encountered in applications are more likely to be generated
by  a multi-modal distribution and a low temperature model with many fixed points 
should be more relevant. In such a case we assume that most of the correlations
are already captured by the definition of single beliefs fixed points and the residual correlations
is contained in the co-beliefs of each fixed point. For a multi-modal distribution 
with $q$ modes with weight ${w_k,k=1\ldots q}$ and a pair of variables $(s_i,s_j)$ we indeed have
\begin{align*}
\chi_{ij} &= \sum_{k=1}^q w_k \cov(s_i,s_j\vert k) + 
\sum_{k=1}^q w_k ({\mathbb E}(s_i\vert k)-{\mathbb E}(s_i))({\mathbb E}(s_j\vert k)-{\mathbb E}(s_j))\\[0.2cm]
&\egaldef \chi_{ij}^{intra}+\chi_{ij}^{inter},
\end{align*}
where the first term is the average intra cluster susceptibility while the second 
is the inter cluster susceptibility. All the preceding approach can then 
be followed by replacing the single Bethe susceptibility and 
higher order moments in equations (\ref{eq:gblr},\ref{eq:Fisher})
in the proper way by their multiple BP fixed point counterparts. 
For the susceptibility coefficients, the inter cluster susceptibility coefficients
$\chi^{inter}$ are given directly from the single variable belief fixed points. 
The intra cluster susceptibilities $\chi^{k}$ are 
treated the same way as the former Bethe susceptibility. This means that 
the co-beliefs of fixed points $k\in\{1,\ldots q\}$ are entered in formula
(\ref{eq:invchis}) which by inversion yields the $\chi^{k}$'s, these in turn 
leading to $\chi^{intra}$ by superposition. Higher order moments 
are obtain by simple superposition. Improved models could be then searched along the 
direction indicated by this natural gradient.

\section{Weights propagation on Bethe-dual graphs}\label{sec:dwp}
\subsection{Low temperature expansion }
When the local fields are zero, which in many cases may be obtained with a proper definition of spin 
variables in a given inference problem, a traditional way 
to deal with the low temperature regime is given by the high coupling expansion. This is obtained by 
rewriting 
\begin{equation}\label{eq:highJ}
e^{J_{ij}s_is_j} = \cosh(J_{ij})(1+\tanh(J_{ij})s_is_j).
\end{equation}
Using this simple identity the partition function rewrites
\[
Z(\J) = Z_0\times \sum_{\{\tau_{ij}\in\{0,1\}\}}\prod_{ij}\bigl(\bar \tau_{ij}+
\tau_{ij}\tanh(J_{ij})\bigr)\prod_i\ind{\sum_{j\in \partial i}\tau_{ij}=\ 0\mod 2},
\]
with 
\[
Z_0 = \prod_{(ij)}\cosh(J_{ij}).
\]
The summation over bond variables $\tau_{ij}\in\{0,1\}$, corresponds 
to choosing one of the 2 terms in the factor (\ref{eq:highJ}).
The summation over spin variables then selects bonds configurations 
having an even number of  bonds $\tau_{ij}=1$ attached to each vertex $i$.
From this condition it results that the paths formed by these bonds
must be closed. The contribution of a given path is simply the product
of all bond factor $\tanh(J_{ij})$ along the path. As such the partition function 
is expressed as 
\[
Z(\J) = Z_0\times Z_{loops}
\]
with 
\[
Z_{loops} \egaldef \sum_{\ell}Q_\ell,
\]
where the last sum runs over all possible closed loops $\G_\ell$, i.e. subgraphs  for which each 
vertex has an even degree, including the empty graph and 
\[
Q_\ell \egaldef \prod_{(ij)\in\E_\ell}\tanh(J_{ij}),
\]
where $\E_\ell$ denotes the set of edges involved in loop $\G_\ell$.
This is a special case of 
the loop expansion around a belief propagation fixed point proposed by Chertkov and Chernyak 
in~\cite{chertkov1}. In their case $Z_0$ is replaced by $Z_{Bethe}$ and the loop corrections
runs over all generalized loops, i.e. all subgraphs containing no vertex with degree $1$. 

\begin{figure}[ht]
\centerline{\resizebox*{0.7\textwidth}{!}{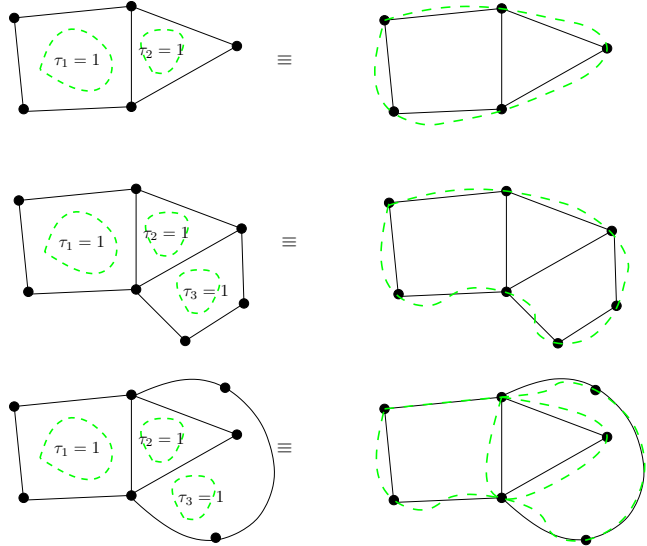}}
\caption{\label{fig:loops} Loops generated from basic cycles combinations.}
\end{figure}

\begin{figure}[ht]
\centerline{\resizebox*{\textwidth}{!}{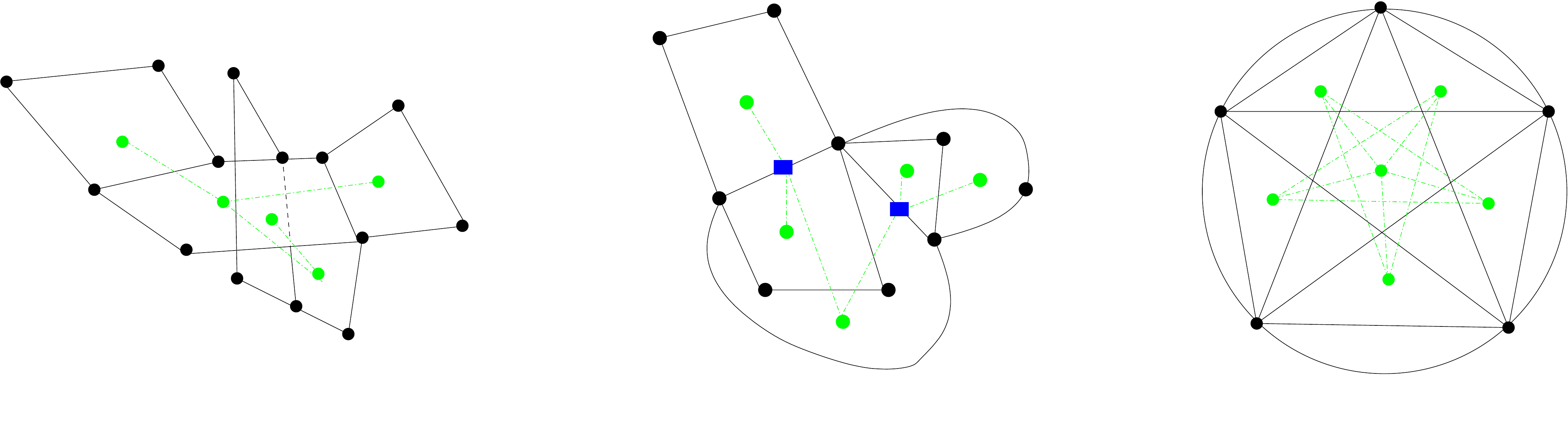}}
\caption{\label{fig:graphs}Examples of pairwise loopy graphs along with one possible dual graph. 
A planar graph with pairwise singly connected 
dual graph (a). A planar pairwise graph with dual three wise  factor graph (b). 
The complete $K_5$ (non-planar) graph and a dual planar pairwise graph (c).}
\end{figure}

If the graph has no loop, the partition function 
reduces to $Z_0$. If there are loops and $k(\G)$ connected components in the graph, 
we may define a set $\{\G_c,c=1,\ldots, C(\G)\}$ of independent cycles,
with $C(\G)=\vert\E\vert-\vert\V\vert + k(\G)$ the so-called cyclomatic number~\cite{Berge} of graph $\G$.
Spanning the set $\{0,1\}^{C(\G)}$ yields all possible loops with the convention that edges are counted
modulo 2 for a given cycle superposition (see Figure~\ref{fig:loops}). 
The partition function can therefore be written as a sum over dual
binary variables $\tau_c\in\{0,1\}$ attached to each cycle $c\in\{1,\ldots \vert C\vert\}$:
\begin{equation}\label{eq:dualmeasure}
Z_{loops} = \sum_{\tau}Q_{\Gs}(\tau),
\end{equation}
where $Q_{\Gs}(\tau)$ represents the weight for any loop configuration specified by $\{\tau_c\}$ on the 
dual (factor-)graph $\Gs$ formed by the cycles. For instance,
when the primal graph $\G$ is a $2$-d lattice, the dual one is also $2$-d and the  
Kramers-Wannier duality allows one to express the partition function at the dual 
coupling $J^\star\egaldef\log(\tanh(J))$ of the associated Ising model on this graph, with spin variable 
$s_c=2\tau_c-1$ attached to each plaquette representing an independent cycle $c$.
For general primal graph $\G$, different cases as illustrated on Figure~\ref{fig:graphs} 
can then be considered  by increasing levels of complexity, 
depending on the properties of $\Gs$. The  vertices of this graph, corresponding to the elements of the cycles basis 
are connected via edges or more generally factors which correspond to common edges in $\G$ shared by these cycles.

If there exists a  basis of disjoint cycles sharing no link in common,
the partition function then factorizes as
\[
Z_{loops} = Z_1 \egaldef \prod_{c=1}^{C(\G)}(1+Q_c),
\]
with 
\[
Q_c \egaldef \prod_{(ij)\in \ell_c}\tanh(J_{ij}),
\]
the weight attached to each cycle $c$. 

If one cannot find such a cycle basis, 
but still assuming there exists a basis such that each link belongs to at most 2 cycles and each cycle 
has a link in common with at most one other cycle, the partition function then reads
\begin{align}
Z_{loops} &= \sum_{\tau}\prod_{c=1}^{C(\G)}(\bar\tau_c+\tau_cQ_c)
\prod_{c,c'}\bigl(\bar\tau_c\bar\tau_{c'}+\tau_c\bar\tau_{c'}+\bar\tau_c\tau_{c'}+
\tau_c\tau_{c'}Q_{cc'}\bigr),\label{eq:Zdual}\\[0.2cm]
&= Z_1\prod_{cc'}\Bigl(1+\frac{Q_c Q_{c'}(Q_{cc'}-1)}{(1+Q_c)(1+Q_{c'})}\Bigr)\egaldef Z_1Z_2,\label{eq:Z2}
\end{align}
where
\[
Q_{cc'}\egaldef \Bigl(\prod_{(ij)\in\G_c\cap\G_{c'}}\tanh(J_{ij})\Bigr)^{-2}.
\]
When the dual graph, i.e. the graph of loops has higher interactions levels, these expressions  
constitute the first and second orders of approximation of a systematic 
cluster expansion taking into account cycle clusters of any size. The more general case 
where some links are common to more than 2 cycles at a time, leads to models 
with higher interaction order than pairwise factors. 
Since the interaction between cycles variables involves 
$\tanh(J_{ij})$ factors, we expect this dual cluster approximation 
to work better when the primal couplings get stronger.
To get the linear response, i.e. the susceptibility matrix, we derive the log partition function
with respect to the couplings $J_{ij}$.
The $0$'th order simply reads:
\[
\frac{\partial \log(Z_0)}{\partial J_{ij}} = \tanh(J_{ij}).
\] 
The first order reads:
\[
\frac{\partial \log(Z_1)}{\partial J_{ij}} = \frac{1-\tanh^2(J_{ij})}{\tanh(J_{ij})}
\sum_{c,(ij)\in\ell_c}\frac{Q_c}{1+Q_c}.
\] 
At second order different terms arise depending on whether $(ij)$ is part of one or two cycles  at a time.
\begin{align*}
\frac{\partial \log(Z_2)}{\partial J_{ij}} &= \frac{1-\tanh^2(J_{ij})}{\tanh(J_{ij})}\Bigl(
\sum_{c,c',\G_c\cap\G_{c'}\ne\emptyset,\atop (ij)\notin\G_c\cap\G_{c'}}
\frac{Q_cQ_{c'}(Q_{cc'}-1)}{(1+Q_c)(1+Q_c+Q_{c'}+Q_cQ_{c'}Q_{cc'})}\\[0.2cm]
&-\sum_{c,c'\atop (ij)\in\ell_c\cap\ell_c'}
\frac{Q_cQ_{c'}}{1+Q_c+Q_{c'}+Q_cQ_{c'}Q_{cc'}}\bigl(
\frac{1+Q_cQ_{cc'}}{1+Q_c}+\frac{1+Q_{c'}Q_{cc'}}{1+Q_{c'}}
\bigr)\Bigr).
\end{align*}
This last expression involves terms of different approximation order at low temperature, 
since $Q_{cc'}-1\approx o(1)$ in that case.
To be consistent let us keep only the second one. Summing up to order one all the preceding 
contributions gives finally a 
set of constraints to be solved of the form
\[
\tanh(J_{ij})+\frac{1-\tanh^2(J_{ij})}{\tanh(J_{ij})}R_{ij} = \hat \chi_{ij},
\]
with 
\[
R_{ij} = 
\begin{cases}
\DD \frac{Q_c}{1+Q_c}\hspace{3.5cm}\  if\ \exists !\ c,\ s.t. (ij)\in\G_c\\[0.2cm]
\DD \frac{Q_c+Q_{c'}}{1+Q_c+Q_{c'}+Q_cQ_{c'}Q_{cc'}}\qquad (ij)\in\G_c\cap\G_{c'}.
\end{cases}
\]
This quantity $R_{ij}$ is build solely on the loops containing the link $(ij)$
restricted to the subset of basic cycles and pair-combinations of basic cycles.

\subsection{Pairwise cycle weights propagation}
This simple cluster expansion might break down rapidly when independent cycles 
start to accumulate to form large sized connected components in the dual graph. Nevertheless, 
if this graph  remains  singly connected, we can set up a message passing procedure 
to compute the exact weights. Let us first restrict the 
discussion to the case where there exists a cycle basis such that the dual cycle graph $\Gs$ is pairwise.
From Mac Lane's planarity criterion this is actually equivalent to having $\G$ planar.

Since the sign of $Q_{\Gs}(\tau)$ is not guaranteed to be positive, 
there is possibly no probability interpretation for these weights.
Nevertheless, we can proceed analogously to ordinary belief propagation.
First define the single and pair cycle's weights:
\begin{align*}
q_c &\egaldef \frac{1}{Z_{loops}}\sum_{\tau}\tau_c Q_{\Gs}(\tau)\\[0.2cm]
q_{cc'} &\egaldef \frac{1}{Z_{loops}}\sum_{\tau}\tau_c\tau_{c'} Q_{\Gs}(\tau).
\end{align*}
From~(\ref{eq:Zdual}) we have
\begin{align}
\chi_{ij} &= \frac{\partial\log(Z_0)}{\partial J_{ij}} + 
\sum_c \frac{\partial\log(Z_{loops})}{\partial Q_c}\frac{\partial Q_c}{\partial J_{ij}}
+\sum_{cc'} \frac{\partial\log(Z_{loops})}{\partial Q_{cc'}}\frac{\partial Q_{cc'}}{\partial J_{ij}}\nonumber\\[0.2cm]
&=\tanh(J_{ij})+\sum_c \frac{q_c}{Q_c}\frac{\partial Q_c}{\partial J_{ij}}
+\sum_{cc'} \frac{q_{cc'}}{Q_{cc'}}\frac{\partial Q_{cc'}}{\partial J_{ij}}\nonumber\\[0.2cm]
&= \tanh(J_{ij})+\frac{1-\tanh^2(J_{ij})}{\tanh(J_{ij})}
\Bigl(\sum_{c,\atop (ij)\in\G c}q_c - 2\sum_{cc',\atop (ij)\in \G_c\cap\G_{c'}}q_{cc'}\Bigr).\label{eq:xijpwdual}
\end{align}
Then, to compute the cycles weights, 
the corresponding message passing procedure involves messages of the 
form
\[
m_{c'\to c}(\tau_c) = (1-m_{c'\to c})\bar\tau_c+m_{c'\to c}\tau_c,
\]
which update rules are given by 
\[
m_{c\to c'} = \frac{1+r_{c\to c'}Q_cQ_{cc'}}{2+r_{c\to c'}Q_c(1+Q_{cc'})},
\]
where 
\[
r_{c\to c'} = \prod_{c''\in\partial c\backslash c'}\frac{m_{c''\to c}}{1-m_{c''\to c}},
\]
$\partial c$ representing the neighborhood of $c$ in $\Gs$.
Finally, letting
\[
\nu_{c\to c'}\egaldef\frac{m_{c\to c'}}{1-m_{c\to c'}},
\]
leads to the following cycle weights propagation update rules:
\begin{align*}
\nu_{c\to c'} &\longleftarrow \frac{1+r_{c\to c'}Q_cQ_{cc'}}{1+r_{c'\to c}Q_c},\\[0.2cm]
r_{c\to c'} &\longleftarrow \prod_{c''\in\partial c\backslash c'}\nu_{c''\to c}.
\end{align*}
From these messages, we obtain the following expressions for the cycles weights:
\begin{align}
q_c &= \frac{Q_cr_c}{1+Q_cr_c}\label{eq:qc}\\[0.2cm]
q_{cc'} &= \frac{Q_cQ_{c'}Q_{cc'}r_{c'\to c}r_{c\to c'}}
{1+Q_cr_{c\to c'}+Q_{c'}r_{c'\to c}+Q_cQ_{c'}Q_{cc'}r_{c'\to c}r_{c\to c'}},\label{eq:qcc}
\end{align}
with 
\[
r_c \egaldef \prod_{c'\in\partial c}\nu_{c'\to c}.
\]
Another useful expression resulting from  belief propagation equation is the partition function 
in terms of the single and pairwise beliefs normalizations. Introducing also
\[
s_c\egaldef \prod_{c'\in\partial c}(1+\nu_{c'\to c})\qquad\text{and}\qquad
s_{c'\to c}\egaldef \prod_{c''\in\partial c'\backslash c}(1+\nu_{c''\to c'}),
\]
we have 
\[
Z_{loops} = \prod_cZ_c\prod_{(cc')\in\cal T}\frac{Z_{cc'}}{Z_cZ_{c'}}.
\] 
with 
\begin{align}
Z_c &= \frac{1+Q_cr_c}{s_c}\label{eq:localzc}\\[0.2cm]
Z_{cc'} &= \frac{1+Q_cr_{c\to c'}+Q_{c'}r_{c'\to c}+Q_cQ_{c'}Q_{cc'}r_{c'\to c}r_{c\to c'}}
{s_{c\to c'}s_{c'\to c}}.\label{eq:localzcc}
\end{align}

\subsection{General cycle weights propagation}
Note that any two cycles having vertices but no edges in common do not interact, so it is maybe not
that obvious that a graph  having a singly connected dual graph must be planar. However 
since the complete graph $K_5$ shown on Figure~\ref{fig:graphs}.c and also the bipartite graph $K_{3,3}$ have 
non-planar dual graphs, from Kuratowski characterisation of planar graphs, this is indeed likely to be the case.
For planar graphs, exact methods have been proposed in the literature
based on Pfaffian's decompositions of the partition 
function~\cite{GlJa,chertkov4} with a computational cost of $O(N^3)$. At least, for the subclass of factor 
graph that we consider, the computational cost becomes linear in the number of cycles which scales anyway like  
$O(N)$ for planar graphs. As far as exact determination of $Z$
is concerned, the pairwise cycle weights propagation described in the preceding section
should be suitable in all relevant cases. However, in practice, finding the proper cycle basis 
might not be always an easy task. Also, by analogy with loopy belief propagation, 
we don't want to limit ourselves to exact cases, and propagating weights on loopy dual graphs could 
lead possibly to interesting approximate results even for non-planar graphs.    

So let us consider the case where some edges are shared by more than two cycles. The dual factor 
graph is constructed by associating one factor to each such edge in addition to the  ones already shared by exactly two 
cycles. The dual loop partition function then reads
\begin{equation}
Z_{loops} = \sum_{\tau}\prod_{c=1}^{C(\G)}(\bar\tau_c+\tau_cQ_c)
\times\prod_{e\in\E}
\left[\sum_{k=0}^{d^\star(e)}
\delta\bigl(k - \sum_{c,\G_c\ni e}\tau_c\bigr)\tanh(J_e)^{-2\lfloor k/2\rfloor}\right],\label{eq:Zloops}
\end{equation}
where $\lfloor x\rfloor$ denotes the entire part of $x$,
$e$ indexes any edge in the original graph $\G$ with $J_e$ the corresponding coupling, 
while $d^\star(e)$ is the degree of the factor associated to $e$
in the dual graph $\Gs$, that is the number of cycles containing $e$. 
In this expression the factor $\tanh(J_e)^{-2\lfloor k/2\rfloor}$ is there to compensate
for overcounting the edge factor $\tanh(J_e)$ when $k$ cycles containing this edge are taken into account. 
Note that if two or more edges are shared 
exactly by the same set of cycles, they should be gathered into a single factor, with $\tanh(J_e)$
simply replaced by the product of hyperbolic tangents corresponding to these edges in the above formula.

From the expression (\ref{eq:Zloops}), the susceptibility coefficient associated to any edge $e\in\E$, reads
\begin{equation}
\chi_e = \tanh(J_e) + \frac{1-\tanh^2(J_e)}{\tanh(J_e)}\Bigl(
\sum_{c\in e}q_c - \sum_{k=2}^{d^\star(e)}2\lfloor \frac{k}{2}\rfloor q_{e,k}\Bigr),\label{eq:linearesponse1}
\end{equation}
where in this case joint weights of interest are given by 
\[
q_{e,k} \egaldef \frac{1}{Z}\sum_{\tau} Q_{\Gs}(\tau)\delta\bigl(k - \sum_{c,\G_c\ni e}\tau_c\bigr).
\]
This generalizes the expression  (\ref{eq:xijpwdual}) obtained for the pairwise dual graph
to arbitrary dual factor graph.

The corresponding message passing algorithm allowing one to compute the weights $q_c$
is now expressed in terms of messages $\nu_{e\to c}$ and $r_{c\to e}$ 
(defined as before from the original messages $m_{e\to c}$)  
respectively from factors to nodes and from nodes to factors.
The update rules read
\begin{align}
\nu_{e\to c} &\longleftarrow \frac{1+\sum_{k=1}^{d^\star(e)-1}
\sum_{c_1,\ldots c_{k}\backslash c\atop e\in \G_{c_i}}\prod_{i=1}^k Q_{c_i}r_{c_i\to e}\ \tanh(J_e)^{-2\lfloor (k+1)/2\rfloor}}
{1+\sum_{k=1}^{d^\star(e)-1}
\sum_{c_1,\ldots c_{k}\backslash c\atop e\in \G_{c_i}}
\prod_{i=1}^k Q_{c_i}r_{c_i\to e}\ \tanh(J_e)^{-2\lfloor k/2\rfloor}},\label{eq:gcwpup}\\[0.2cm]
r_{c\to e} &\longleftarrow \prod_{e'\ni c\backslash e}\nu_{e\to c}.\nonumber
\end{align}
In terms of the messages, after adapting the definition of $r_c$, the expression (\ref{eq:qc}) 
of $q_c$ remains valid 
while  (\ref{eq:qcc}) generalizes to the new weights:
\[
q_{e,k} = \frac{\sum_{c_1,\ldots c_{k}\atop e\in \G_{c_i}}\prod_{i=1}^k Q_{c_i}r_{c_i\to e}\ \tanh(J_e)^{-2\lfloor (k+1)/2\rfloor}}
{1+\sum_{k=1}^{d^\star(e)}
\sum_{c_1,\ldots c_{k}\atop e\in \G_{c_i}}
\prod_{i=1}^k Q_{c_i}r_{c_i\to e}\ \tanh(J_e)^{-2\lfloor k/2\rfloor}}
\]
which coincides with (\ref{eq:qcc}) for $k=2$.
Concerning local partition functions, (\ref{eq:localzc}) remains unchanged after adapting the definition 
of $r_c$ and $s_c$, while (\ref{eq:localzcc}) generalizes to 
\[
Z_e = \frac{1+ \sum_{k=1}^{d^\star(e)}
\sum_{c_1,\ldots c_{k}\atop e\in \G_{c_i}}\prod_{i=1}^k Q_{c_i}r_{c_i\to e}\ \tanh(J_e)^{-2\lfloor (k+1)/2\rfloor}}
{\prod_{i=1}^k s_{c_i\to e}}
\]
so that we finally have
\[
Z_{loops} = \prod_e\frac{Z_e}{\prod_{c\in e}Z_c}\prod_c Z_c,
\]
when the dual graph is singly connected.

\subsection{Extended pairwise dual-graph and dual weight propagation}
When the degree of factors get larger, the combinatorial burden to evaluate $q_{e,k}$ and 
associated messages for $k\gg 1$, becomes intractable. Coming back to (\ref{eq:Zloops}) let us remark first the 
following simplification in the way to write each factor:
\begin{align*}
\sum_{k=0}^{d^\star(e)}
\delta\bigl(k - \sum_{c,e\in\G_c}\tau_c\bigr)t_e^{-2\lfloor k/2\rfloor} &= \frac{1}{2}\Bigl[
\prod_{c,e\in\G_c}(\bar\tau_c+\tau_ct_e^{-1})+\prod_{c,e\in\G_c}(\bar\tau_c-\tau_ct_e^{-1})\\[0.2cm]
&+ t_e\Bigl(\prod_{c,e\in\G_c}(\bar\tau_c+\tau_ct_e^{-1})-\prod_{c,e\in\G_c}(\bar\tau_c-\tau_ct_e^{-1})\Bigr)\Bigr],\\[0.2cm]
= \frac{1}{2}\Bigl[\prod_{c,e\in\G_c}&(\bar\tau_c+\tau_ct_e^{-1})(1+t_e)+
\prod_{c,e\in\G_c}(\bar\tau_c-\tau_ct_e^{-1})(1-t_e)\Bigr],
\end{align*}
after separating the odd and even part in $k$,
with the notation $t_e\egaldef \tanh(J_e)$. This suggests the introduction of an additional 
binary variable $\sigma_e\in\{-1,1\}$ associated to each factor $e$,  such that the loop partition function now 
reads
\begin{equation}
Z_{loops} = \sum_{\tau,\sigma} Q_{\G^\star}(\tau,\sigma),\label{eq:Zdwp}
\end{equation}
with 
\begin{equation}
Q_{\G^\star}(\tau,\sigma) \egaldef \prod_{c=1}^{C(\G)}(\bar\tau_c+\tau_cQ_c)
\times\prod_{e\in\E}\frac{1+\sigma_e t_e}{2}
\prod_{c,e\in\G_c}(\bar\tau_c+\tau_c\frac{\sigma_e}{t_e}),\label{def:Qtausigma}
\end{equation}
i.e. expressing it as a sum over cycles and edges binary variables, 
of a joint weight measure corresponding to an extended pairwise factor graph, 
containing cycle-edges interactions. 
Concerning the susceptibility this new formulation leads to a simpler
expression. Indeed, for any edge $(ij)=e\in\G$, 
after deriving $Z(\J)$ with respect to $J_e$ and arranging the terms
we finally obtain
\begin{equation}
\chi_{e} = 2q_e -1,\label{eq:linearesponse2} 
\end{equation}
with
\[
q_e \egaldef \frac{1}{Z_{loops}} \sum_{\tau,\sigma}\frac{1+\sigma_e}{2} Q_{\G^\star}(\tau,\sigma).
\]
In the case where $e$ is part of a larger factor $f$ containing other edges shared exactly by the 
same set of cycles, this formula should be slightly modified.
The susceptibility $\chi_e$ is then expressed in terms of the weight $q_f$ associated to this factor and reads:
\[
\chi_e = \frac{t_e^2-t_f^2}{t_e(1-t_f^2)}+\frac{t_f}{t_e}\frac{1-t_e^2}{1-t_f^2}\bigl(2q_f-1),
\]
where $t_f$ is now the product of $\tanh(J_{e'})$ of all edges $e'$ represented by this factor, including $e$.
Note that there is no approximation at this point.
$2q_e-1$ represents the  dual "magnetization'' associated to variables $\sigma_e$.
$\Gs$ has the same structure as before, except that factors have been replaced by vertices. 

Assuming $\Gs$ singly connected we may again settle a message passing procedure in order to compute these weights. 
We have now to distinguish between messages $m_{c\to e}(\sigma_e)$ sent by vertex's cycles to vertex's edges
and $m_{e\to c}(\tau_c)$  sent by edges to cycles vertices.
Letting 
\[
\nu_{e\to c} \egaldef \frac{m_{e\to c}(\tau_c=1)}{m_{e\to c}(\tau_c=0)}\qquad\text{and}\qquad
\nu_{c\to e} \egaldef \frac{m_{c\to e}(\sigma_e=1)}{m_{c\to e}(\sigma_c = -1)},
\] 
we come up with the following update rules:
\begin{equation}
\begin{cases}
\DD \nu_{e\to c} \longleftarrow \frac{r_{e\to c}e^{2J_e}-1}{r_{e\to c}e^{2J_e}+1}\ \coth(J_e),\\[0.4cm]
\DD \nu_{c\to e} \longleftarrow \frac{1+r_{c\to e} Q_c\coth(J_e)}{1-r_{c\to e} Q_c\coth(J_e)},\\[0.4cm]
\DD r_{e\to c} \longleftarrow \prod_{c'\in\partial e\backslash c}\nu_{c'\to e}.\\[0.4cm]
\DD r_{c\to e} \longleftarrow \prod_{e'\in\partial c\backslash e}\nu_{e'\to c}.
\end{cases}\label{eq:dwpupdate}
\end{equation}
After convergence we get for the edge weights:
\[
q_e = \frac{e^{2J_e}\ r_e}{1+e^{2J_e}\ r_e},\qquad\text{with}\qquad r_e \egaldef \prod_{c\in\partial e}\nu_{c\to e}.
\]
Finally to compute the dual partition function we need as before to compute local ones which 
read:
\[
\begin{cases}
\DD Z_c = \frac{1+Q_cr_c}{s_c}\\[0.4cm]
\DD Z_e = \frac{e^{-J_e}+e^{J_e}r_e}{2s_e\cosh(J_e)}\\[0.4cm]
\DD Z_{ec} = 
\frac{e^{-J_e}+e^{J_e}r_{e\to c}-Q_ce^{-J_e} \coth(J_e)r_{c\to e}+Q_ce^{J_e}\coth(J_e)r_{e\to c}r_{c\to e}}{2s_{e\to c}s_{c\to e}\cosh(J_e)},
\end{cases}
\]
with now
\begin{align*}
&s_c\egaldef \prod_{e\in\partial c}(1+\nu_{e\to c}),\qquad
s_e\egaldef \prod_{c\in\partial e}(1+\nu_{c\to e}),\qquad
r_{c\to e}\egaldef \prod_{e'\in\partial c\backslash e}\nu_{e'\to c},\\[0.2cm]
&s_{c\to e}\egaldef \prod_{e'\in\partial c\backslash e}(1+\nu_{e'\to c}),\qquad\text{and\qquad}
s_{e\to c}\egaldef \prod_{c'\in\partial e\backslash c}(1+\nu_{c'\to e}),
\end{align*}
to obtain 
\[
Z_{loops} = \prod_{(e,c)\in\E^\star}\frac{Z_{ec}}{Z_eZ_c}\prod_{c\in\G^\star}Z_c\prod_{e\in\G^\star}Z_e.
\]
In practice when the connectivity $d^\star(e)$ of an edge variable $e$ is not greater than $2$, 
the corresponding variables $\sigma_e$ maybe summed out beforehand,  
so that finally the joint weight measure associated to $Z_{loops}$
involves cycle-cycle and cycle-edges interactions given in (\ref{eq:Zdual}) and (\ref{def:Qtausigma}).

\subsection{Linear response theory}
Using this extended pairwise dual model and the corresponding fixed point of 
DWP, we can derive the linear response theory for any pair of nodes $i\in\G$ and $j\in\G$, 
when the dual graph $\Gs \equiv \T$ forms a tree.

\begin{figure}[ht]
\centerline{\resizebox*{0.5\textwidth}{!}{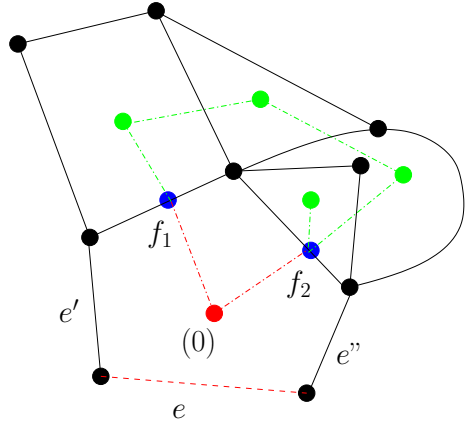}}
\caption{\label{fig:lineresp} Example of dual extended graph for computing $\chi_e$ for a given edge $e\notin\E$.}
\end{figure}

First, for any edge $e\in\E$, either of expression 
(\ref{eq:xijpwdual}), (\ref{eq:linearesponse1}) and (\ref{eq:linearesponse2}) can be used in principle
to determine $\chi_e$, except that the first one requires a good choice of the cycle basis, while the others two 
need to care less about it, as long as the dual graph remains singly connected. 

If $e\notin\E$, then let us give the index $0$
to the new independent cycle which is formed by adding $e$ to the initial graph $\G$ with some 
arbitrary coupling $J_e$ (see Figure~\ref{fig:lineresp}). 
The corresponding susceptibility then reads
\[
\chi_e = \frac{d\log Z_{loops}(J_e)}{dJ_e}\Big\vert_{J_e=0},
\] 
where, from (\ref{eq:Zdwp},\ref{def:Qtausigma}), 
\[
Z_{loops}(J_e) = \sum_{\tau,\sigma,\tau_0} Q_{\G^\star}(\tau,\sigma)\bigl(\bar\tau_0+\tau_0Q_0(J_e)\bigr)
\prod_{e'\in\G_0\backslash e}\bigl(\bar\tau_0+\tau_0\frac{\sigma_{e'}}{t_{e'}}\bigr),
\]
with cycle factor 
\[
Q_0(J_e) \egaldef \prod_{(ij)\in\G_0} \tanh(J_{ij}).
\]
Let 
\[
Q_{0\backslash e} \egaldef Q_0(J_e)/\tanh(J_e),
\]
the cycle weight, where the edge $e$ is not taken into account, which by definition is independent of $J_e$.
We have now
\begin{align*}
\chi_e &= \frac{Q_{0\backslash e}}{Z_{loops(0)}}\sum_{\tau,\sigma}Q_{\G^\star}(\tau,\sigma)
\prod_{e'\in\G_0\backslash e}\frac{\sigma_{e'}}{t_{e'}}\\[0.2cm]
&\egaldef <\prod_{e'\in\G_0\backslash e}\sigma_{e'}>_{Q_{\G^\star}},
\end{align*}
where this last quantity represents ``pseudo moment'' of $\sigma_{e'}$ variables  under the 
weight measure $Q_{\G^\star}$ given in (\ref{def:Qtausigma})).

Various cases may be considered, depending on how the additional cycle node is connected to $\Gs$.
In absence of connection, all variables $\sigma_{e'}$ in the above expression are independent and free, 
i.e. $2q_{e'}-1 = \tanh(J_{e'})$ so we get 
\[
\chi_e = Q_{0\backslash e}.
\]
Next, if the extra cycle variable is connected to a single factor $f$, the susceptibility reads
\[
\chi_e = \frac{(2q_f-1)}{t_f} Q_{0\backslash e}.
\]
When more than one factor $f_1,\ldots,f_k$ are involved, as in the example of Figure~\ref{fig:lineresp},
corresponding variables $\sigma_{f_i}$ are not independent and
\[
\chi_e \approx Q_{0\backslash e}\prod_{i=1}^k \frac{2q_{f_i}-1}{t_{f_i}},
\]
can lead to a poor approximation if the dependencies between these variables are not small. 
Nevertheless, it is still possible from the analysis of Section~\ref{sec:cumulant} 
to take into account these ``correlations''.
The explicit formulas given in this latter section make only use of the factorization~(\ref{eq:ansatz}) 
of the joint measure in addition to the tree structure property of the factor graph. They are limited to 
$4$-points susceptibility coefficients, but could be in principle generalized to any cluster size.
They are still formally valid in the present context, where the  
knowledge of the DWP fixed point gives directly exact 
dual pairwise weight's ``susceptibilities'' $\chi_{cc'}^{\star}$ or $\chi_{ce}^{\star}$, 
for any pair $(c,c')\in\E^\star$ or $(c,e)\in\E^\star$ forming an edge in $\G^\star$.
Therefore, in principle the complete determination of the linear response 
can be done with DWP, with a computational cost not exceeding $O(N^2)$. 
The limitation here is simply the size of the largest cycles, which then possibly requires 
to have explicit expressions of dual susceptibilities of corresponding cluster size, 
which in effect is arbitrarily limited  to $4$ in the present work.

\section{$L_0$ norm penalized sparse inverse estimation algorithm}\label{sec:penal}
We propose here to use the Doubly Augmented Lagrange (DAL)
method~\cite{IusemDAL,EcksteinMOR,DongJSC} to solve the 
penalized log-determinant programming in (\ref{eq:logdetopt}). For
a general problem defined as follows:
\begin{equation}\label{eq:dalbasicpro}
\min_x F(x) = f(x) + g(x)
\end{equation}
where $f(x)$ and $g(x)$ are both convex. DAL  splits the combination
of $f(x)$ and $g(x)$ by introducing a new auxiliary variable $y$.
Thus, the original convex programming problem can be formulated as :
\begin{equation}\label{eq:dalbasicsplit}
\begin{aligned}
\min_{x,y} F(x) = f(x) + g(y) \\
\text{s.t. } \; \;  {x - y = 0}
\end{aligned}
\end{equation}
Then it advocates an augmented Lagrangian method to the extended cost
function in (\ref{eq:dalbasicsplit}). Given penalty parameters $\mu$
and $\gamma$, it minimizes the augmented Lagrangian function
\begin{equation}\label{eq:dalbasicsfunc}
L(x,y,\nu,\tilde{x},\tilde{y})
  = f(x) + g(y) + \langle \nu, x-y \rangle
    + \frac{\mu}{2} \|x-y\|_{2}^2
    + \frac{\gamma}{2} \|x-\tilde{x}\|_{2}^2
    + \frac{\gamma}{2} \|y-\tilde{y}\|_{2}^2
\end{equation}
where $\tilde{x}$ and $\tilde{y}$ are the prior guesses of $x$ and $y$
that can obtained either from a proper initialization or the estimated
result in the last round of iteration in an iterative update
procedure. Since optimizing jointly with respect to $x$ and $y$ is
usually difficult, DAL optimizes $x$ and $y$ alternatively.
That gives the following iterative alternative update algorithm with
some simple manipulations:
\begin{equation}\label{eq:dalaltfunc}
\begin{aligned}
x^{k+1} &= \min_x f(x) + \frac{\mu}{2}\|{x - y^k + {\tilde{\nu}}^k}\|_{2}^2
  + \frac{\gamma}{2}\|x-x^k\|_{2}^2 \\
y^{k+1} &= \min_y g(y) + \frac{\mu}{2}\|{x^{k+1} - y + {\tilde{\nu}}^k}\|_{2}^2 + \frac{\gamma}{2}\|y-y^k\|_{2}^2 \\
{\tilde{\nu}}^{k+1} &= {\tilde{\nu}}^k + x^{k+1} - y^{k+1} 
\end{aligned}
\end{equation}

where $\tilde{\nu} = {\frac{1}{\mu}}{\nu}$. As denoted in~\cite{DongJSC} and~\cite{IusemDAL}, 
DAL  improves basic augmented Lagrangian optimization by performing additional
smooth regularization on estimations of $x$ and $y$ in successive
iteration steps. As a result, it guarantees not only the
convergence of the scaled dual variable $\tilde{\nu}$, but also that of the proximal
variables $x^k$ and $y^k$, which could be divergent in basic augmented
Lagrangian method.

We return now to the penalized log-determinant programming in sparse
inverse estimation problem, as seen in (\ref{eq:logdetopt}). The
challenge of optimizing the cost function is twofold.
Firstly, the exact $L_0$-norm penalty is non-differentiable, making
it difficult to find an analytic form of gradient for optimization.
Furthermore, due to the log-determinant term in the cost function, it
implicitly requires that any feasible solution to the sparse approximation $A$ of the precision matrix should be strictly
positive definite. The gradient of the log-determinant term is given
by $\hat C - A^{-1}$, which is not continuous in the positive definite
domain and makes it impossible to obtain any second-order derivative
information to speed up the gradient descent procedure. We hereafter
use $S_{++}$ as the symmetric positive definite symmetric matrices that form the
feasible solution set for this problem. By applying DAL to the
cost function (\ref{eq:logdetopt}), we can derive the following
formulation:
\begin{equation}\label{eq:dalopt}
\begin{aligned}
\hat{J}(A,Z,\tilde{A}, \tilde{Z},\nu) 
=& -\log\det(A) + \Tr(\hat C A) + \lambda P(Z) + \langle \nu , A-Z \rangle\\
& + \frac{\mu}{2} \|A-Z\|_{2}^2 
  + \frac{\gamma}{2} \|A-\tilde{A}\|_{2}^2
  + \frac{\gamma}{2} \|Z-\tilde{Z}\|_{2}^2 \\
\text{s.t.} & \;\; {A, Z} \in S_{++}
\end{aligned}
\end{equation}
where $Z$ is the auxiliary variable that has the same dimension as
the sparse inverse estimation $A$. $\tilde{A}$ and $\tilde{Z}$ are the estimated values of $A$ and
$Z$ derived in the last iteration step. The penalty parameter
$\gamma$ controls the regularity of $A$ and $Z$. By optimizing $A$ and $Z$ alternatively, the DAL procedure can be
easily formulated as an iterative process as follows, for some $\delta > 0$:
\begin{equation}\label{eq:daliter}
\begin{aligned}
 A^{k+1} &=\argmin_A  -\log\det(A) + \Tr(\hat C A) + \lambda P({Z}^{k}) + \langle {\nu}^{k} , A-{Z^k} \rangle \\
&+ \frac{\mu}{2} \|A-{Z^k}\|_{2}^2 + \frac{\gamma}{2}\|A-{A^k}\|_{2}^2 \\
Z^{k+1} &=\argmin_Z \lambda P(Z) + \langle \nu^{k} , A^{k+1}-Z\rangle
  + \frac{\mu}{2} \|A^{k+1}-Z\|_{2}^2 \\
&+ \frac{\gamma}{2}\|Z-{Z^k}\|_{2}^2 \\
& {\nu}^{k+1} = {\nu}^{k} + \delta( A^{k+1} - Z^{k+1}) \\
\text{s.t.} & \; \; A^{k+1}, Z^{k+1} \in S_{++}
\end{aligned}
\end{equation}

By introducing the auxiliary variable $Z$, the
original penalized maximum likelihood problem is decomposed into two
parts. The first one is composed mainly by the convex log-determinant
programming term. Non-convex penalty is absorbed into the left part.
Separating the likelihood function and the penalty leads to
the simpler sub-problems of solving log-determinant programming using
eigenvalue decomposition and $L_0$ norm penalized sparse learning 
alternatively. Each sub-problem contains only one single variable,
making it applicable to call gradient descent operation to search
local optimum. Taking $\tilde{\nu} = {\frac{1}{\mu}}{\nu}$, we can derive the following
scaled version of DAL for the penalized log-determinant programming:
\begin{equation}\label{eq:dalscale}
\begin{aligned}
A^{k+1} &=\argmin_A  -\log\det(A) + \Tr({\hat C}A) 
  + \frac{\mu}{2} \bigl\|A-Z^{k} + \tilde{\nu}^k\bigr\|_{2}^2 
  + \frac{\gamma}{2} \bigl\|A-A^{k}\bigr\|_{2}^2 \\
Z^{k+1} &=\argmin_Z \frac{\mu}{2} \bigl\|A^{k+1}-Z + \tilde{{\nu}}^{k}\bigr\|_{2}^2
  +  \frac{\gamma}{2} \bigl\|Z-Z^{k}\bigr\|_{2}^2 + \lambda P(Z) \\
\tilde{{\nu}}^{k+1} &= \tilde{{\nu}}^{k} + A^{k+1} - Z^{k+1} \\
\text{s.t.} & \; \; {A^{k+1}, Z^{k+1}} \in S_{++}
\end{aligned}
\end{equation}

To attack the challenge caused by non-differentiability of the exact $L_0$ norm penalty, we make use of a differentiable
approximation to $L_0$-norm penalty in the cost function $\hat{J}$,
named as "seamless $L_0$ penalty'' (SELO) in~\cite{DickerSELO}. The
basic definition of this penalty term is given as:
\begin{equation}\label{eq:selopenalty}
P_{\text{SELO}}(Z) = \sum_{i,j} {\frac{1}{\log(2)}} \log( 1 + {\frac{|Z_{i,j}|}{|Z_{i,j}|+{\tau}}} )
\end{equation}
where $Z_{i,j}$ denotes individual entry in the matrix $Z$ and ${\tau}
> 0$ is a tuning parameter. As seen in Figure~\ref{fig:L0L1penfuncf}, as
$\tau$ gets smaller, $P(Z_{i,j})$ approximates better the $L_0$ norm
$I(Z_{i,j} \not= 0)$. SELO penalty is differentiable, thus we can
calculate the gradient of $P(Z)$ explicitly with respect to each
$Z_{i,j}$ and make use of first-order optimality condition to search
local optimum solution. Due to its continuous property, it is more
stable than the exact $L_0$ norm penalty in optimization. As proved
in~\cite{DickerSELO}, the SELO penalty has the oracle property with
proper setting of $\tau$. That's to say, the SELO penalty is
asymptotically normal with the same asymptotic variance as the unbiased
OLS estimator in terms of Least Square Estimation problem. Furthermore, 
if we perform local first-order Taylor expansion to the SELO penalty term, we 
can find the intrinsic relation between the SELO penalty between robust 
M-estimator, which explains the stability of SELO against noise. We describe 
this part in the followings. 

The first two steps in (\ref{eq:dalscale}) are performed with the
positive definite constrains imposed on $A$ and $Z$. The minimizing with
respect to $A$ is accomplished easily by performing Singular Vector
Decomposition (SVD). By calculating the gradient of $\hat{J}$ with respect
to A in (\ref{eq:dalscale}), based on the first-order optimality, we
derive:
\begin{equation}\label{eq:svdsol1}
{\hat C} - A^{-1} + {\mu} (A - Z^{k} + \tilde{{\nu}}^{k}) + {\gamma}(A - A^{k}) = 0
\end{equation}

Based on generalized eigenvalue decomposition, it is easy to verify that ${A^{k+1}} = V\diag(\beta){V^{T}}$, 
where $V$ and $\{d_i\}$ are the eigenvectors and eigenvalues of 
${\mu} (Z^k - \tilde{\nu}^k) - {\hat C} + {\gamma}{A^k}$. ${\beta}_{i}$ is defined as:
\begin{equation}\label{eq:svdsol2}
{\beta}_{i} = \frac{d_i + \sqrt[]{{d_i}^2 + 4(\tilde{{\nu}}+{\gamma})}}{2(\tilde{\nu} + {\gamma})}
\end{equation}

Imposing $Z\in S_{++}$ directly in minimizing the cost function with
respect to $Y$ make the optimization difficult to solve. Thus,
instead, we can derive a feasible solution to $Z$ by a continuous
search on $\mu$. Based on spectral decomposition, it is clear that
$X^{k+1}$ is guaranteed to be positive definite, while it is not
necessarily sparse. In contrast, $Z$ is regularized to be sparse while
not guaranteed to be positive definite. $\mu$ is the regularization
parameter controlling the margin between the estimated $X^{k+1}$ and the
sparse $Z^{k+1}$. Increasingly larger $\mu$ during iterations makes the
sequences $\{X^{k}\}$ and $\{Z^{k}\}$ converge to the same point
gradually by reducing margin between them. Thus, with enough iteration
steps, the derived $Z^k$ follows the positive definite constraint and
sparsity constraint at the same time. We choose here to  increase $\mu$
geometrically with a positive factor
${\eta} > 1$ after every $N_{\mu}$ iterations until its value 
achieves a predefined upper bound $\mu_{\max}$. With this
idea, the iterative DAL solution to the $L_0$ norm penalty is given
as:
\begin{equation}\label{eq:dalselo}
\begin{aligned}
A^{k+1} &=\argmin_A  -\log\det(A) + \Tr({\hat C}A) 
  + \frac{\mu}{2} \bigl\|A-Z^{k} + \tilde{{\nu}}^{k}\bigr\|_{2}^2 
  + \frac{\gamma}{2} \bigl\|A-A^{k}\bigr\|_{2}^2, \\
Z^{k+1} &=\argmin_Z  \frac{\mu}{2} \bigl\|A^{k+1}-Z + \tilde{{\nu}}^{k}\bigr\|_{2}^2 
  + \frac{\gamma}{2} \bigl\|Z-Z^{k}\bigr\|_{2}^2 + \lambda P(Z)\\
\tilde{{\nu}}^{k+1} &= \tilde{{\nu}}^{k} + A^{k+1} - Z^{k+1} \\
\mu^{k+1} &= \min\bigl(\mu\,\eta^{\lfloor k/N_\mu\rfloor},\mu_{\max}\bigr).
\end{aligned}
\end{equation}

By alternatively optimizing w.r.t $A$ and the auxiliary matrix $Z$, we reduce 
margin between $A$ and $Z$ gradually, which finally leads to estimation of sparse graph structure 
and maximizing likelihood of the sparse graph simultaneously. As we can find in the second step of 
(\ref{eq:dalselo}), the guess of sparse graph structure is obtained by solving 
penalized least square regression w.r.t $Z$. Besides its analytical convenience, 
this compact functional form also provides a clear view about intrinsic link between 
the SELO penalized sparse inverse and robust M-estimator, which indicates 
superior stability of the approximated L0 norm penalty over lasso penalization. Due to concavity of the SELO penalty
, we can make use of a Taylor expansion of the SELO penalty as its tight upper bound within a local neighborhood. 
Therefore, minimizing the linear upper bound given good initialization point can  
generate the same optimization path as minimizing the penalty term directly.
Based on this equivalency relation, we replace $P(Z)$ in (\ref{eq:dalselo}) with its first order 
Taylor expansion $g$, which reads:

\begin{equation}\label{eq:robustselo}
\begin{aligned}
Z^{k+1} &=\argmin_Z  \frac{\mu}{2} \bigl\|Z - A^{k+1} + \tilde{{\nu}}^{k}\bigr\|_{2}^2 
  + \frac{\gamma}{2} \bigl\|Z-Z^{k}\bigr\|_{2}^2 + \lambda \sum_{i,j} {g(Z_{i,j})}\\
\end{aligned}
\end{equation}

where $g(Z_{i,j})$ is defined as
\[
 g(Z_{i,j}) = P(Z_{i,j}^{\text{old}}) + \frac{\tau sgn(Z_{i,j}^{\text{old}})}{\log(2) 
(\tau + \|Z_{i,j}^{\text{old}}\|) (\tau + 2\|Z_{i,j}^{\text{old}}\|)} (Z_{i,j} - Z_{i,j}^{\text{old}})
\]

$sgn(Z_{i,j})$ is the sign of the scalar value $Z_{i,j}$ and $Z_{i,j}^{\text{old}}$ is a given initializing 
guess of $Z_{i,j}$, which is a constant term. Minimizing the first-order 
Taylor expansion w.r.t $Z$ is equal to solve a weighted linear regression, with the weight of each $Z_{i,j}$ 
$w_{i,j} = \frac{\tau sgn(Z_{i,j}^{\text{old}})}{\log(2) (\tau + \|Z_{i,j}^{\text{old}}\|) (\tau + 2\|Z_{i,j}^{\text{old}}\|)}$. 
The entry-wise weight value declines to zero quickly when the magnitude of the corresponding $Z_{i,j}$ increases. With such 
weight configuration, only $Z_{i,j}$ approaching to zero has strong influence to the gradient direction, which 
pursues a self-adaptive pruning of entries with small magnitudes. $Z_{i,j}$ with large magnitudes are free from
unnecessary sparsity penalization. This weight setting is intrinsically consistent with the basic idea of robust 
M-estimator. According to ~\cite{ChoQUIC}, robust M-estimator is designed to avoid 
penalizing outliers inducing large bias to the model. Through under-weighting outlier training data, the derived M-estimator 
gains stability against noisy data. In our case, $Z_{i,j}$ with underlying large non-zero magnitude encodes critical 
correlation between random variables. Forcing these terms to be sparse thus results in artifacts in the graph structure 
estimation.  In this sense, the SELO penalty benefits from the consistent design to the robust statistics and 
obtains stable estimation of sparse correlation structure in the graph. In fact, the weight setting in the first 
order Taylor expansion is a close approximate to the robust M-estimator with its $\psi$ function defined by Cauchy distribution. 
It verifies the intrinsic robustness of SELO penalty in a further step. In contrast, the first order expansion 
of the $L1$ norm penalty $\|Z\|$ gives constant weight to all entries in the linear term. Therefore, compared with 
the SELO penalty, $L1$ norm penalty is likely to result in bias in graph structure estimation. 

In the second step of (\ref{eq:dalselo}), we calculate the gradient of
the cost function with respect to $Z$ and achieve the local
minimum by performing the first-order optimum condition on it.
Therefore, the updated value of each entry of $Z$ is given by a root
of a cubic equation, as defined below:
\begin{equation}\label{eq:dalzupdate}
\begin{aligned}
\text{if} & \; {Z_{i,j} > 0}, \; Z_{i,j}\text{ is the positive root of} \\
& 2{Z_{i,j}}^{3} + (3{\tau} - 2{{\theta}_{i,j}}){Z_{i,j}}^{2} + 
({\tau}^2 - 3{\tau}{\theta_{i,j}}){Z_{i,j}} - {{\tau}^{2}}{\theta_{i,j}} + \frac{{\lambda}{\tau}}{{\mu} + {\gamma}} = 0\\
\text{if} & \; {Z_{i,j} < 0}, \; Z_{i,j}\text{ is the negative root of} \\
& 2{Z_{i,j}}^{3} - (3{\tau} + 2{{\theta}_{i,j}}){Z_{i,j}}^{2} + 
({\tau}^2 + 3{\tau}{\theta_{i,j}}){Z_{i,j}} - {{\tau}^{2}}{\theta_{i,j}} - \frac{{\lambda}{\tau}}{{\mu} + {\gamma}} = 0\\ 
&\text{else} \; {Z_{i,j} = 0}
\end{aligned}
\end{equation}
where $Z_{i,j}$ is one single entry of $Z$ and 
\[
\theta_{i,j} =\frac{\gamma Z_{i,j}^k + \mu(A_{i,j}^{k+1} + {\tilde{\nu}}^k)}
                  {\mu +\gamma}.
\]
Solving the cubic equations can be done rapidly using
Cardano's formula within a time cost $O(n^2)$. Besides, the spectral
decomposition procedure has the general time cost $O(n^3)$. Given the
total number of iterations $K$, theoretical computation
complexity of DAL is $O(K{n}^3)$. For our experiments,  we initialize $\mu$ to $0.06$, the multiplier factor $\eta$ to 
$1.3$ and the regularization penalty parameter $\gamma$ to $10^{-4}$.
To approximate the $L_0$ norm penalty, $\tau$ is set to be $5\cdot 10^{-4}$. In
our experiment, to derive the Pareto curve of the optimization result,
we traverse different values of $\lambda$. Most learning
procedures converge with no more than $K=500$ iteration steps.

To validate performance of sparse inverse estimation 
based on the $L_0$ norm penalty, we involve an alternative sparse inverse matrix learning method using $L_1$ norm penalization 
for comparison. Taking $P(A)$ in (\ref{eq:logdetopt}) to be
the $L_1$ matrix norm of $A$, we strengthen conditional dependence structure between random variables 
by jointly minimizing the negative log likelihood function and the $L_1$ norm penalty of the inverse
matrix. Since $L_1$ norm penalty is strictly convex, we can use a
quadratic approximation to the cost function to search for the global
optimum, which avoids singular vector decomposition with complexity of
$O(p^3)$ and improves the computational efficiency of this solution to
$O(p)$,  where $p$ is the number of random variables in the GMRF model.
This quadratic approximation based sparse inverse matrix learning is given
in~\cite{ChoQUIC}, named as QUIC. We perform it directly on the
empirical covariance matrix with different settings of the
regularization coefficient $\lambda$. According to works in compressed sensing, the equality between $L_1$ norm
penalty and $L_0$ norm penalty holds if and only if the design matrix satisfies restricted isometry property.
However, restricted isometry property is sometimes too strong in
practical case. Furthermore, to our best knowledge, there is no similar necessary condition guaranteeing 
equivalence between $L_1$ and $L_0$ norm
penalty in sparse inverse estimation problem. Therefore, in our case, $L_1$ norm penalized log-determinant programming
is highly likely to be biased from the underlying sparse correlation structure in
the graph, which leads to much denser inverse matrices.

\section{Experiments}\label{sec:experiments}
In this section, various solutions based on the different methods
exposed before are compared. 
\paragraph{Inverse Ising problem}
Let us start with the inverse Ising problem. The first set of
experiments illustrates how the linear-response approach exposed in Section~\ref{sec:prelim} 
works when the underlying model to be found is itself an Ising model.
\begin{figure}[ht]
\centering
\includegraphics*[width=0.49\columnwidth]{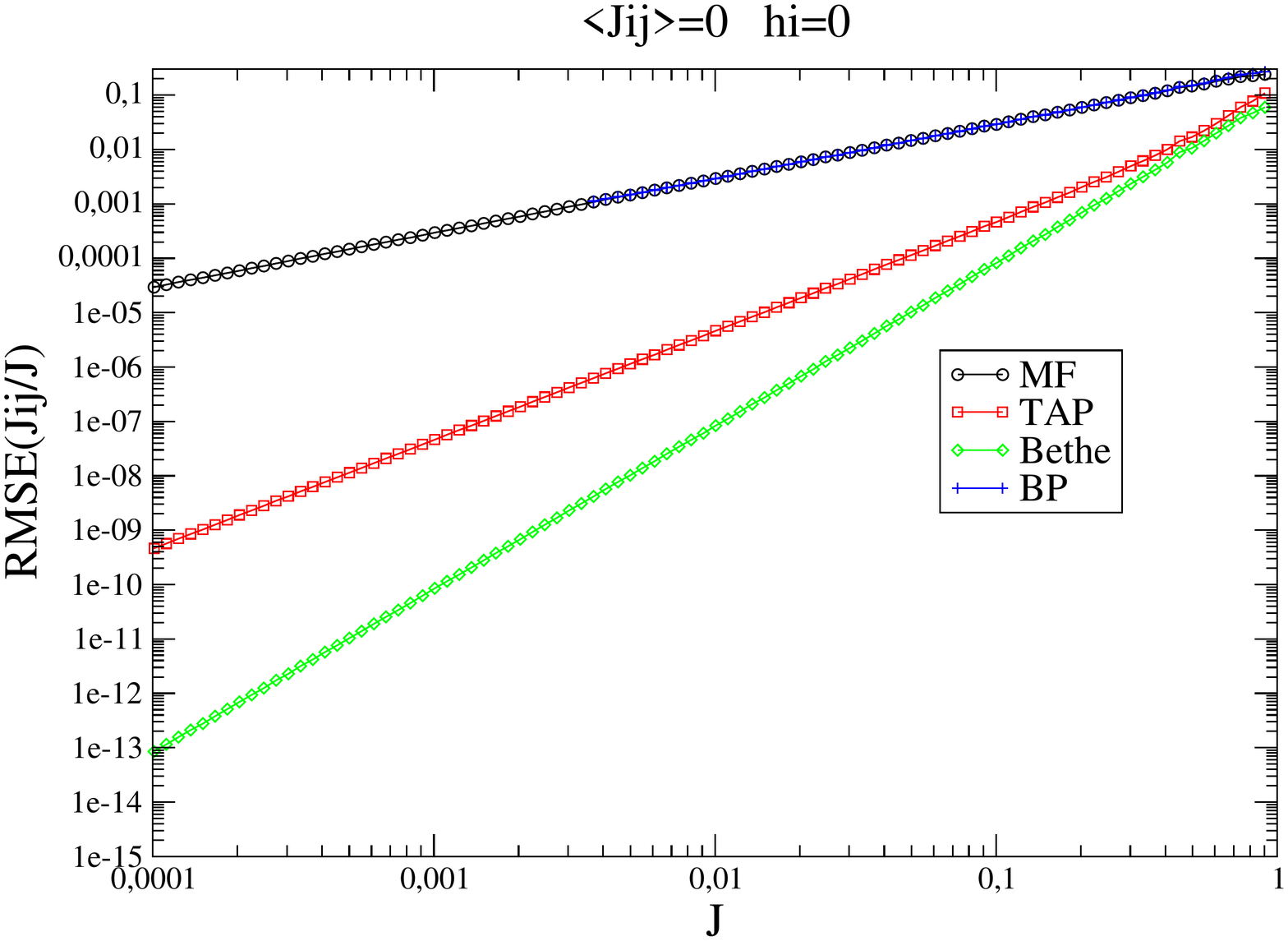}
\includegraphics*[width=0.49\columnwidth]{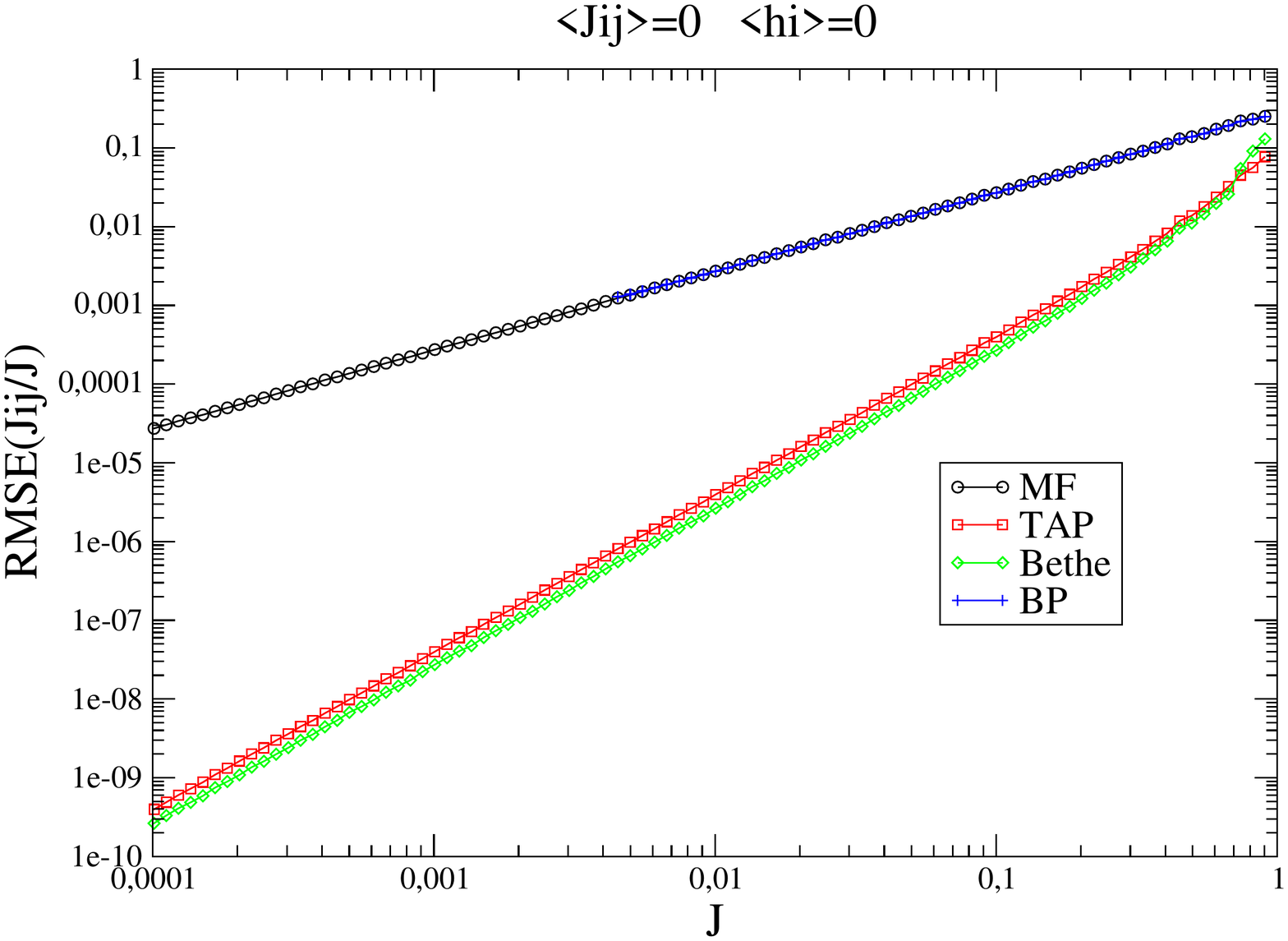}\\
\hfill(a)\hfill\hfill(b)\hfill\hbox{}\\[0.5cm]
\includegraphics*[width=0.49\columnwidth]{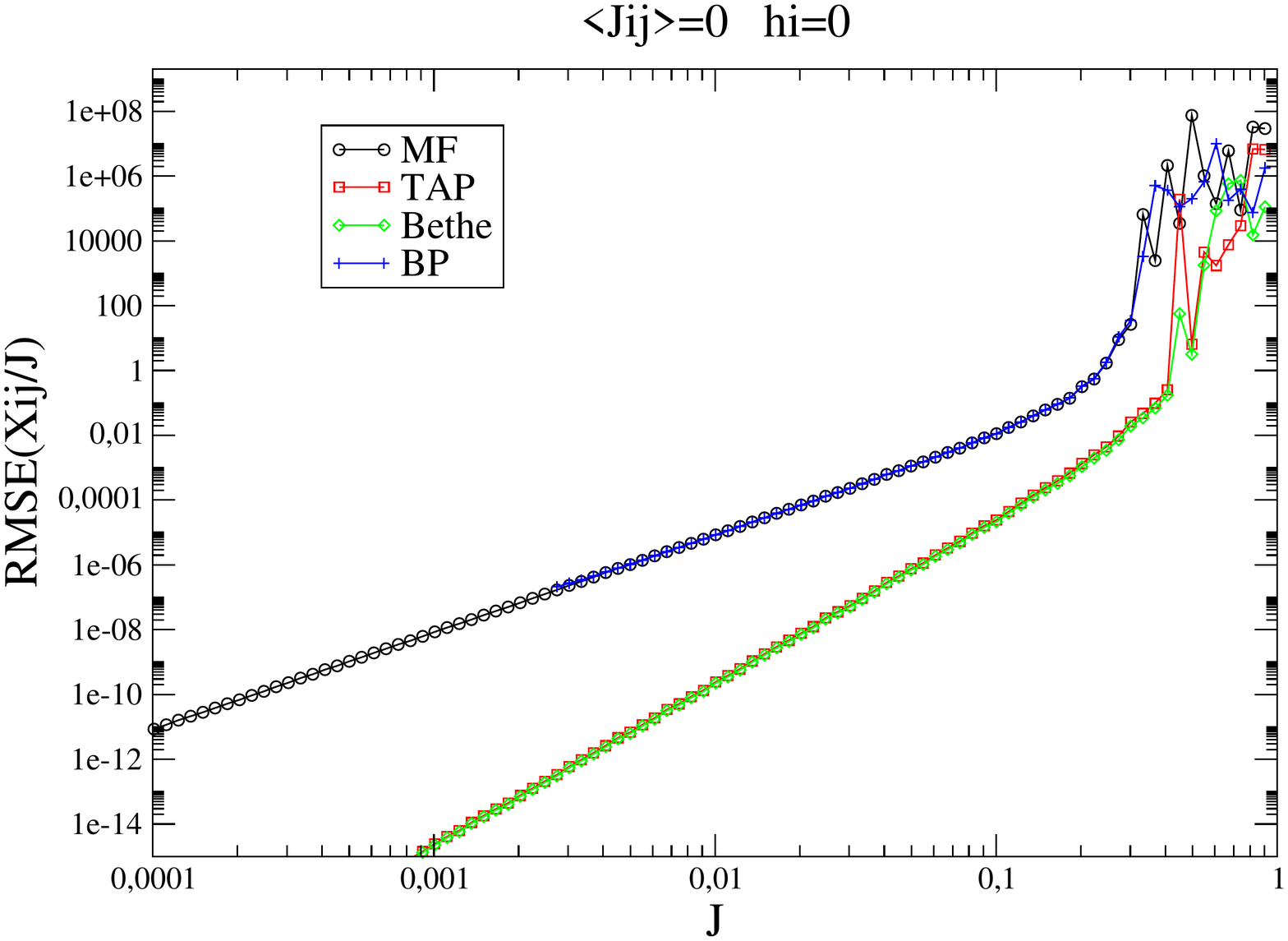}
\includegraphics*[width=0.49\columnwidth]{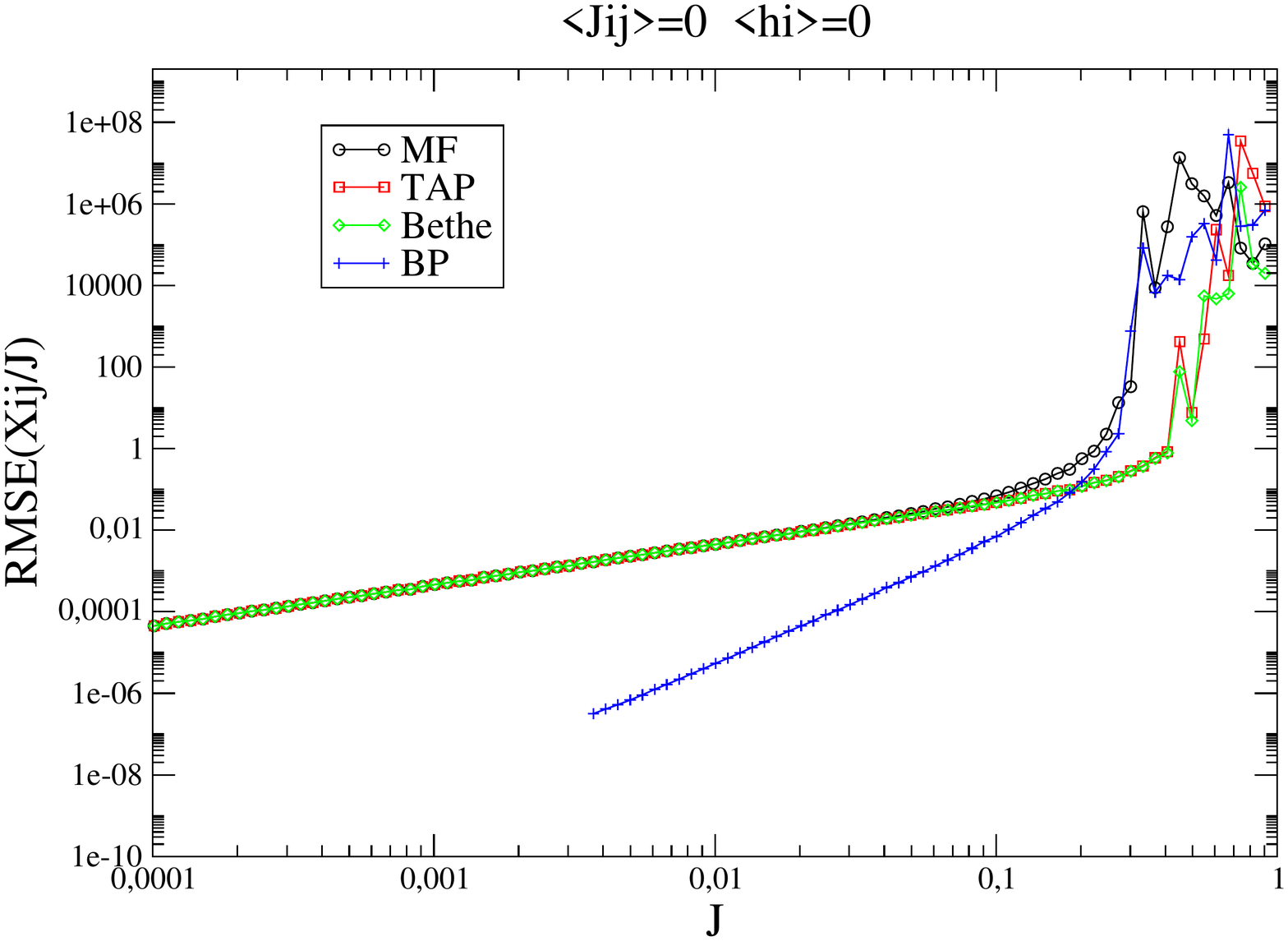}\\
\hfill(c)\hfill\hfill(d)\hfill\hbox{}\\
\caption{\label{fig:invising} Comparison between various approximate
  solutions to the inverse Ising problem. RMSE errors as a function of
  the temperature are plotted in (a) and (b) for the couplings $J_{ij}$,
  in (c) and (d) for the susceptibility matrix $\chi_{ij}$ obtained from
  the corresponding BP fixed point. Local fields $h_i$ are zero in (a)
  and (c) and finite but zero in average in (b) and (d).}
\end{figure}
The quality of the solution can then be assessed directly by comparing 
the couplings $J_{ij}$ found with the actual ones. Figure~\ref{fig:invising} 
are obtained by generating at random $10^3$ Ising models  
of small size $N = 10$ either with no local fields ($h_i=0,\forall i=1\ldots N$)
or with random centered ones $h_i=U[0,1]-1/2$  
and with couplings $J_{ij} = \frac{J}{\sqrt{N/3}}(2*U[0,1]-1)$, centered with variance $J^2/N$,
 $J$ being the common rescaling factor corresponding to the inverse temperature.
A glassy transition is expected at $J=1$.
The couplings are then determined using  (\ref{eq:JMF}),
(\ref{eq:JTAP}), (\ref{eq:Jij}) and (\ref{eq:JBETHE}) respectively for the 
mean-field, TAP, BP and Bethe (equivalent to susceptibility propagation) solutions.
Figure~\ref{fig:invising}.a shows that the Bethe approximation yields the most precise
results in absence of local fields while it is equivalent to TAP when a local field 
is present as shown on Figure~\ref{fig:invising}.b.
Since we want to use these methods in conjunction with BP we have also compared 
the BP-susceptibilities they deliver. To do that, we simply run BP to get a set 
of belief and co-beliefs in conjunction with equation (\ref{eq:invchis}) which 
after inversion yields a susceptibility matrix to be compared with 
the exact ones. The comparison  shown on Figure~\ref{fig:invising}.c
indicates that Bethe and TAP yield the best results in absence of local field,
but are less robust when compared to the more naive BP method  when 
local fields are present as seen on Figure~\ref{fig:invising}.d. This is due to 
the fact that BP delivers exact beliefs when model (\ref{eq:Jij},\ref{eq:hi})
is used, which is not necessarily the case for other methods when the local fields
are non-vanishing. It is actually not a problem of accuracy but of BP compatibility 
which is raised by this plot.

\paragraph{Sparse inverse models}
Let us now test the IPS based approach proposed in Section~\ref{sec:onelink} to build a model link by link
for comparison with more conventional optimization schema based on $L_0$ and $L_1$ penalizations. 
\begin{figure}
\centering
\resizebox{0.48\columnwidth}{!}{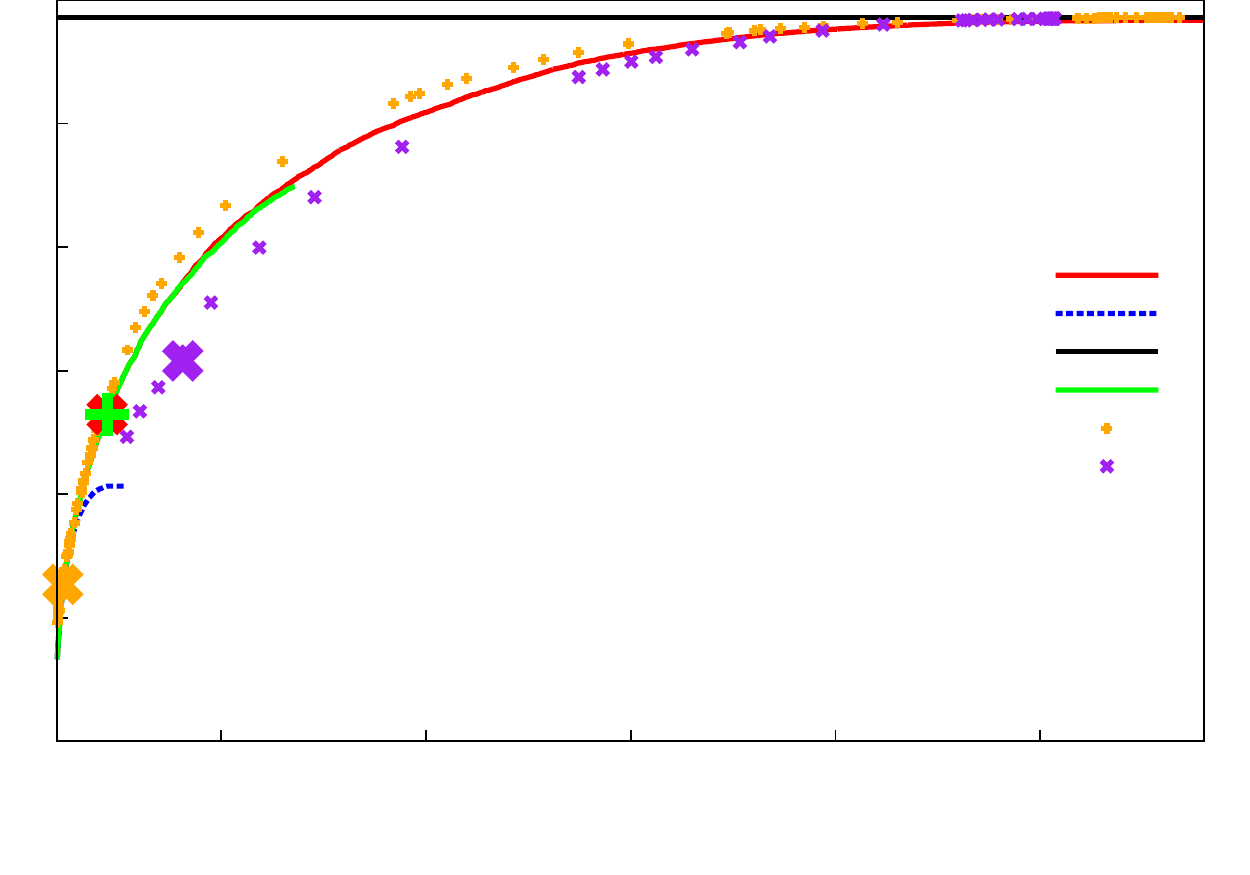}\hfill\resizebox{0.48\columnwidth}{!}{
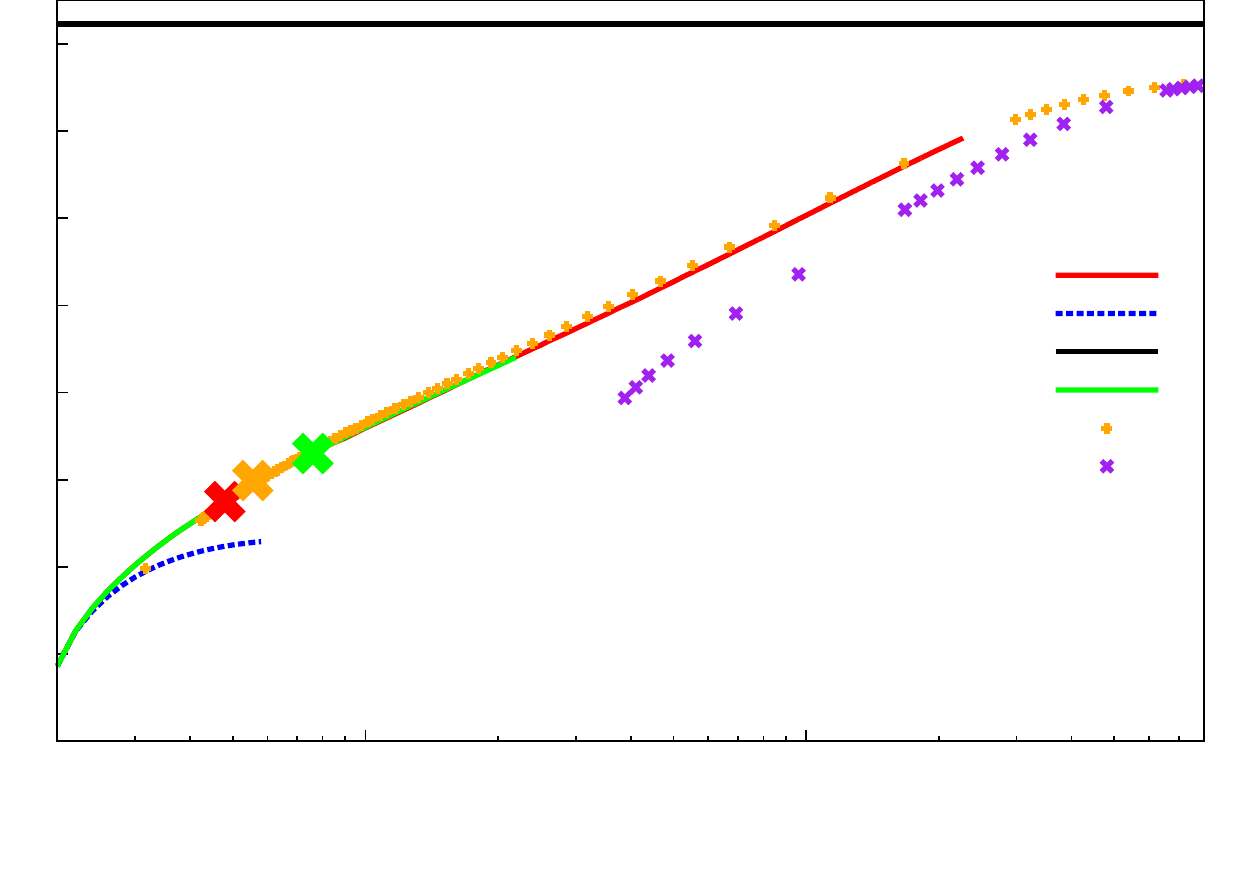}
\caption{\label{fig:invGauss}
Comparison of the greedy information based MRF inference method with $L_0$ and $L_1$ norm penalized
optimizations. (a) corresponds to the Sioux-Fall data of size $N=60$. (b) corresponds to the IAU data of size $N=1000$. $<K>$ is 
the ratio of links to nodes $N$. The log likelihood on the $y$ axes is unnormalized and corresponds to (\ref{def:LLcov}).
The reference (black legend) is the log likelihood given by the full inverse covariance matrix.
Different versions of IPS are tested with respect to graph constraints explained in the text 
(plain greedy, WS, $\rho({\cal R})<1$). The end of GaBP compatibility for each algorithm is indicated by $\times$'s.}
\end{figure}

We show the results of tests only for the Gaussian case where the link modification can be 
treated exactly. In the Ising case the method can be used only marginally to propose 
little correction on the maximum spanning tree or any other sparse model. 
In general we cannot expect the method to be able to compete with the inverse susceptibility propagation
schema i.e. what we call here the Bethe inverse model (\ref{eq:JBETHE}). The reason is 
that the LL gain given by one link is more costly to assess than in the Gaussian case
and it is also only approximate. So the stability of the schema is more difficult 
to control when many links have to be added because the condition of the validity of the 
Bethe approximation are not controlled without paying an additional computational price.
For the Gaussian case instead the situation is much more favorable because the gain can be computed exactly 
with low computational cost even when the graph is dense. The test we show on Figure~\ref{fig:invGauss}
are done on simulated data produced by the traffic simulator METROPOLIS~\cite{dPMa}, 
our original motivation for this work being related to traffic inference~\cite{FuLaFo,FuHaLa}.
The first set corresponds to a small traffic network called Sioux-Fall consisting of 72 links, 
from which we extract the $N=60$ most varying ones (the other one being mostly idle).
The second set (IAU) is obtained for a large scale albeit simplified network of the Paris agglomeration of size
13626 links, out of which  we extracted a selection of the $N=1000$ most varying ones.
Each sample data is a $N$-dimensional vector of observed travel times $\{\hat t_i,i=1\ldots N\}$, 
giving a snapshot of the network at a given time in the 
day. The total number of samples is $S=3600$ for Sioux-Falls and $S=7152$ for IAU, obtained by 
generating many days of various traffic scenarios. Then for each link the travel time distribution 
is far from being Gaussian, having heavy tails in particular. So to deal with normal variables (when taken individually)
we make the following standard transformation:
\begin{equation}\label{def:Gaencode}
y_i = F_{Gauss}^{-1}\hat F_i(t_i),\qquad\forall i=1\ldots N
\end{equation}
which map the travel time $t_i$ to a genuine Gaussian variable $y_i$, 
where $\hat F_i$ and $F_{Gauss}$ are respectively the empirical cdf of $t_i$ and of 
a centered normal variable. The input of the different algorithms under study is then 
the covariance matrix $\cov(y_i,y_j)$. This mapping will actually be important in the next 
section when using the devised MRF for inference tasks.

Figure~\ref{fig:invGauss} displays the comparison between various methods.
Performances of the greedy method are  comparable to the $L_0$ penalized optimization.

To generate one solution both methods are comparable also in term of computational cost, 
but the greedy is faster in very sparse regime, and since it is incremental, it generate
a full Pareto subset for the cost of one solution. On this figure we see also that the
$L_1$ method is simply not adapted to this problem. From the
figures, we can see that the estimated inverse matrix derived based on
$L_1$ norm penalty needs distinctively more non-zero entries to
achieve similar log-likelihood level as the $L_0$ penalty, indicating
its failure of discovering the underlying sparse structure, the thresholding of small 
non-zero entries being harmful w.r.t. positive definiteness. The reason might be 
that is adapted to situations where a genuine sparse structure exists, which is not the 
case in the present data. 

As expected WS is a
very strict constraint and stops to early with very sparse low likelihood models. Relaxing this
constraints into $\rho(R') < 1$ yields better GaBP compatibility, but provides no
guarantee about it. The justification about the constraints $\rho(R') < 1$ is based on the convergence
of the series $\sum_k R^k$ \cite{MaJoWi}. 

\paragraph{Dual weight propagation for Inverse Ising:}
we have performed some numerical checks of the DWP equations presented in Section~\ref{sec:dwp}.
These experiments are done on a sparse bipartite random Ising model to  test the accuracy of 
the partition function estimation on loopy-dual graphs. To be able to compare with the true 
value of the partition function we have considered a sparse bipartite graph, with a reduced 
number ($10$) of variables on the top layer so that complete enumeration of these variables state
can be done. The links are chosen randomly with the constraint that the graph be connected and 
that the degree of bottom layer's variables do not vary by more than one unit. 
Couplings are independent centered random variables with absolute mean $J$.

For each experimental point, the cycle basis is chosen with a simple greedy heuristic, starting first with the fundamental cycles
of a random spanning tree; these cycles are then randomly mixed two by two by linear combinations, 
such as to reduce the mean size of the basic cycles; the procedure stops when a local minimum is obtained. 
Note indeed that the dual cyclomatic number is given by 
\[
C(\G^\star) = \sum_{c=1}^{C(\G)}d^\star(c)-C(\G)-\vert\E\vert+k(\G^\star),
\] 
with $d^\star(c)$ the degree of a cycle node in the dual graph, i.e. its size in terms of edges,
assuming that no two edges are gathered to form a single factor. So reducing the mean size of basic cycles
contributes to reduce the cyclomatic number $C(\G^\star)$ of the dual graph.

\begin{figure}
\centering
\includegraphics[width=0.49\textwidth]{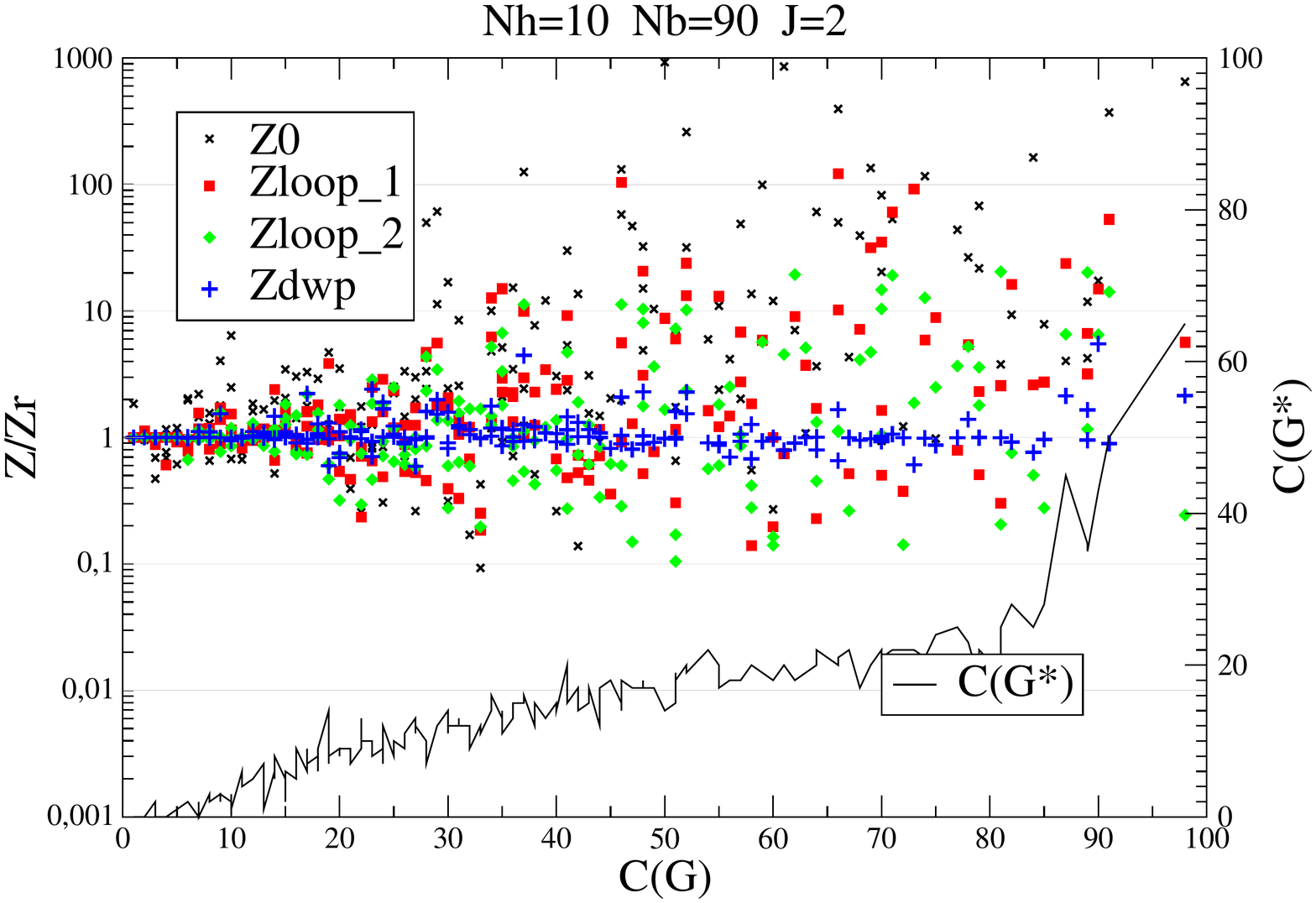}
\includegraphics[width=0.49\textwidth]{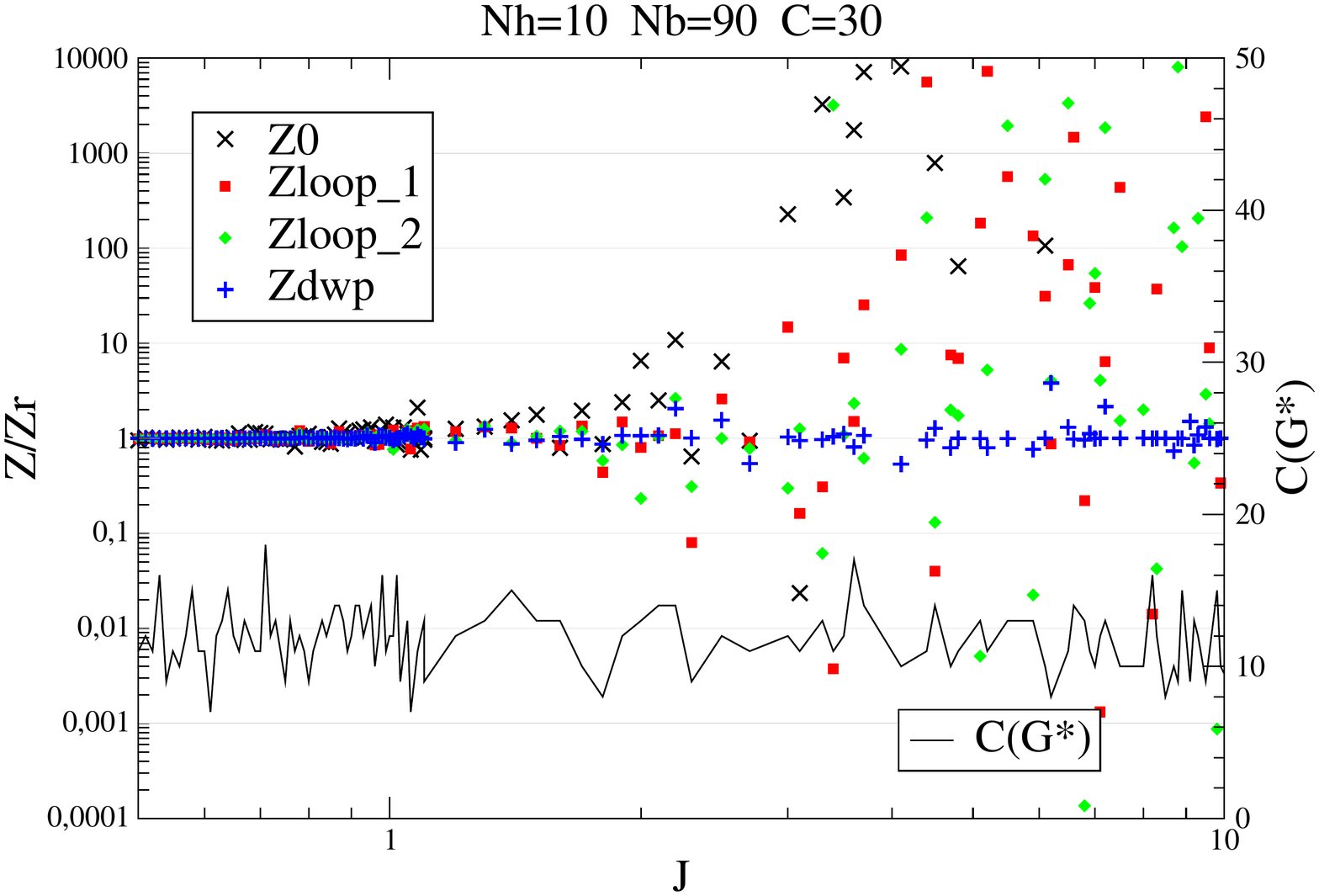}\\[0.2cm]
(a)\hspace{6cm}(b)
\includegraphics[width=0.49\textwidth]{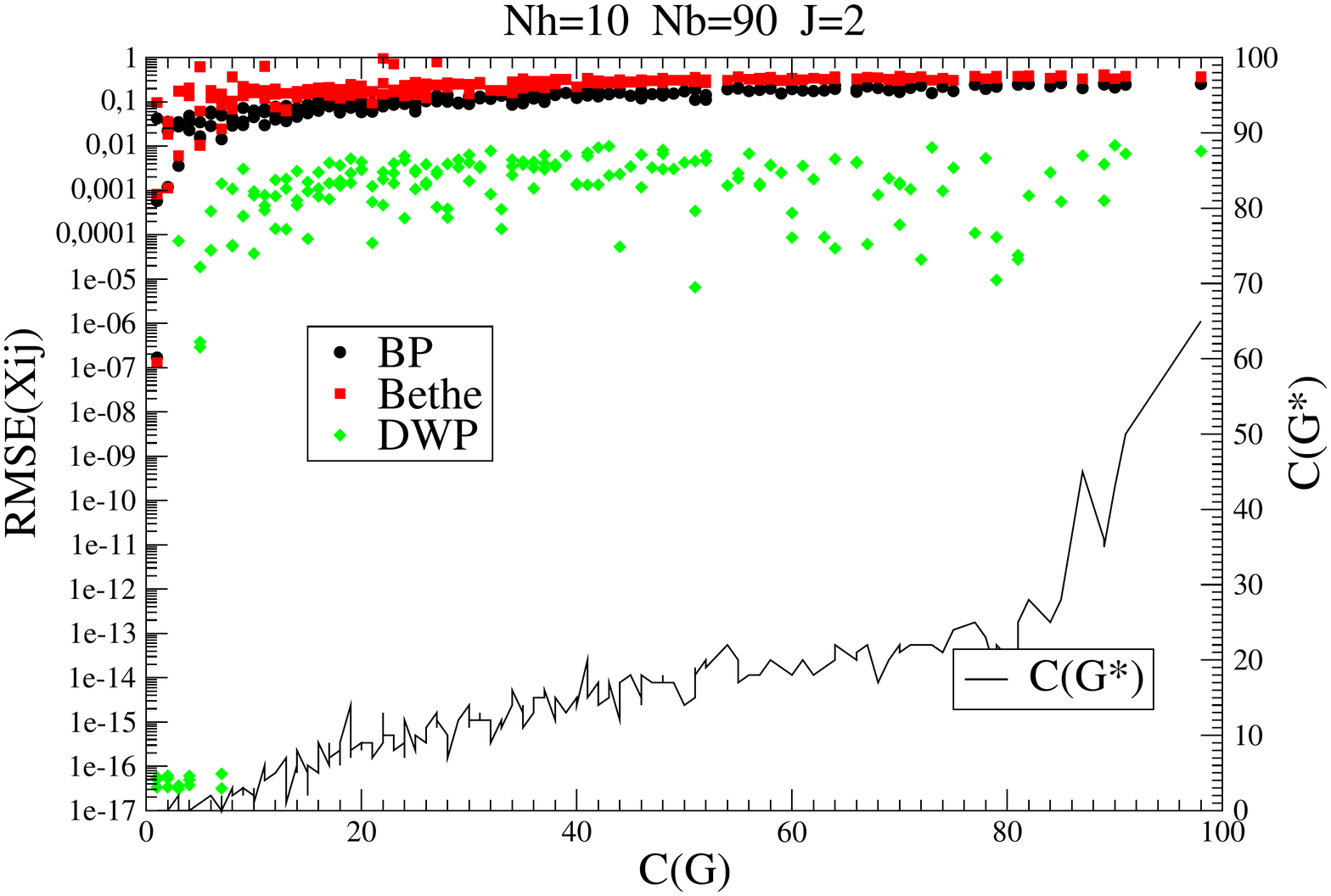}
\includegraphics[width=0.49\textwidth]{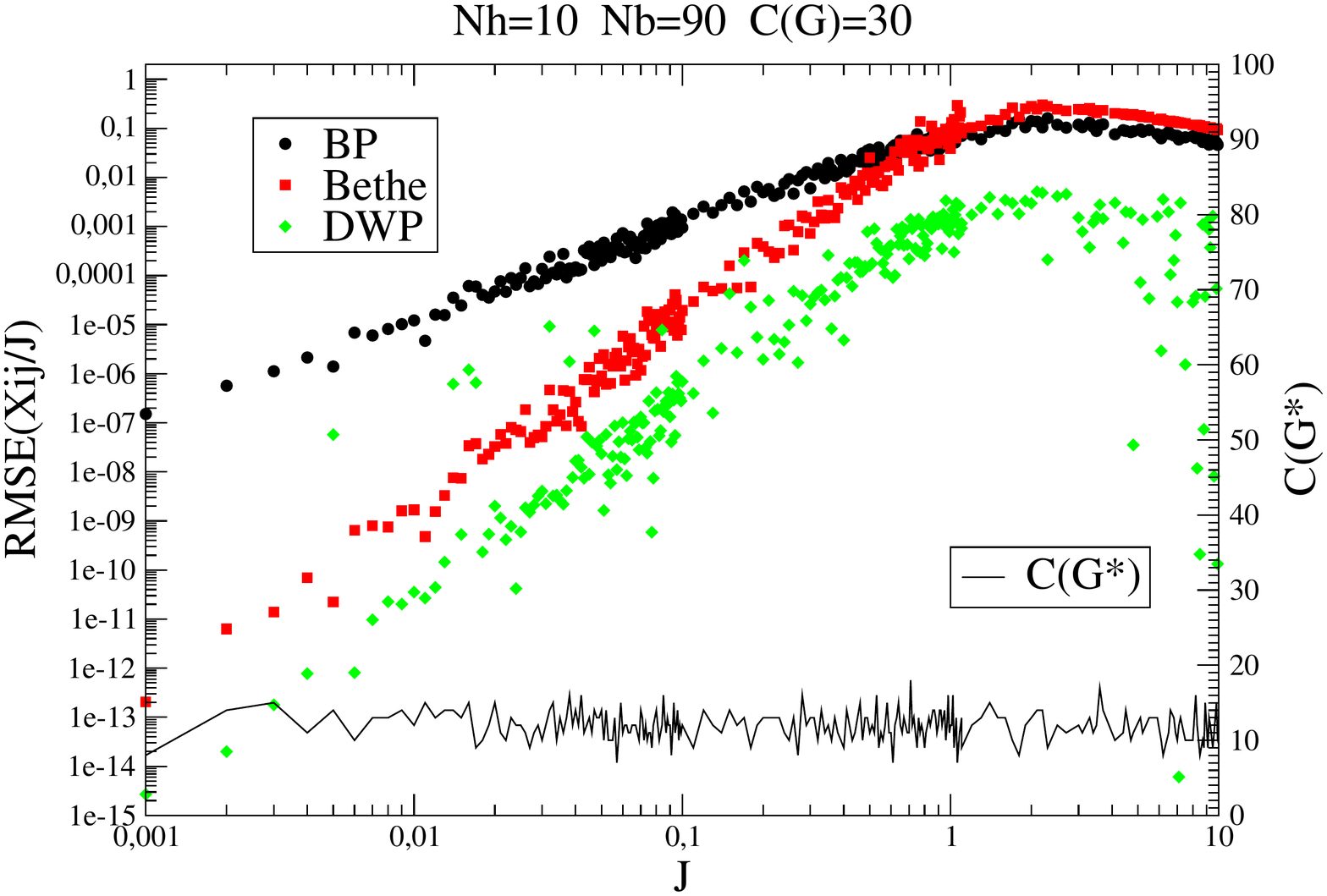}\\[0.2cm]
(c)\hspace{6cm}(d)
\includegraphics[width=0.49\textwidth]{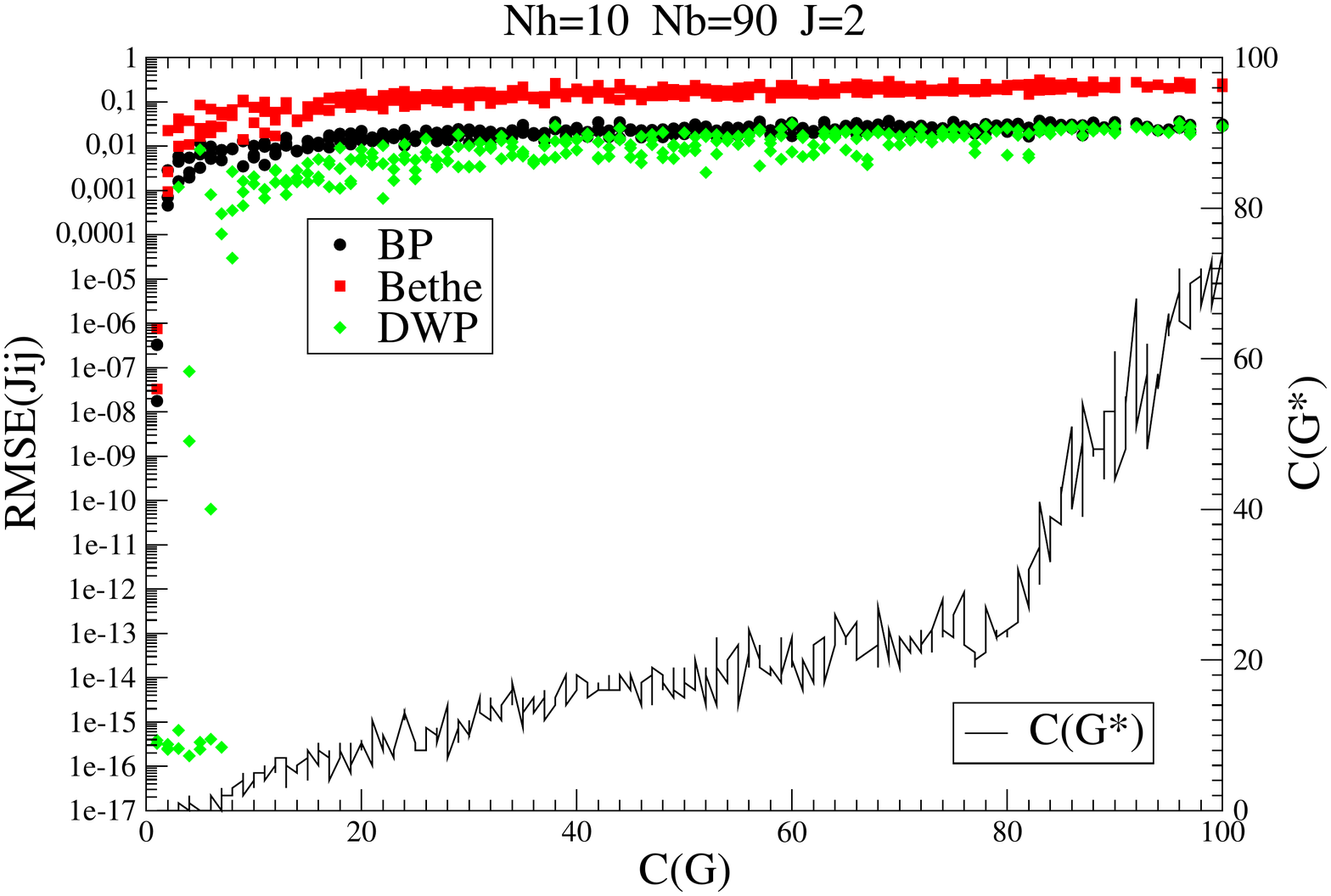}
\includegraphics[width=0.49\textwidth]{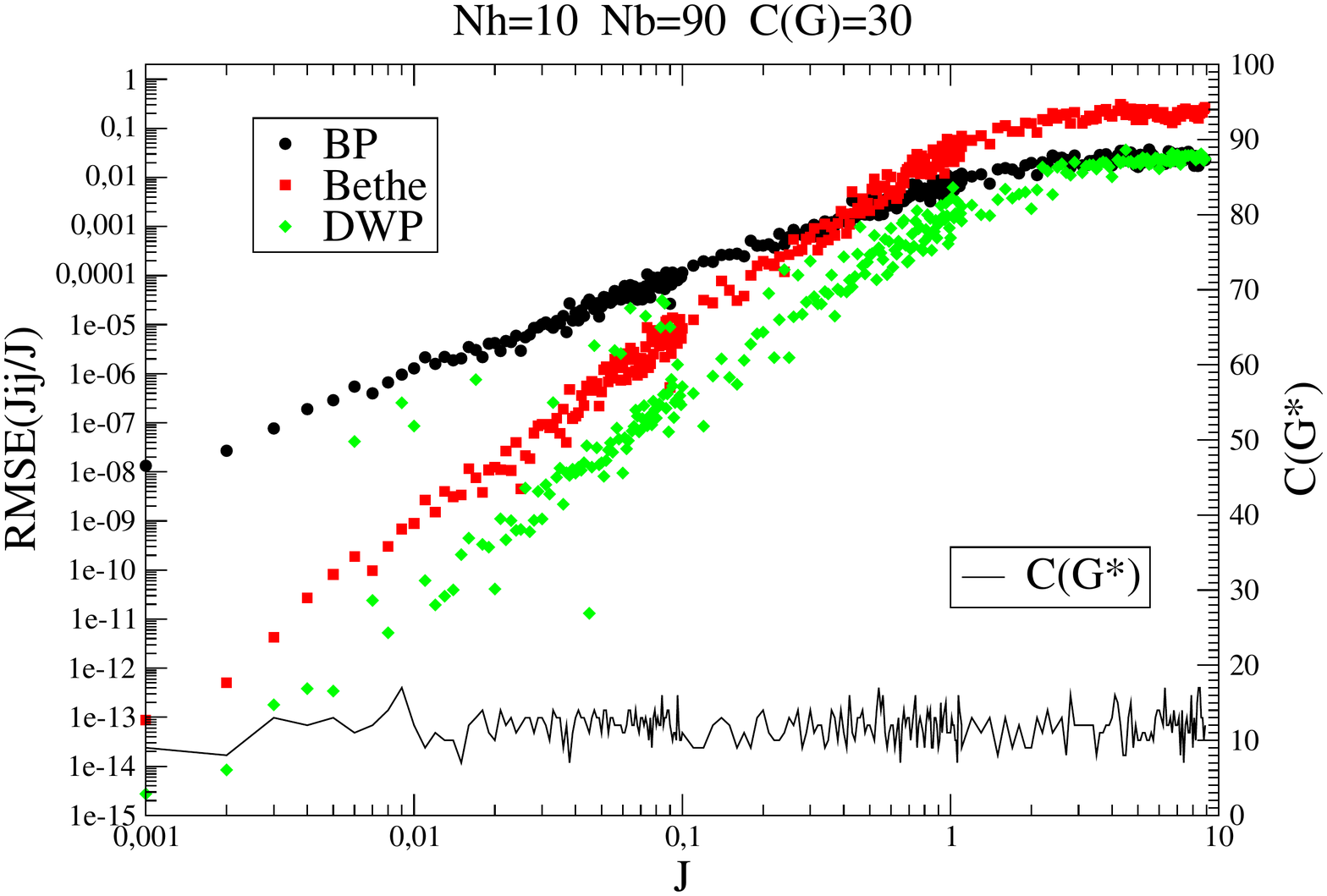}\\[0.2cm]
(e)\hspace{6cm}(f)
\caption{\label{fig:bipartite}Partition function (a) and (b), 
error on the susceptibility coefficients (c) and (d) and 
on the couplings (e) and (f) for of a sparse bipartite random Ising model 
with $10$ variables on the top layer and $90$ on the bottom one. 
On (a), (c) and (e) the number of
fundamental cycles in the primal graph is varied with fixed mean absolute 
coupling $J=2$. On (b), (d) and (f) the coupling is varied with fixed number of primary cycles $C(\G)=30$. 
Dual cyclomatic number $C(\G^\star)$ is also plotted. The experiment is repeated $3$ times for 
each value. Points are plotted only when DWP converges. Fluctuations correspond to different 
choice of the primal cycle basis. On (a) and (b)
$Z_r$ is the reference exact partition function. The value 
$Z_{dwp}$ obtained by dual weight propagation (DWP) is compared to  $Z_0$, $Z_1$ and $Z_2$ corresponding 
respectively to no loop, independent cycles and pair of cycles approximations. 
On (b), (c), (d) and (e) errors on the values 
obtained by dual weight propagation (DWP) are compared to bare one (BP) and to the error made by 
the Bethe estimation.} 	
\end{figure}

Figure~\ref{fig:bipartite} show results concerning the partition function estimation. 
Different levels of approximation in loop contributions given in Section~\ref{sec:dwp} are 
compared to the one obtained with DWP, when varying either the primal cyclomatic number $C(\G)$ or the mean coupling $J$.
The convergence and the results delivered by DWP are very sensitive to the quality of the cycle basis which is used, 
and this is reflected on  Figure~\ref{fig:bipartite}, where
the fluctuations that we see are mainly due the difference between cycles basis. 
Nevertheless, we can see on these plots that when it converges, DWP delivers rather satisfactory values, 
very often close to the true one, and never beyond a factor of $5$, which represent a few percent of missing information
for a system of size $100$. As expected results becomes very accurate in the low temperature regime (high value of $J$)
as seen on Figure~\ref{fig:bipartite}.b. even though convergence of DWP seems more delicate in this regime.
On the same figure, the use of DWP for the IIP, 
is illustrated on the same bipartite sparse model. Figure~\ref{fig:bipartite}.c and d
shows that the precision on susceptibility coefficients corresponding to edges of $\G$ when the couplings
$J_{ij}$ are given. The comparison is made with the bare BP approximation ($\chi_{ij} = \tanh(J_{ij})$) and 
the Bethe one (equivalent to susceptibility propagation) explained at the beginning of
this section and based on (\ref{eq:invchis}). As seen, the precision can be increased by several orders of 
magnitude both a low and large temperature with relatively loopy graphs. We can expect this 
to be reflected in the quality of the solutions of the IIP problem delivered by DWP. 
As seen on Figure~\ref{fig:bipartite}.e and f, performing a gradient descent on the log likelihood, 
which in principle is strictly valid in absence of dual loop, yields comparatively rather deceptive improvements 
at very low temperature (high $J$),
when compared again with direct BP and Bethe based methods, except when $C(\G^\star)$ is small enough.
Beyond possible numerical problems, this is probably possibly due to an insufficient 
convergence rate of DWP during the gradient descent, which has then 
to be interrupt to far away from the good solution, when $C(\G^\star)$ is to high. This problem should be 
addressed in future work. 

\begin{figure}[ht]
\centering
\includegraphics*[width=0.49\columnwidth]{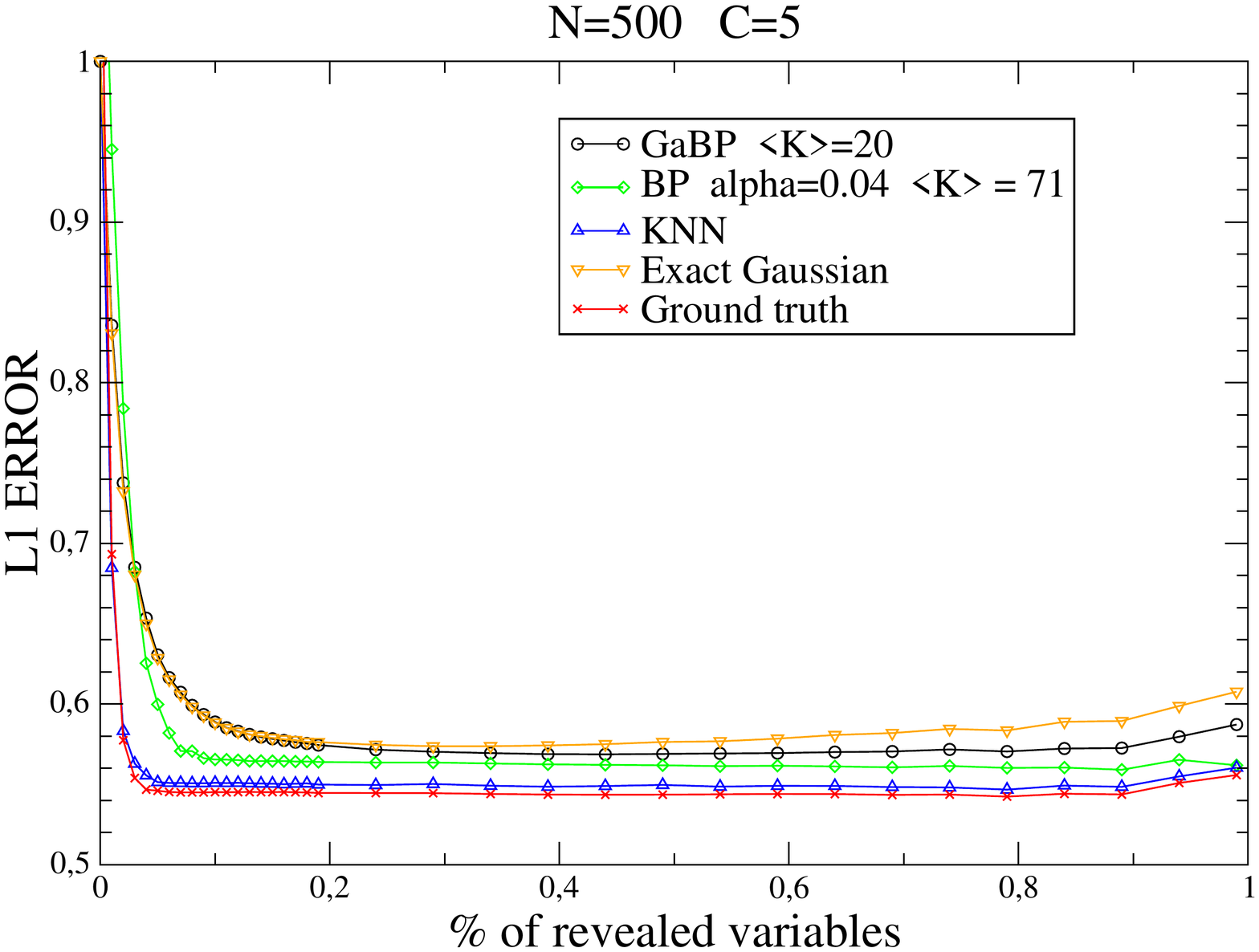}
\includegraphics*[width=0.49\columnwidth]{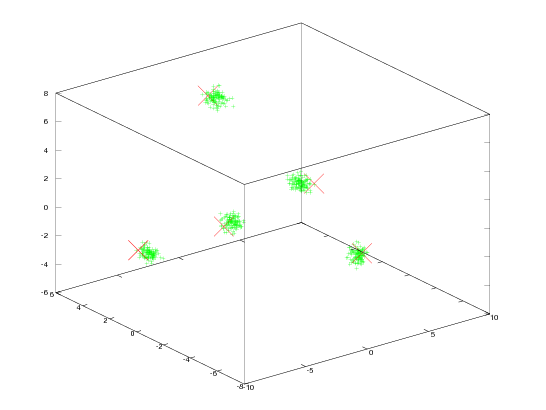}
\hspace{4cm}(a)\hspace{4cm}(b)
\caption{\label{fig:multimodes}Comparison of decimation curves in the case of a multi-modal distribution with
five cluster for  $N=500$ variables (a). Projection of the dataset in the $3$-d dominant PCA space
along with the corresponding BP fixed points projections obtained with the Ising model.}
\end{figure}
\paragraph{Inverse models for inference:}
we turn now to experiments related to the original motivation of this work, which is to use calibrated model for some
inference tasks. The experiments goes as follows: we have an historical data set consisting 
of a certain number of samples, each one being a  $N$-dimensional variable vector, say travel times, 
which serves to build these models~\footnote{In fact the pairwise MRF models exploit only pairwise 
observations but for sake of comparison with a \textsc{knn} predictor we generate complete historical sample data.}.
Given a sample test data we want to infer the $(1-\rho)N$ hidden variables when a certain fraction $\rho$ 
of the variables are revealed. In practice we proceed gradually on each test sample by revealing one by one the 
variables in a random order and plot as a function of $\rho$ the $L_1$ error made by 
the inference model on the hidden variables. 
Both for Ising and Gaussian MRF, the inference is not performed in the original variable space, but 
in an associated one obtained through a mapping (a traffic index) using the empirical cumulative distribution of each variable. 
For the Gaussian model the inference is performed in the index space defined previously by (\ref{def:Gaencode}).
For the Ising models we have studied a variety of possible mapping~\cite{MaLaFu} in order to associate
a binary variable to a real one such that a belief associated to a binary state can be converted back into 
a travel time prediction. Without entering into the details (see~\cite{MaLaFu} for details), 
to define in practice this binary state $\sigma_i$,
either we make use of the median value $x_i^{1/2} = F_i^{-1}(1/2)$ in the distribution of $x_i$ for all $i=1\ldots N$:
\[
\sigma_i  = \ind{x_i>x_i^{1/2}}\qquad (i).
\]
Either we perform a soft mapping using the cdf:
\[
P(\sigma_i=1) = \hat F_i(x_i)\qquad (ii),
\]
the last one having the advantage of being functionally invertible if $\hat F_i^{-1}$ is defined, 
while the former one being inverted using Bayes rule. The data we are considering are ``low temperature''
data in the sense that correlations are too strong for an Ising model with one single fixed 
point. This is reflected in the fact that none of the basic methods given in the Section~\ref{sec:prelim} 
is working. To overcome this we use a simple heuristic which consists in to add a parameter  $\alpha\in[0,1]$ 
in the BP model like e.g. in (\ref{eq:bpalpha}) or to multiply the $J_{ij}$ by $\alpha$ 
for the MF, TAP and Bethe models, the local field being consequently modified owing to their 
dependency on the $J_{ij}$.
Concerning the factor-graph we have considered various graph selection procedures. All are based on 
the mutual information given empirically between variables. A global/local threshold can be 
used to construct the graph, the parameter being the mean/local connectivity $K$; 
the MST can be used conveniently as a backbone and additional links are obtained through
the thresholding selection procedures. These two parameter $\alpha$ and $K$ are 
calibrated such as to optimize the performance for each type of model so that fair comparisons can 
be made afterward.
\begin{figure}[ht]
\centering
\includegraphics*[width=0.49\columnwidth]{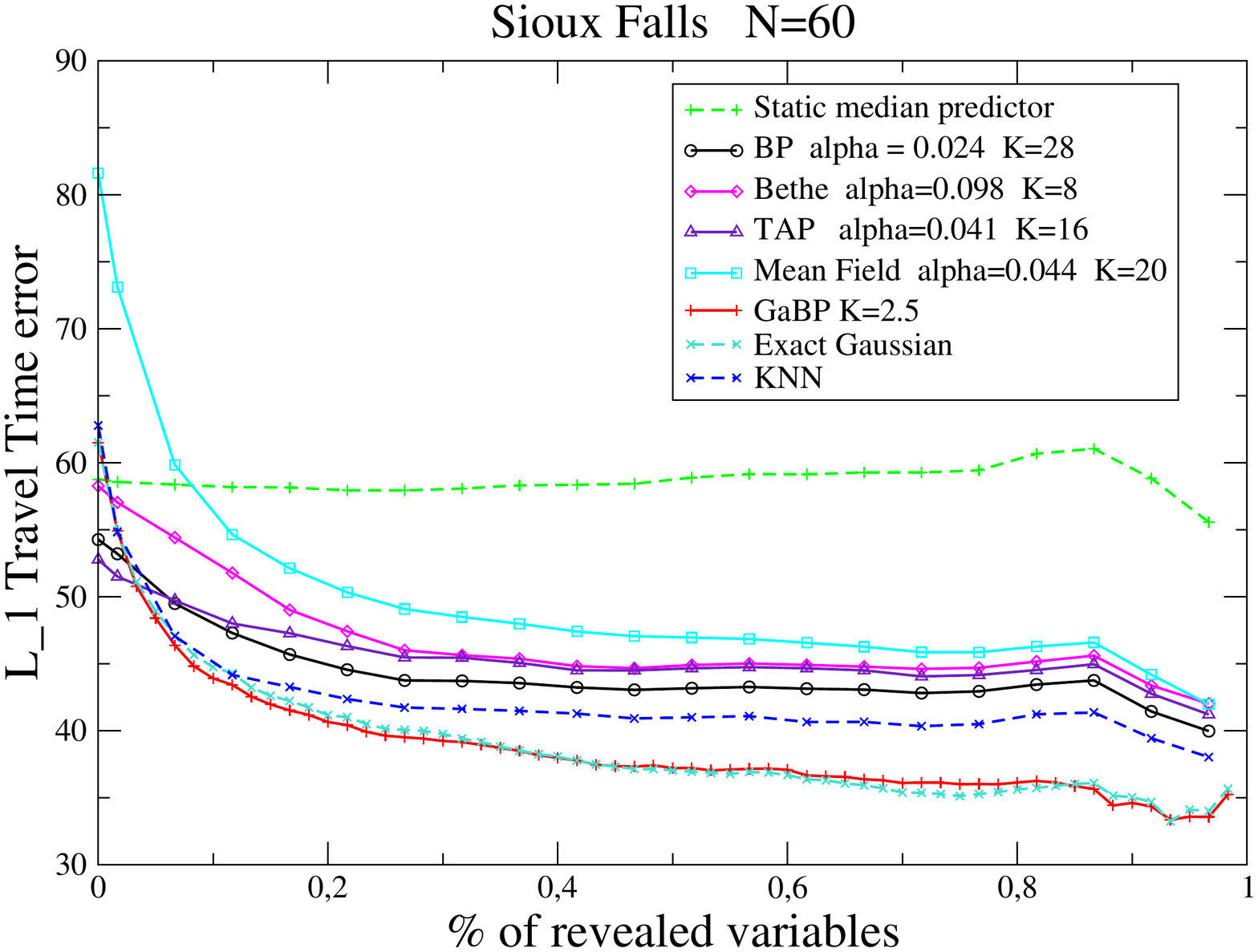}
\includegraphics*[width=0.49\columnwidth]{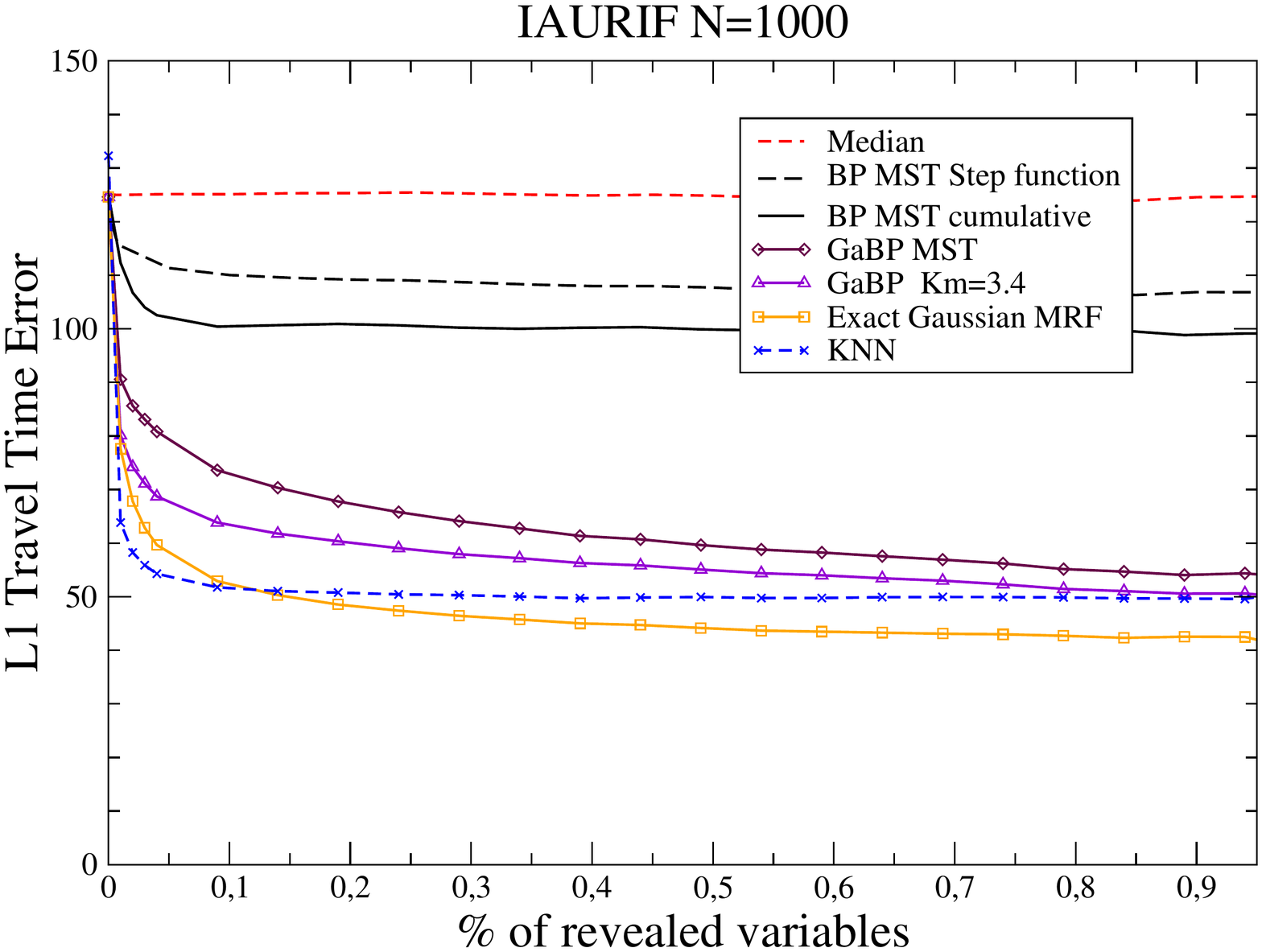}
\hspace{3cm}(a)\hspace{6cm}(b)
\caption{\label{fig:metrodecim}Comparison of decimation curves between various MRF for Sioux-Falls data (a) and 
IAU data (b)}
\end{figure}
One important difference between the Ising model and the Gaussian one is that multiple fixed points 
may show up in the Ising case while only a single one, stable or not stable, is present in the Gaussian case.
This can be an advantage in favor of the Ising model when the data have well separated clusters. 
Figure~\ref{fig:multimodes} illustrates this point. The data are sampled from a distribution containing $5$
modes, each one being a product form over $N$ random bimodal distributions attached to each link.
On Figure~\ref{fig:multimodes}.a which displays the error as a function of the fraction of revealed variables
we see that the Ising model obtained with (\ref{eq:bpalpha}), encoded with the median value (i),  
gives better prediction than the exact Gaussian model or the approximated GaBP
compatible one. Indeed  Figure~\ref{fig:multimodes}.b shows a projection of the data 
in the most relevant $3$-d PCA space along with the projected position of BP  
fixed points (given by their sets of beliefs) delivered by the Ising model.
As we see, the model is able to attach one BP fixed point 
to each component of the distribution. Ideally we would like a perfect calibration
of the Ising model in order that these fixed points be located at the center of each cluster.
The method proposed in Section~\ref{sec:perturbation} could help to do this, but has not been 
implemented yet.
On Figure~\ref{fig:multimodes}.a we see also that the \textsc{knn} predictor performs optimally in this case, 
since the error curve coincides exactly with the one given by the hidden generative model of the data (ground truth).
Figure~\ref{fig:metrodecim} shows how the different models compare on the data generated by the traffic 
simulator. On the Sioux-Falls network, the Gaussian model gives the best results, and a sparse version 
obtained with the greedy algorithm of section~\ref{sec:greedyalgo} reach the same level of performance
and outperforms \textsc{knn}. The best Ising model is obtained with the (\ref{eq:bpalpha}) with type (ii)
encoding. For IAU  the full Gaussian model is also competitive w.r.t \textsc{knn}, but the best 
sparse GaBP model is not quite able to follow. In fact the correlations are quite high
in this data, which explain why the best Ising model shows very poor performance. The best 
Ising model in that case corresponds to the plain BP model with type (ii) encoding and MST graph.   

\section{Conclusion}
This paper is based on the observation that  in many cases, the Bethe approximation can be
a good starting point for building inverse models from data observations. We have developed
here three different ways of perturbing such a mean-field solution. One of them (see Section~\ref{sec:onelink}),
based on IPS, is valid both for binary and Gaussian 
variables, and leads to an efficient algorithm in the Gaussian case to 
generated sparse approximation models compatible with BP. The additional 
requirement that the model be compatible with BP for large scale applications 
discards dense models and simplifies in a way the search space on model selection.
This method should be also used in complementary to the two others (natural gradient and 
DWP) for the Ising case, because the susceptibility evaluation step which was considered too expensive
and not precise enough with the Bethe method, can be done efficiently in 
$O(N^2)$ thanks to DWP in the very sparse regime. Hence, constructing the graph with IIP, while 
keeping it DWP-compatible could lead to interesting approximate solutions.
This point will certainly be considered for future work. Concerning 
DWP, various possible improvements will be addressed in future studies, 
concerning respectively the cycle basis choice, the DWP update scheduling itself and the gradient descent, 
to render this method efficient for a broader class of problems as the simple one considered here.      
Its extension to problems will local fields should also be addressed experimentally.

\paragraph{Acknowledgments} This work was supported by the French National Research Agency (ANR) grant N° ANR-08-SYSC-017.

\bibliographystyle{acm}
\bibliography{invising}

\end{document}